\documentclass[final,leqno]{siamltex}
\usepackage{amsfonts}
\usepackage{amsmath}
\usepackage{amssymb}
\usepackage{bbm}
\usepackage{graphicx}
\usepackage{float}
\usepackage{subfigure}
\usepackage{booktabs}
\usepackage{setspace}
\usepackage[toc,page]{appendix}
\usepackage{enumitem}
\usepackage{comment}
\usepackage{multicol}
\usepackage{multirow}
\usepackage{color}
\usepackage[table]{xcolor}
\usepackage{url}

\newtheorem{assumption}[theorem]{Standing Assumption}
\newtheorem{example}[theorem]{Example}
\newtheorem{remark}[theorem]{Remark}
\newcommand{\al}{{\alpha_\lambda}}
\newcommand{\bl}{{\beta_\lambda}}

\newcommand{\xscol}{x^\star_{b,\lambda}}
\newcommand{\xscal}{x^\star_{f,\lambda}}

\usepackage{color}

\newcommand{\E}{\mathbb E}

\newcommand{\Tau}{\mathcal T}
\newcommand{\1}{\mathbbm{1}}
\newcommand{\ol}[1]{{\overline{#1}}}
\newcommand{\ul}[1]{{\underline{#1}}}

\title{Callable convertible bonds under liquidity constraints and hybrid priorities
\thanks{We thank both referees for the very helpful comments and suggestions. GL's research partially
supported by National Natural Science Foundation of China (No. 12171169) and
Laboratory of Mathematics for Nonlinear Science, Fudan University.}}

\author{David Hobson,\ \ Gechun Liang,\ \ Edward Wang\thanks{Department of Statistics, University of Warwick, Coventry, CV4 7AL, U.K.
Email adress: {\tt d.hobson@warwick.ac.uk;}\ \ {\tt g.liang@warwick.ac.uk;}\ \ {\tt
e.wang.5@warwick.ac.uk}}}

\begin{document}

\maketitle

\begin{abstract}
This paper investigates the callable convertible bond problem in the presence of a liquidity constraint modelled by Poisson signals. We assume that neither the bondholder nor the firm has absolute priority when they stop the game simultaneously, but instead, a proportion $m\in[0,1]$ of the bond is converted to the firm's stock and the rest is called by the firm. The paper thus generalizes the special case studied in [G. Liang and H. Sun, Dynkin games with Poisson random intervention times, SIAM Journal
on Control and Optimization, 57 (2019), pp. 2962–2991.] where the bondholder has priority ($m=1$), and presents a complete solution to the callable convertible bond problem with liquidity constraint. The callable convertible bond is an example of a Dynkin game, but falls outside the standard paradigm since the payoffs do not depend in an ordered way upon which agent stops the game. We show how to deal with this non-ordered situation by introducing a new technique which may be of interest in its own right, and then apply it to the bond problem.

\end{abstract}



\pagestyle{myheadings} \thispagestyle{plain} \markboth{}{Callable convertible bonds}

%%%%%%%%%%%%%%%%%%%%%%%%%%%%%%%%%%%%%%%%%%%%%%%%%%%%%%%%%%%
%%%%%%%%%%%%%%%%%%%%%%%%%%%%%%%%%%%%%%%%%%%%%%%%%%%%%%%%%%
\section{Introduction}

The holder of a perpetual bond receives a coupon from the firm indefinitely. If the bond is convertible, the bondholder has the additional opportunity to exchange the bond for a fixed number of units of the firm's stock at a moment of the bondholder's choosing. If the convertible bond is callable then the firm has the option to call the bond on payment of a fixed surrender price to the bondholder. The problem of pricing the callable convertible bond involves finding the value of the bond and the optimal stopping rules for conversion (by the bondholder) and call (by the firm). %{\bf References?}

The callable convertible bond is an example of a {zero-sum} Dynkin game. A Dynkin game is a game played by two players, each of whom chooses a stopping time in order to maximise their expected payoff. Reversing the sign of one of the objective functions we can think of one player as a maximiser and the other as a minimiser. The two players are then denoted by the \textsc{sup} player and the \textsc{inf} player. The game is zero-sum if in these new coordinates the objective functions of the two players are the same, in which case we may imagine that
the game involves a payment from one player to the other. The goal of the \textsc{inf} player is to minimise the payment made to the \textsc{sup} player and the goal of the \textsc{sup} player is to maximise this quantity. If {the \textsc{sup} player} is first to stop (at $\tau$) then the payment is $L_\tau$; if {the \textsc{inf} player} is first to stop (at $\sigma$) then the payment is $U_\sigma$; under a tie ($\tau = \sigma$) then the payment is $M_\tau$.
For stopping rules $\tau$ and $\sigma$, the expected value of the payment is $J= J(\tau,\sigma) = J^{U,M,L}(\tau,\sigma) = \E[ L_\tau \1_{\{\tau < \sigma\}} +U_\sigma \1_{\{\sigma < \tau\}} + M_\tau \1_{\{\tau = \sigma\}}]$. The objective of the \textsc{sup} player is to maximize $J$ whilst the objective of the \textsc{inf} player is to minimize $J$ and this leads to two problem values $\ol{v} = \inf_\sigma \sup_\tau J(\tau,\sigma)$ and $\ul{v}= \sup_\tau \inf_\sigma J(\tau,\sigma)$ (respectively the upper and lower value) depending on whether we take the perspective of the \textsc{inf} or the \textsc{sup} player. Trivially $\ul{v} \leq \ol{v}$: of great importance is whether $\ul{v}=\ol{v}$ in which case the Dynkin game is said to have a value $v=\ul{v}=\ol{v}$. See Cvitanic and Karatzas \cite{cvitanic1996backward} and Ekstr{\"o}m and Peskir \cite{Ekstrom2008SICON} for general treatments using the backward stochastic differential equation approach and the Markovian setup, respectively, and Kifer \cite{Kifer_2} for an extensive survey with applications to financial game options. One of the main ideas is to find a saddle point, i.e. a pair of stopping times $(\tau^*,\sigma^*)$ such that (i) $J(\tau,\sigma^*) \leq J(\tau^*,\sigma^*)$ for all $\tau$ and (ii) $J(\tau^*,\sigma^*) \leq J(\tau^*,\sigma)$ for all $\sigma$; then it is straightforward to show that the game has a value and $(\tau^*,\sigma^*)$ are optimal for the \textsc{sup} and the \textsc{inf} player, respectively.

Historically, the relative order of the payoff processes $L$, $U$ and $M$ in a Dynkin game has been important in proving that the game has a value.  Under the standard assumption $L\equiv M\leq U$, Bismut \cite{bismut1977probleme} proved that a Dynkin game has a value under the Mokobodzki condition which states that there exist two supermartingales whose difference lies between $L$ and $U$. The need for the Mokobodzki condition was later removed by Lepeltier and Maingueneau \cite{lepeltier1984jeu}. Nonetheless, the literature often assumes that the order condition $L\leq M\leq U$ holds. Under this order condition, Ekstr{\"o}m and Peskir \cite{Ekstrom2008SICON} proved the existence of game value and saddle point in a Markovian setup, given that the payoff processes have quasi-left continuous sample paths. See also Kifer \cite[Theorem 1]{Kifer_2} for a statement of the existence result under the order condition. Subsequently, the order condition was 
relaxed by Stettner \cite[Theroem 3]{Stettner1982}, and  
further relaxed by Merkulov \cite[Chapter 5]{merkulov2021value} who assumed that the \textit{generalised order condition} $L\wedge U\leq M\leq L\vee U$ holds.

When the order condition fails it may be that the {candidate} optimal strategy involves both agents wanting to stop at the same moment and then for a solution to exist we should allow agents to use randomised strategies, see Laraki and Solan~\cite{laraki2005value}, Rosenberg et al~\cite{rosenberg2001stopping} and Touzi and Vieille~\cite{touzi2002continuous}.
Randomised strategies allow players to `hide' their stopping time from the opposing player, thus players can avoid stopping simultaneously and there is no need to assume even the generalised order condition. In fact, provided that players are allowed to use randomised strategies, it was proved in \cite{laraki2005value}, \cite{rosenberg2001stopping} and \cite{touzi2002continuous} that the game has a value under fairly general conditions on the payoffs and no assumption on the relative order. One important result in this context is that the game is always terminated (i.e. at least one player chooses to stop) at or before the first moment $t$ that $L_t > U_t$, irrespective of whether randomised stopping times are allowed or not.

In general, the information structure is a crucial element in Dynkin games. See, for example, De Angelis et al~\cite{DeAngelis2021_a},  De Angelis et al~\cite{DeAngelis2021_b} for the treatment of asymmetric/incomplete information and Bayraktar and Yao~\cite{Bayraktar2017} for a robust version of a Dynkin game in which the players are ambiguous about their probability model.
Moving beyond the classical set-up in a different direction, Liang and Sun \cite{Liang_Sun} introduced a constrained Dynkin game in which the players' stopping strategies are constrained to be event times of an independent Poisson process. The constrained Dynkin game is an extension of the Dynkin game in the same way that the constrained optimal stopping problem of Dupuis and Wang \cite{dupuis2002optimal} %, see also \cite{Hobson}, \cite{HZ}, \cite{lempa2012optimal} and \cite{liang2015stochastic}
is an extension of a classical optimal stopping problem. Constrained optimal stopping problems, where the constraint is modelled by only allowing stopping at the event times of an independent Poisson process have also been considered by Lempa \cite{lempa2012optimal}, Liang \cite{liang2015stochastic}, Hobson and Zeng \cite{HZ}, Hobson \cite{Hobson} and Lempa and Saarinen \cite{lempa2023zero}.
When the problem involves a time-homogeneous payoff, a time-homogeneous diffusion process and an infinite time-horizon, the use of a Poisson process to determine the set of possible stopping times maximizes the tractability of the constrained problem.
%Then  Liang and Sun~\cite{Liang_Sun} proved existence of a constrained game value under the order condition $L=M \leq U$ on the payoffs.

The Dynkin game formulation of the perpetual callable convertible bond was first introduced in
S{\^\i}rbu et al \cite{si2004perpetual} (see also S{\^\i}rbu and Shreve \cite{si2006}
for the finite horizon counterpart). S{\^\i}rbu et al reduced the problem from a
Dynkin game to {two optimal stopping problems}, and discussed when call
precedes conversion and vice versa. A key element of the problem specification in \cite{si2004perpetual} is that the firm's stock price is calculated endogenously. Extensions of this convertible bond model include, for example,
Bielecki et al~\cite{bielecki2008arbitrage} which considered the
problem of the decomposition of a convertible bond into a bond
component and an option component,
Cr{\'e}pey and Rahal~\cite{crepey2011pricing} which studied the convertible bond with call
protection, and
Chen et al~\cite{chen2013nonzero} which added the tax benefit and bankruptcy
cost to the convertible bond.

In contrast to the aforementioned works on convertible bonds, Liang and Sun \cite{Liang_Sun} introduced a constraint on the players' stopping strategies by assuming that both bondholder and firm may only stop at times which are event times of an independent Poisson process. The idea is that the existence of liquidity constraint may restrict the players' abilities to stop at arbitrary stopping times. Liang and Sun \cite{Liang_Sun} assumes that the firm's share price is exogenous, and then the analysis is an extension of the constrained optimal stopping problem of Dupuis and Wang \cite{dupuis2002optimal}.
In the same way that the classical callable convertible problem is related to a classical Dynkin game, the bond problem with liquidity constraint can be related to a Dynkin game with liquidity constraint (i.e. a condition that the stopping times take values in the event times of a Poisson process).

One of the difficulties in the callable convertible bond problem is that, although it can be recast in the standard form of a Dynkin game, the payoff process $U$ (corresponding to calling the bond at the surrender price) does not necessarily dominate the payoff process $L$ (corresponding to converting the bond) --- when the stock price is large the conversion value is larger than the surrender price. As a result the general existence results for Dynkin games (see Lepeltier and Maingueneau \cite{lepeltier1984jeu} or Kifer \cite{Kifer_2}) and especially the existence theorem in the Poisson constrained case (Liang and Sun \cite[Theorem 2.3]{Liang_Sun}) do not apply. Motivated by the intuition for the unconstrained case that one or other player should terminate the game at or before the first time that the condition $L \leq U$ fails (or, in the Poisson case, the first event time $T_k$ of the Poisson process at which $L_{T_k} > U_{T_k}$), our approach is to fix one player's stopping rule to be to stop at the first opportunity at which the order condition fails and to address the optimal stopping problem for the other player in this scenario. The solution to this optimal stopping problem (i.e. the value of the problem and the optimal stopping rule) can be combined with the stopping rule for the first player to give a candidate solution for the game (i.e. the value of the game, together with the stopping rules for the two players). It remains to check that this candidate solution for the game is indeed the solution. This will not always be the case. Nonetheless,
under some additional structural conditions, including the fact that the generalised order condition holds, our first main result, Theorem~\ref{theorem:reduction1}, states that the value of the optimal stopping problem agrees with the value of the original Dynkin game. In particular, the result can be applied in the Poisson constrained case and the structural conditions are satisfied in the callable convertible bond problem (with or without the liquidity constraint). The callable convertible bond problem with liquidity constraint can be divided into many subcases depending on the relative values of the various parameters, but it turns out that all the cases can be solved in this way, although sometimes we must assume that the bondholder is the first player and sometimes the firm is the first player.

The above-mentioned approach has parallels with the approach in S{\^\i}rbu and Shreve \cite{si2006}. In fact, the argument in \cite{si2006} of reducing the game of interest to optimal stopping problems can be seen as an example of our result, as all the conditions in our Theorem~\ref{theorem:reduction1} are verified in the justification of the reduction in \cite{si2006}. Nonetheless, although the topic (callable convertible bonds) and the approach (reduce the game to an optimal stopping problem, and then check that the solution of the stopping problem is a saddle point of the game) are similar,
our setup is very different to that in S{\^\i}rbu and Shreve \cite{si2006} and therefore the results and conclusions are very different.
One difference is that in our model, the stock price is specified exogenously and does not depend on the value of the bond. In contrast, in \cite{si2006}, the stock price process is assumed to be endogenous, and hence its drift depends on the value of the bond.
In many ways this makes the model in \cite{si2006} more sophisticated. The second and more significant difference is that we have the added feature of a liquidity constraint so that stopping is only allowed at Poisson event times. One impact of the liquidity constraint is that unlike in \cite{si2006} where the calculation of the game value is a triviality in the stopping region of either player, we need to compute the value function on the whole state space, and this makes the problem more complicated. A third difference is that S{\^\i}rbu and Shreve~\cite{si2006} assume that the bondholder has priority (so that the bondholder is given a final chance to exercise at the time when the firm calls the bond). We consider more general forms of priority, which include the S{\^\i}rbu and Shreve \cite{si2006} set-up as a special case.

Our results about {the} Dynkin game under {the} generalised order condition are related to results of Merkulov \cite[Chapter 5]{merkulov2021value}. For a Dynkin game which satisfies the generalised order condition, Merkulov \cite[Chapter 5]{merkulov2021value} proved that the game of interest can be reduced to an auxiliary Dynkin game which satisfies the order condition, in the sense that the two games have the same value and saddle point. Then, if the game of interest has a Markovian structure, one can make use of the result in Ekstr{\"o}m and Peskir \cite{Ekstrom2008SICON} to conclude the existence of {a} value and saddle point for a Dynkin game under {the} generalised order condition. Nonetheless, this is purely an existence result, and does not necessarily help in the search for the optimal strategies and/or game value. In contrast, our focus is on reducing the game and its value to an optimal stopping problem for which it is often possible to obtain explicit solutions. Although this reduction cannot always be done, when the conditions which make our approach feasible are satisfied (for example in the case of the callable convertible bond) we are able to give the solution to the game (and not just to prove the existence of a solution). The distinctive feature of our work however, is that we consider the Poisson constrained case, which is beyond the scope of \cite[Chapter 5]{merkulov2021value}.

The callable convertible bond under liquidity constraint was studied in Liang and Sun \cite{Liang_Sun} but, unfortunately,
part of the analysis in \cite{Liang_Sun} is incorrect. Liang and Sun~\cite{Liang_Sun} reduced the original problem from the state space $x\in(0,\infty)$ to $x \in (0,\bar{x}^{\lambda}]$,
and argued first that the order condition $L\leq M\leq U$ would hold in the domain $x\in(0,\bar{x}^{\lambda}]$, and second that the game would stop at the earliest Poisson arrival time after the firm's stock price process $X^x$ exceeds $\bar{x}^{\lambda}$. (Here $\bar{x}^{\lambda}$ is determined endogenously.)
However, the claim that the game stops at the earliest Poisson arrival time \textit{after} the firm's stock price process $X^x$ first exceeds some threshold is false. Instead, the game stops at the first Poisson arrival {\em whilst} the firm's stock price process $X^x$ is {above} some threshold. As a result, the analysis for the optimal stopping strategies of the convertible bond in \cite{Liang_Sun} is void.
One goal of this study is to correct the analysis of \cite{Liang_Sun} and to provide a correct solution to the callable convertible bond problem with liquidity constraint.

One common assumption in callable convertible bond models is that the bondholder has priority when both agents stop simultaneously, so $M=L$. This is indeed the case studied in Liang and Sun \cite{Liang_Sun}. This assumption simplifies the bondholder's optimal stopping problem. More specifically, if the firm attempts to stop at the same time as the bondholder, the bondholder's payoff remains unchanged, and they still receive the same number of shares of the firm's stock. In this paper, we extend this assumption to allow for mixed priority when both agents stop simultaneously. We assume that in the case of simultaneous stopping a proportion $m\in[0,1]$ of the bond is converted and the rest is called by the firm who pays the surrender price\footnote{Implicit in our model is the idea that both the bondholder and the firm can observe the actions of the other party, and if the other party calls or converts then they have an opportunity to respond. However, at the Nash equilibrium this ability to respond is not useful, since each player can predict when the other agent will act.}. Then $M$ is a convex combination of $L$ and $U$. This results in a symmetric role for the bondholder and the firm, meaning that if the firm attempts to preempt the bondholder's stopping action by stopping simultaneously, the bondholder's payoff will change, and vice versa. Furthermore, we study the impact of the proportion parameter $m$ on the callable convertible bond value and demonstrate that the value increases with $m$ {(Proposition~\ref{prop:increase})} by establishing a new parity result between two Dynkin games with different simultaneous payoffs ({Theorem \ref{theorem_value_2}}).

In the classical case of a callable convertible bond (no liquidity constraint, priority for the bondholder) it is standard to assume that
the drift of the stock price is dominated by the interest rate (as otherwise the problem is ill-posed). Whilst the condition of a dominated drift is commonly used in risk-neutral pricing problems that involve strictly positive dividends, allowing for a general drift is more applicable to pricing problems in incomplete markets. An interesting observation from our analysis is that, when the drift of the stock price is greater than the interest rate but remains in a reasonable range (as shown in Figure {\ref{fig_sketch6}}, Table \ref{table1}), there exists a critical threshold $\hat{m}$ for the proportion parameter $m$. When $m\in[\hat{m},1]$ it is never optimal for the bondholder to unilaterally convert the bond, regardless of the bond's surrender price, although the bondholder may still convert in response to a call by the firm. This finding contrasts with the case where the drift of the stock price is dominated by the interest rate. In that case, for a large surrender price the threshold strategies used by the bondholder and firm are such that conversion always precedes call (as shown in Figure  {\ref{fig_sketch5}}, Table \ref{table1}).

The structure of the rest of the paper is as follows.
In Section 2, we will present two results for a general Dynkin game. The first is a sufficient condition to reduce a Dynkin game to an optimal stopping problem for one of the players, and the second is a parity result between two Dynkin games with different simultaneous payoffs.
{The remainder of the paper is about the callable convertible bond problem.} We define the problem and provide some auxiliary algebraic results in Section 3. These results will be instrumental in solving two auxiliary optimal stopping problems. Specifically, a bondholder's optimal stopping problem will be addressed in Section 4, and a firm's problem will be explored in Section 5. These solutions will in turn be used to construct saddle points for the callable convertible bond problem {with liquidity constraint} in Section 6.  Finally, in Section 7, we will analyse the dependence of the bond value on the proportion parameter $m$ and the Poisson rate $\lambda$ and show that as $\lambda$ increases to infinity we recover the classical case of the callable convertible bond without liquidity constraint. We also analyse the dependence of the optimal stopping strategy on the proportion parameter $m$: we find that although the payoffs are not ordered in $m$ the game value is monotonic in $m$.

\section{Dynkin games without the order condition}

In this section, we state and prove two main results. First, we give a sufficient condition under which a Dynkin game can be reduced to an optimal stopping problem for one of the players. Since optimal stopping problems are typically much easier to solve than Dynkin games this can lead to a great simplification. Second, we prove a parity result between two Dynkin games with different simultaneous payoffs. In future sections these results will be applied to the callable convertible bond problem.

We work on a general filtered probability space $(\Omega,\mathcal{F},\mathbb{G}=\{\mathcal{G}_t\}_{t\geq 0},\mathbb{P})$ satisfying the \emph{usual} condition{s} (i.e. right continuous and complete filtration). We work in the classical formulation of a zero-sum Dynkin game: let $L=(L_t)_{t \geq 0}$, $U=(U_t)_{t \geq 0}$ and $M =(M_t)_{t \geq 0}$ be {right-continuous} $\mathbb{G}$-\emph{adapted} non-negative payoff processes\footnote{In the main body of the paper when we discuss the convertible bond problem, the payoffs are continuous stochastic processes, and the right-continuity and measurability conditions are trivially satisfied.}.

Let $\Tau$ be a subset of the set of $\mathbb{G}$-stopping times. The idea is that agents cannot choose any $\mathbb{G}$-stopping time but rather are restricted to elements of $\Tau$. We consider two cases: without and with constraint. When there is no constraint the player(s) may choose any $\mathbb{G}$-stopping time provided it is bounded above by the $\mathbb{G}$-stopping time $\eta$, where $\eta$ is the terminal time of the game (which may be infinite). {When there is a constraint, it takes the form that the player(s) are restricted to using stopping times which coincide with those of a pre-specified family of stopping times. The motivating example is that the player(s) are constrained to use $\mathbb{G}$-stopping times, which take values in the event times of a Poisson process.}

\begin{assumption}\label{assumption:tau}
    The set $\Tau$ satisfies: either,

    (1) $\Tau$ is the set of all $\mathbb{G}$-stopping times taking values in $[0,K]$.

    or,

    (2) $\Tau=\{\tau:\tau \textrm{ is a $\mathbb{G}$-stopping time and } \tau(\omega)=T_k(\omega)\,\textrm{for some }1\leq k\leq K\}$ where $\{T_k\}_{1\leq k\leq K}$ is a given, nondecreasing sequence of $\mathbb{G}$-stopping times .

   Here, we allow $K$ to be finite or $+\infty$.
%(1) $\Tau$ contains a largest element (i.e. there exists $\eta\in \Tau$ such that $\eta \geq \tau$ for any $\tau \in \Tau$).

%(2) For any $\tau,\sigma\in \Tau$ and $A\in\mathcal{G}_{\tau\wedge \sigma}$, $\tau\1_A+\sigma\1_{\Omega\setminus A}\in\Tau$.
\end{assumption}

Note that for $\mathbb{G}$-stopping times $\gamma_1$ and $\gamma_2$ and $A\in\mathcal{G}_{\gamma_1\wedge\gamma_2}$, $\gamma=\gamma_1\1_A+\gamma_2\1_{A^c}$ is also a $\mathbb{G}$-stopping time. Therefore, given $\gamma_1, \gamma_2\in \Tau$, we have $\gamma \in \Tau$. Also note that, $\Tau$ contains a largest element, which we denote by $\eta$. The stopping time $\eta$ is the terminal time of the game, and we define the terminal payoff of the game by a $\mathcal{G}_\eta$-measurable random variable $M_\eta$. Thus, if neither player stops before $\eta$ then the payoff is $M_\eta$. Note that we do not assume that $M_\eta = \lim_{t \uparrow \eta} M_t$, or even that this limit exists.

Given a stopping time $\gamma \in \Tau$ we define $\Tau^\gamma = \{ \tau \in \Tau, \tau \leq \gamma \}$. We have $\Tau = \Tau^\eta$.

Assumption \ref{assumption:tau} covers the following examples: the set of all stopping times taking values in $[0,K]$; the set of all integer-valued stopping times taking values in $[0,K]$; the set of all stopping times taking values in $\{T_k; 1 \leq  k \leq K\}$ where $T_k$ are the event times of a Poisson process. Here we allow $K$ to be finite or infinite. %We will formulate the Poisson process example more rigorously in the next section when we focus on the Poisson case.

Define the payoff of the game by
\begin{equation}\label{eq:dynkinpayoff}
    R^{U,M,L}(\tau,\sigma) =  L_\tau \1_{ \{ \tau < \sigma \} } +U_\sigma \1_{ \{ \sigma < \tau \} } + M_\tau \1_{ \{ \tau = \sigma<\eta \} }+M_\eta \1_{\{\tau=\sigma=\eta\}}
\end{equation}
for $\sigma,\tau\in \Tau$. The expected payoff is denoted by
$J^{U,M,L}(\tau,\sigma) = \E[ R^{U,M,L}(\tau,\sigma)]$. The upper and lower values are defined as
\[ \ol{v} \equiv \ol{v}^{U,M,L}(\Tau) = \inf_{\sigma \in \Tau} \sup_{\tau \in \Tau} J^{U,M,L}(\tau,\sigma), \hspace{8mm} \ul{v} = \ul{v}^{U,M,L}(\Tau) = \sup_{\tau \in \Tau} \inf_{\sigma \in \Tau} J^{U,M,L}(\tau,\sigma). \]
If $\ul{v} = \ol{v}$ then we say the game $J^{U,M,L}$ has a value $v = v^{U,M,L}(\Tau)$ where $v = \ul{v}=\ol{v}$.

The following condition is often assumed in the literature.
\begin{definition}
We say that the payoff processes $L,M,U$ (and the corresponding Dynkin game with payoff $R^{U,M,L}$) satisfy the {\em order condition} if $L \leq M \leq U$.
\end{definition}

One reason why this assumption is common in the literature is that when the order condition fails, there are simple examples which show that the game may not have a value.

\begin{example}
    Suppose that $L_t =e^{-rt}, M_t = 0$ and $U_t = 2e^{-rt}$, where $r$ is a positive constant {and suppose $\Tau$ is the set of all stopping times}. Given any $\sigma$, the \textsc{sup} player can choose a stopping strategy $\tau_\sigma=\1_{\{\sigma=0\}}$, such that $J^{U,M,L}(\tau_\sigma,\sigma)\geq 1$, {where} equality holds if and only if $\sigma>0$ with probability $1$. Therefore $\ol{v} = \inf_\sigma \sup_\tau  J^{U,M,L}(\tau, \sigma) =1$. On the other hand, given any $\tau$, the \textsc{inf} player can choose a stopping strategy $\sigma_\tau=\tau$, such that $J^{U,M,L}(\tau,\sigma_\tau)=0$. Therefore $\ul{v}=\sup_\tau\inf_\sigma   J^{U,M,L}(\tau, \sigma) =0$.
\end{example}

A weaker condition than the order condition is the generalised order condition.
Although the order condition is violated in the callable convertible bond problem, the generalised order condition is satisfied. Sometimes the generalised order condition provides sufficient structure that it is possible to prove useful and interesting results.

\begin{definition}\label{definition_GOC}
We say that the payoffs processes $L,M,U$ (and the corresponding Dynkin game with payoff $R^{U,M,L}$) satisfy the {\em generalised order condition} if $L\wedge U \leq M \leq L\vee U$.
\end{definition}

It is important to understand Dynkin games outside the paradigm where the order condition holds. Even if the general theory fails in this case, we may still hope that when the generalised order condition holds the game has a value.
Indeed, the pragmatic approach to finding a solution is to try to find a saddle point, i.e. to find a pair $(\tau^*,\sigma^*) \in \Tau \times \Tau$ such that $J^{U,M,L}(\tau, \sigma^*) \leq J^{U,M,L}(\tau^*,\sigma^*) \leq J^{U,M,L}(\tau^*, \sigma)$ for all $(\tau,\sigma) \in \Tau \times \Tau$. Then, %(suppressing $\Tau$ and $U,L$ from the notation)
\[ \ol{v}^{U,M,L}(\Tau) \leq \sup_{\tau \in \Tau} J^{U,M,L}(\tau,\sigma^*) \leq J^{U,M,L}(\tau^*,\sigma^*) \leq \inf_{\sigma \in \Tau} J^{U,M,L}(\tau^*,\sigma) \leq \ul{v}^{U,M,L}(\Tau). \]
Since trivially $\ul{v}^{U,M,L}(\Tau) \leq \ol{v}^{U,M,L}(\Tau)$, we conclude that $\ul{v}^{U,M,L}(\Tau) = \ol{v}^{U,M,L}(\Tau)$ and the game has a value\footnote{The existence of a saddle point gives a direct proof of the existence of a game value, but leaves us no nearer to finding the game value unless we can identify the optimisers $\sigma^*$ and $\tau^*$.}.

Our first key result provides a sufficient condition to reduce a Dynkin game to an optimal stopping problem, in which one of the players stops at the first opportunity when the order condition fails and the other player considers the optimal stopping problem which arises if they assume that their opponent is using this rule. We do not assume that the order condition is satisfied, but only that the generalised order condition holds.

Given $U,M,L$ which satisfy the generalised order condition and the set of stopping times $\Tau$ (with the largest element $\eta$), we want to define a random variable $\gamma$, which denotes the first time such that $L\geq U$ holds under the constraint defined by $\Tau$. In the unconstrained case, if $\Tau= \Tau^K$ is the set of all $\mathbb{G}$-stopping times taking values in $[0,K]$, we take $\gamma=\inf\{t\geq 0:\,L_t-U_t\geq 0\}\wedge K$. In the constrained case, if $\Tau=\Tau^{\eta}$ is defined by the nondecreasing sequence $\{T_k\}_{1\leq k\leq K}$, we define $\gamma$ by $\gamma(\omega)=\inf\{T_k(\omega):\, 1\leq k\leq K, L_{T_k(\omega)}(\omega)\geq U_{T_k(\omega)}(\omega)\}\wedge \eta(\omega)$. There are two possible cases for each $\omega$: either $L_{T_k(\omega)}(\omega)\geq U_{T_k(\omega)}(\omega)$ holds for some $1\leq k\leq K$ and then  $\gamma(\omega)=T_k(\omega)$, or $L_{T_k(\omega)}(\omega)< U_{T_k(\omega)}(\omega)$ holds for every $k$, in which case the infimum is taken over an empty set and $\gamma(\omega)=\eta(\omega)$.
\begin{lemma}\label{lem:gamma}
 Suppose that the payoff processes $U,M,L$ satisfy the generalised order condition. Then, the random variable $\gamma$ is in $\Tau$ and satisfies the following properties:

(1) $L_{\gamma}\geq M_\gamma\geq  U_{\gamma}$ on the set $\{\gamma<\eta\}$.

(2) For any $\tau \in \Tau$,  on the set $\{\tau<\gamma\}$, $L_\tau\leq M_\tau\leq U_\tau$ holds with $L_\tau<U_\tau$.
\end{lemma}

\begin{proof}
Firstly, let $\Tau$ be the set of all $\mathbb{G}$-stopping times taking values in $[0,K]$. Then $\gamma=\inf\{t\geq 0:\,L_t\geq U_t\}\wedge K$, and $\inf\{t\geq 0:\,L_t\geq U_t\}$ is a stopping time by the Debut theorem (see Jacod and Shiryaev \cite[Theorem I.1.27]{jacod2013limit}). Therefore $\gamma$ is a stopping time which takes value in $[0,K]$ and is hence in $\Tau$. Property (1) follows by the right continuity of $L$ and $U$ and the generalised order condition, and Property (2) follows by the definition of $\gamma$.

Next, let $\Tau$ be the set of all stopping times taking values in $\{T_k; 1 \leq k \leq K\}$ where $\{T_k\}_{1\leq k\leq K}$ is a sequence of nondecreasing $\mathbb{G}$-stopping times. Then we have $\gamma=T_N\wedge \eta$ where $N=\inf\{1 \leq n \leq K:\, L_{T_n}\geq U_{T_n}\}$.

Note that, $T_N$ is a $\mathbb{G}$-stopping time. We have: \[\{T_N\leq t\}=\bigcup_{k=1}^{K}(\{N=k\}\cap\{T_k\leq t\})=\bigcup_{k=1}^{K}\left((\bigcap_{n=1}^{k-1}\{L_{T_n}<U_{T_n}\})\cap\{L_{T_k}\geq U_{T_k}\}\cap\{T_k\leq t\}\right).\]
Observe that $(\bigcap_{n=1}^{k-1}\{L_{T_n}<U_{T_n}\})\cap\{L_{T_k}\geq U_{T_k}\}\in \mathcal{G}_{T_k}$, and hence $\{T_N\leq t\}$ can be written as a countable union of sets in $\mathcal{G}_t$, which implies that $T_N$ is a $\mathbb{G}$-stopping time. Therefore $\gamma$ is a $\mathbb{G}-$stopping time and takes values in $\{T_k; 1 \leq k \leq K\}$, hence $\gamma \in \Tau$.

Property (1) follows by the definition of $\gamma$ and the generalised order condition. Property (2) follows by the definition of $\gamma$.
\end{proof}

Recall that $\Tau^\gamma = \{ \tau \in \Tau : \tau \leq \gamma \}$ and that on $\tau = \sigma = \eta$ the payoff is $M_\eta$.

\begin{theorem}\label{theorem:reduction1}
Suppose that the payoff processes $U,M,L$ satisfy the generalised order condition. Define $V_{L\wedge M}:=\sup_{\tau\in\mathcal{T}^\gamma}\E[(L\wedge M)_\tau\1_{\{\tau<\eta\}}+M_\eta \1_{\{\tau=\eta\}}]$. Assume that the supremum is attained. In particular, assume that $\tau^* \in \Tau^\gamma$ is such that $V_{L\wedge M} = \E[(L\wedge M)_{\tau^*}\1_{\{\tau^*<\eta\}}+M_\eta\1_{\{\tau^*=\eta\}}]$.

Suppose further that $U_\sigma\geq \E[(L\wedge M)_{\tau^*}\1_{\{\tau^*<\eta\}}+M_\eta\1_{\{\tau^*=\eta\}} | \mathcal{G}_\sigma ]$ on the set $\{\sigma<\tau^*\}$. Then, the game value $v^{U,M,L}(\mathcal{T})$ exists and is equal to $V_{L\wedge M}$, and $(\tau^*,\gamma)$ is a saddle point.
\end{theorem}

\begin{proof}
First, observe that $V_{L \wedge M} = J^{U,M,L}(\tau^*, \gamma)$. Indeed,
\begin{eqnarray*}
    {J^{U,M,L}(\tau^*,\gamma)}    &=& \E[L_{\tau^*}\1_{\{\tau^*<\gamma\}}+M_{\gamma}\1_{\{\tau^*=\gamma \leq \eta\}}+U_{\gamma}\1_{\{\gamma<\tau^*\}}]\\
    &=& \E[L_{\tau^*}\1_{\{\tau^*<\gamma\}}\1_{\{\tau^*<\eta\}}+M_{\gamma}\1_{\{\tau^*=\gamma<\eta\}}+M_\eta\1_{\{\tau^*=\gamma=\eta\}}]\\
    &=& \E[(L\wedge M)_{\tau^*}\1_{\{\tau^*<\eta\}}+M_\eta\1_{\{\tau^*=\eta\}}]=V_{L\wedge M},
\end{eqnarray*}
where the second equality holds since $\tau^*\leq \gamma$. For the third equality, $L_{\tau^*}\1_{\{\tau^*<\gamma\}}\1_{\{\tau^*<\eta\}}=(L\wedge M)_{\tau^*}\1_{\{\tau^*<\gamma\}}\1_{\{\tau^*<\eta\}}$ by Property (2) of Lemma \ref{lem:gamma}, and $M_{\gamma}\1_{\{\tau^*=\gamma<\eta\}}=(L\wedge M)_{\gamma}\1_{\{\tau^*=\gamma<\eta\}}$ by Property (1) of Lemma \ref{lem:gamma}.

To prove that the game value exists and is equal to $V_{L \wedge M}$ we check the saddle point property. Take any $\tau, \sigma\in\Tau$. On the one hand, by the assumption that $U_\sigma\geq \E[(L\wedge M)_{\tau^*}\1_{\{\tau^*<\eta\}}+M_\eta\1_{\{\tau^*=\eta\}} | \mathcal{G}_\sigma ]$,
    \begin{eqnarray*}
        J^{U,M,L}(\tau^*,\sigma)&=&\E[L_{\tau^*}\1_{\{\tau^*<\sigma\}}+M_{\tau^*}\1_{\{\tau^*=\sigma<\eta\}}+U_\sigma\1_{\{\sigma<\tau^*\}}+M_\eta\1_{\{\tau^*=\sigma=\eta\}}]\\
        &\geq &\E[(L\wedge M)_{\tau^*}\1_{\{\tau^*<\sigma\}}+(L\wedge M)_{\tau^*}\1_{\{\tau^*=\sigma<\eta\}}+U_\sigma\1_{\{\sigma<\tau^*\}}+M_\eta\1_{\{\tau^*=\sigma=\eta\}}]\\
        &\geq &\E[(L\wedge M)_{\tau^*}\1_{\{\tau^*\leq \sigma,\,\tau^*<\eta\}}+\E[(L\wedge M)_{\tau^*}\1_{\{\tau^*<\eta\}}+M_\eta\1_{\{\tau^*=\eta\}} | \mathcal{G}_\sigma ]\1_{\{\sigma<\tau^*\}}\\&&\hspace{80mm}+M_\eta\1_{\{\tau^*=\sigma=\eta\}}]\\
        &\geq &\E[\E[(L\wedge M)_{\tau^*}\1_{\{\tau^*<\eta\}}+M_\eta\1_{\{\tau^*=\eta\}} | \mathcal{G}_\sigma ]]\\
        &=& V_{L\wedge M}.
    \end{eqnarray*}

On the other hand, by Properties (1) and (2) of Lemma \ref{lem:gamma}, %(where we apply property (2) with $\tau$):
    \begin{eqnarray*}
        J^{U,M,L}(\tau,\gamma)&=&\E[L_{\tau}\1_{\{\tau<\gamma\}}+M_{\gamma}\1_{\{\tau=\gamma<\eta\}}+U_{\gamma}\1_{\{\gamma<\tau\}}+M_\eta\1_{\{\tau=\gamma=\eta\}}]\\
        &=&\E[(L\wedge M)_{\tau}\1_{\{\tau<\gamma\}}+(L\wedge M)_{\gamma}\1_{\{\tau=\gamma<\eta\}}+U_{\gamma}\1_{\{\gamma<\tau\}}+M_\eta\1_{\{\tau=\gamma=\eta\}}]\\
        &=&\E[(L\wedge M)_\tau\1_{\{\tau\leq \gamma,\,\tau<\eta\}}+U_{\gamma}\1_{\{\gamma<\tau\}}+M_\eta\1_{\{\tau=\gamma=\eta\}}]\\
        &\leq &\E[(L\wedge M)_\tau\1_{\{\tau\leq \gamma,\,\tau<\eta\}}+(L\wedge M)_{\gamma}\1_{\{\gamma<\tau\}}+M_\eta\1_{\{\tau=\gamma=\eta\}}]\\
        &\leq &\E[(L\wedge M)_{\{\tau\wedge\gamma\}}\1_{\{\tau\wedge \gamma<\eta\}}+M_\eta\1_{\{\tau\wedge \gamma=\eta\}}]\leq V_{L\wedge M},
    \end{eqnarray*}
where the final inequality follows from the fact that $\tau\wedge \gamma\in \Tau$ (by Assumption \ref{assumption:tau}) and the definition of $V_{L\wedge M}$. Therefore we have $J^{U,M,L}(\tau,\gamma)\leq J^{U,M,L}(\tau^*,\gamma)\leq J^{U,M,L}(\tau^*,\sigma)$, %This is exactly the definition of a saddle point, hence
and the proof is complete.
\end{proof}

\begin{remark}\label{remark:reduction}
In Theorem \ref{theorem:reduction1}, we assumed the existence of an optimal strategy for the optimal stopping problem {$\sup_{\tau\in\mathcal{T}^\gamma}\E[(L\wedge M)_\tau\1_{\{\tau<\eta\}}+M_\eta\1_{\{\tau=\eta\}}]$}. If $\Tau$ satifies case (1) of Assumption \ref{assumption:tau} and the process $L\wedge M$ is quasi-left continuous with $ M_\eta=(L\wedge M)_\eta$, the existence of an optimal strategy can be proved via a Snell envelope argument (see Lamberton \cite[Section 2.3.3]{Lamberton2009}).
\end{remark}

The conditions in Theorem~\ref{theorem:reduction1} can be simplified if we consider a perpetual problem ($\eta = \infty$) and a Markovian structure and discounted payoffs.

\begin{corollary}\label{cor_reduction1}
    Suppose that $\Tau=\{\tau:\tau \textrm{ is a }\mathbb{G}\textrm{-stopping time},\,\tau(\omega)=T_k(\omega)$, \textrm{for some }$1\leq k\leq \infty\}$ where $\{T_k\}_{1\leq k\leq \infty}$ are event times of a Poisson process independent of $X$ and $T_\infty=\infty$.

   Suppose that the payoff processes can be written as $L_t=e^{-rt}l(X^x_t), M_t=e^{-rt}m(X^x_t)$, $U_t=e^{-rt}u(X^x_t)$, where $X^x$ is a time-homogeneous diffusion with state space given by an interval $I\subset \mathbb{R}$, $X^x_0=x\in I$, $r>0$ is a constant and $l,m,u$ are continuous functions with $l\wedge u\leq m\leq l\vee u$. Suppose that\footnote{Note that we do not assume that $M_\infty=0$ a.s., just that $M_\infty=0$ a.s. on the set $\{\gamma=\infty\}$.} $M_\infty=0$ a.s. on the set $\{\gamma=\infty\}$.

   Suppose further that the optimal stopping problem $\sup_{\tau\in\mathcal{T}^\gamma}\E[e^{-r\tau}(l\wedge m)(X^x_\tau)\1_{\{\tau<\infty\}}]$ has a value function $V$.  Assume that the optimal strategy $\tau^*\in \Tau^\gamma$ exists and is of the form $\tau^*(\omega)=\inf\{T_k(\omega):\,1\leq k\leq \infty,\, X^x_{T_k(\omega)}(\omega)\in A\}$ for some closed set $A$. % as the stopping region.

   If $u(x)\geq V(x)$ holds on $I\setminus A$, then the game value $v^{U,M,L}(\Tau)$ exists and is equal to $V(x)$, and $(\tau^*,\gamma)$ is a saddle point.
\end{corollary}

\begin{proof}
     By Theorem~\ref{theorem:reduction1} it suffices to check that $U_\sigma\geq \E[(L\wedge M)_{\tau^*}\1_{\{\tau^*<\infty\}}+M_\infty\1_{\{\tau^*=\infty\}} | \mathcal{G}_\sigma ]$ holds on $\{\sigma<\tau^*\}$ for any $\sigma \in \Tau$. Note that we have $\tau^*\leq \gamma$, therefore $\{\tau^*=\infty\}\subset \{\gamma=\infty\}$ and $M_\infty=0$ a.s. on the set $\{\tau^*=\infty\}$ by assumption. Hence it suffices to check that $U_\sigma\geq \E[(L\wedge M)_{\tau^*}\1_{\{\tau^*<\infty\}} | \mathcal{G}_\sigma ]$ holds on $\{\sigma<\tau^*\}$ for any $\sigma \in \Tau$.

Observe that on the set $\{\sigma<\tau^*\}$ we have $X^x_\sigma\in I\setminus A$. By the strong Markov property:
\begin{eqnarray*}
    e^{-r\sigma}V(X^x_\sigma)\1_{\{\sigma<\tau^*\}}&=&e^{-r\sigma}\E^{X^x_\sigma}[e^{-r\tau^*}(l\wedge m)(X_{\tau^*})\1_{\{\tau^*<\infty\}}]\1_{\{\sigma<\tau^*\}}\\
    &=& e^{-r\sigma}\E^{x}[(e^{-r\tau^*}(l\wedge m)(X_{\tau^*})\1_{\{\tau^*<\infty\}})\circ\theta_\sigma|\mathcal{G}_{\sigma}]\1_{\{\sigma<\tau^*\}}\\
    &=& \E^x[e^{-r\tau^*}(l\wedge m)(X^x_{\tau^*})\1_{\{\tau^*<\infty\}}|\mathcal{G}_\sigma]\1_{\{\sigma<\tau^*\}},
\end{eqnarray*}
where $\theta_\sigma$ is the shift operator on the canonical space.

Then on $\sigma < \tau^*$, $X^x_\sigma\in I\setminus A$ and $u \geq V$ so that
\[ U_\sigma = e^{-r \sigma} u(X_\sigma) \geq e^{-r \sigma}V(X_\sigma) = \E[(L \wedge M)_{\tau^*} \1_{\{\tau^*<\infty\}}|\mathcal{G}_\sigma] \]
as required.
\end{proof}

We call $A$ the stopping region of the optimal stopping problem and $I\setminus A$ the continuation region.

In the convertible bond problem which is the focus of later parts of the paper, the payoff processes can be written in a similar form to these stated in Corollary \ref{cor_reduction1}, but with an extra constant $p$. %For future use, we record the following result:
\begin{corollary}\label{cor_reductionc1}
    Suppose that the payoff processes are of the form $L_t=e^{-rt}l(X^x_t)+p, M_t=e^{-rt}m(X^x_t)+p, U_t=e^{-rt}u(X^x_t)+p$, where $p$ is a constant, and $r, X^x,l,u,m,\Tau,\gamma$ are as in Corollary \ref{cor_reduction1}, with $M_\infty =p$ a.s. on the set $\{\gamma=\infty\}$.

    Suppose that the problem $\sup_{\tau\in\mathcal{T}^\gamma}\E[(e^{-r\tau}(l\wedge m)(X^x_\tau)+p)\1_{\{\tau<\infty\}}+M_\infty\1_{\{\tau=\infty\}}]$ has a value function $V$, and assume that the optimal strategy $\tau^*\in \Tau$ exists and is of the form $\tau^*(\omega)=\inf\{T_k(\omega):\,1\leq k\leq \infty,\, X^x_{T_k(\omega)}(\omega)\in A\}$ for some closed set $A$. %as the stopping region.

    If $u(x)\geq V(x)-p$ holds on $I\setminus A$, then the game value $v^{U,M,L}(\Tau)$ exists and is equal to $V(x)$, and $(\tau^*,\gamma)$ is a saddle point.
\end{corollary}

%\begin{remark}
%    In the case when $\Tau$ is the collection of all $\mathbb{G}$-stopping times and the processes $M,L$ are quasi-left continuous with $L_\eta\geq M_\eta$, the optimal stopping time $\tau^*$ exists with $A=\{x\in I:V(x)=(l\wedge u)(x)\}$.
%\end{remark}

We can also consider the \textsc{inf} player's optimal stopping problem, and, as for the \textsc{sup} player's problem, the conditions can be simplified under a Markovian structure. The proofs of the following theorem and corollary are omitted since they are similar to those of Theorem \ref{theorem:reduction1} and Corollary \ref{cor_reduction1}.

\begin{theorem}\label{theorem:reduction2}
Suppose that the payoff processes $U,M,L$ satisfy the generalised order condition. Define $V_{U\vee M}:=\inf_{\sigma\in\mathcal{T}^\gamma}\E[(U\vee M)_\sigma\1_{\{\sigma<\eta\}}+M_\eta\1_{\{\sigma=\eta\}}]$. Assume that the infimum is attained. In particular, assume that $\sigma^*\in \Tau^\gamma$ is such that $V_{U\vee M}=\E[(U\vee M)_{\sigma^*}\1_{\{\sigma^*<\eta\}}+M_\eta\1_{\{\sigma=\eta\}}]$.

Suppose further that $L_\tau\leq  \E[(U\vee M)_{\sigma^*}\1_{\{\sigma^*<\eta\}}+M_\eta\1_{\{\sigma^*=\eta\}} | \mathcal{G}_\tau ]$ on the set $\{\tau<\sigma^*\}$. Then, the game value $v^{U,M,L}(\mathcal{T})$ exists and is equal to $V_{U\vee M}$, and $(\gamma,\sigma^*)$ is a saddle point.
\end{theorem}

\begin{corollary}\label{cor_reduction2}
    Suppose that $\Tau=\{\tau:\tau \textrm{ is a }\mathbb{G}\textrm{-stopping time}\,\tau(\omega)=T_k(\omega)$, \textrm{for some }$1\leq k\leq \infty\}$ where $\{T_k\}_{1\leq k\leq \infty}$ are event times of a Poisson process independent of $X$ and $T_\infty=\infty$.

    Suppose that the payoff processes can be written as $L_t=e^{-rt}l(X^x_t), M_t=e^{-rt}m(X^x_t)$, $U_t=e^{-rt}u(X^x_t)$, where $X^x$ is a time-homogeneous diffusion with state space $I\subset \mathbb{R}$, $X^x_0=x\in I$, $r>0$ is a constant and $l,m,u$ are continuous functions with $l\wedge u\leq m\leq l\vee u$. Suppose that $M_\infty=0$ a.s. on the set $\{\gamma=\infty\}$.

    {Suppose further that the optimal stopping problem $\inf_{\sigma\in\mathcal{T}^{\gamma}}\E[e^{-r\sigma}(u\vee m)(X^x_\sigma)\1_{\{\sigma<\infty\}}]$ has a value function $V$}, and assume that the optimal strategy $\sigma^*\in \Tau$ exists and is of the form $\sigma^*(\omega)=\inf\{T_k(\omega):\,1\leq k\leq \infty,\, X^x_{T_k(\omega)}(\omega)\in A\}$ with some closed set $A$ as the stopping region.

   If $l(x)\leq V(x)$ holds on the continuation region $I\setminus A$ of the optimal stopping problem, then the game value $v^{U,M,L}(\Tau)$ exists and is equal to $V(x)$, and $(\gamma,\sigma^*)$ is a saddle point.
\end{corollary}

\begin{corollary}\label{cor_reductionc2}
     Suppose that the payoff processes are of the form $L_t=e^{-rt}l(X^x_t)+p, M_t=e^{-rt}m(X^x_t)+p, U_t=e^{-rt}u(X^x_t)+p$, where $p$ is a constant, and $r, X^x,l,u,m,\gamma$ are as in Corollary \ref{cor_reduction1}, with $M_\infty=p$ a.s. on the set $\{\gamma=\infty\}$.

    Suppose that the problem $\inf_{\sigma\in\mathcal{T}^\gamma}\E[(e^{-r\sigma}(u\vee m)(X^x_\sigma)+p)\1_{\{\sigma<\infty\}}+M_\infty\1_{\sigma=\infty}]$ has a value function $V$, and assume that the optimal strategy $\sigma^*\in \Tau$ exists and is of the form $\sigma^*(\omega)=\inf\{T_k(\omega):\,1\leq k\leq \infty,\, X^x_{T_k(\omega)}(\omega)\in A\}$ with some closed set $A$ as the stopping region.

    If $l(x)\leq V(x)-p$ holds on the continuation region $I\setminus A$ of the optimal stopping problem, then the game value $v^{U,M,L}(\Tau)$ exists and is equal to $V(x)$, and $(\gamma,\sigma^*)$ is a saddle point.
\end{corollary}

Our second key result is a result about parity between two Dynkin games with different simultaneous payoffs. We will establish under certain conditions, and for a Dynkin game with payoff triple $(U,M^1,L)$ which has a saddle point, that it is possible to choose a larger simultaneous payoff $M^2$ without changing the game value. This result will be used in our study of the callable convertible bond to compare the value of two Dynkin games which differ only in the values of the simultaneous payoffs. Note that, although the result will be applied to a game which satisfies the generalised order condition, such a condition is not necessary and the following theorem works for general games.

\begin{theorem}\label{theorem_value_2}
Suppose that $\Tau$ satisfies Assumption \ref{assumption:tau} and $\eta$ is the largest element in $\Tau$. Suppose $M^1\leq M^2$ and the game defined by the payoff processes $U,M^1,L$ and set of stopping times $\Tau$ has a value $v^{U,M^1,L}(\Tau)$, with a corresponding saddle point $(\tau^*,\sigma^*)$. Suppose:

(i) $\{M^1<M^2\}\subset \{M^2\leq U\}$;

(ii) $M_\eta^1=M_\eta^2$ a.s. on the set $\{\sigma^*=\eta\}$.

Then, the game defined by the payoff processes $U,M^2,L$, terminal payoff $M_\eta^2$ and set of stopping times $\Tau$ has a value $v^{U,M^2,L}(\Tau)=v^{U,M^1,L}(\Tau)$.

%Moreover, the same result holds if we replace (ii) by
%(ii)' $\mathbb{P}(\sigma^*=\eta)=0$.
\end{theorem}

\begin{remark}
Clearly a sufficient condition for (ii) is $\mathbb{P}(\sigma^*=\eta)=0$.
\end{remark}

\begin{proof}
 We omit the label $\Tau$ as it remains constant throughout. It suffices to prove $\ol{v}^{U,M^2,L}\leq v^{U,M^1,L}\leq \ul{v}^{U,M^2,L}$.
Indeed, the second inequality is immediate, so it remains to prove the first inequality. Firstly, note that
\begin{eqnarray*}
\ol{v}^{U,M^2,L}=\inf_{\sigma\in\Tau}\sup_{\tau\in\Tau}J^{U,M^2,L}(\tau,\sigma)
\leq  \sup_{\tau\in\Tau}J^{U,M^2,L}(\tau,\sigma^*).
\end{eqnarray*}
If, for any stopping time $\tau\in\Tau$, we can construct a stopping time $\tau'\in\Tau$ (depending on $\tau$) such that
\begin{equation}\label{eq:comparison}
    J^{U,M^2,L}(\tau,\sigma^*)\leq J^{U,M^1,L}(\tau',\sigma^*),
\end{equation}
then we can conclude $\sup_{\tau\in\Tau}J^{U,M^2,L}(\tau,\sigma^*)\leq \sup_{\tau \in \Tau}J^{U,M^1,L}(\tau,\sigma^*)=v^{U,M^1,L}$ and the claim is proven.

We claim that $\tau'= \tau\1_{\{\tau\neq \sigma^*\}}+\tau\1_{\{M^1_{\sigma^*}=M^2_{\sigma^*},\tau=\sigma^*\}}+\eta\1_{\{M^1_{\sigma^*}<M^2_{\sigma^*},\tau=\sigma^*\}}$ is a stopping time for which (\ref{eq:comparison}) holds.
Note that we can rewrite $\tau'$ in the form $\tau'=\eta\1_B+\tau\1_{B^c}$ where $B=\{M^1_{\sigma^*}<M^2_{\sigma^*},\tau=\sigma^*\}$. It is immediate that $B\in\mathcal{G}_{\tau}=\mathcal{G}_{\tau\wedge \eta}$, hence $\tau'\in\Tau$. Moreover, the following properties hold by construction:
\begin{enumerate}[label={(\arabic*)}]
    \item $\{\tau'<\sigma^*\}=\{\tau<\sigma^*\}$, and $\tau=\tau'$ on this set.
    \item $\{\tau'>\sigma^*\}=\{\tau>\sigma^*\}\cup\{M^1_{\sigma^*}<M^2_{\sigma^*},\tau=\sigma^*<\eta\} $, with $\tau'=\tau$ on $\{\tau>\sigma^*\}$ and $\tau'=\eta$ on $\{M^1_{\sigma^*}<M^2_{\sigma^*},\tau=\sigma^*<\eta\}$.
    \item $\{\tau'=\sigma^*\}=\{M^1_{\sigma^*}={M}^2_{\sigma^*},\tau=\sigma^*<\eta\}\cup\{\tau=\sigma^*=\eta\}$, with $\tau'=\tau<\eta$ on $\{M^1_{\sigma^*}={M}^2_{\sigma^*},\tau=\sigma^*<\eta\}$, and $\tau'=\tau=\eta$ on $\{\tau=\sigma^*=\eta\}$.
\end{enumerate}
%Now we make use of the assumptions (i) and (ii).
{Using both (i) and the fact that ${M}^2_{\sigma^*}\leq U_{\sigma^*}$ on the set $\{M^1_{\sigma^*}<{M}^2_{\sigma^*},\sigma^*=\tau<\eta\}$ and (ii) and the fact that $M_\eta^1=M_\eta^2$ holds on the set $\{\tau=\eta=\sigma^*\}$ for the inequality we have that
\begin{eqnarray*}
    R^{U,M^2,L}(\tau,\sigma^*)&=& L_\tau\1_{\{\tau<\sigma^*\}}+U_{\sigma^*}\1_{\{\sigma^*<\tau\}}+{M}^2_{\sigma^*}\1_{\{\sigma^*=\tau<\eta\}}\1_{\{M^1_{\sigma^*}<{M}^2_{\sigma^*}\}}\\
    && \hspace{20mm}+M_\eta^2\1_{\{\sigma^*=\tau=\eta\}}+{M}^2_{\sigma^*}\1_{\{\sigma^*=\tau<\eta\}}\1_{\{M^1_{\sigma^*}={M}^2_{\sigma^*}\}}\\
%    &\leq & L_\tau\1_{\{\tau<\sigma^*\}}+U_{\sigma^*}\1_{\{\sigma^*<\tau\}}+U_{\sigma^*}\1_{\{\sigma^*=\tau<\eta\}}\1_{M^1_{\sigma^*}<{M}^2_{\sigma^*}}\\
%    && \hspace{20mm} +M^2_{\sigma^*}\1_{\{\sigma^*=\tau=\eta\}}+M^1_{\sigma^*}\1_{\{\sigma^*=\tau<\eta\}}\1_{\{M^1_{\sigma^*}={M}^2_{\sigma^*}\}} \\
    & \leq &L_\tau\1_{\{\tau<\sigma^*\}}+U_{\sigma^*}\1_{\{\sigma^*<\tau\}}+U_{\sigma^*}\1_{\{\sigma^*=\tau<\eta\}}\1_{\{M^1_{\sigma^*}<{M}^2_{\sigma^*}\}}\\
    && \hspace{20mm} +M_\eta^1\1_{\{\sigma^*=\tau=\eta\}}+M^1_{\sigma^*}\1_{\{\sigma^*=\tau<\eta\}}\1_{\{M^1_{\sigma^*}={M}^2_{\sigma^*}\}}\\
    &=&L_{\tau}\1_{\{\tau<\sigma^*\}}+U_{\sigma^*}\1_{\{\tau>\sigma^*\}\cup\{M^1_{\sigma^*}<M^2_{\sigma^*}, \tau=\sigma^*<\eta\}} \\
    && \hspace{20mm}+M^1_{\sigma^*}\1_{\{M^1_{\sigma^*}={M}^2_{\sigma^*},\tau=\sigma^*<\eta\}}+M_\eta^1\1_{\{\tau=\sigma^*=\eta\}}\\
    &=&R^{U,M^1,L}(\tau',\sigma^*),
\end{eqnarray*}
{where the last equality follows from the identifications (1), (2) and (3).}}
Then, $J^{U,M^2,L}(\tau,\sigma^*)\leq J^{U,M^1,L}(\tau',\sigma^*)$ follows by taking expectation on both sides.
\end{proof}

In the remaining sections we consider the callable convertible bond problem. In the callable convertible bond problem the order condition fails, but the generalised order condition holds. Theorems~\ref{theorem:reduction1} and \ref{theorem:reduction2} give us a route to finding the game value/optimal strategy/saddle point. Theorem~\ref{theorem_value_2} is used for comparative statics and understanding how the bond value depends on the priority rule for simultaneous stopping.

\section{The callable convertible bond: problem specification and auxiliary results}\label{sec:problem}
\subsection{Specification of the convertible bond problem}

The firm issues callable convertible bonds as perpetuities with a positive constant
coupon rate $c$, and a share of the bond is purchased by an investor at initial time $t=0$. Whilst holding the callable convertible
bond, the investor will continuously receive the coupon rate
from the firm until the contract is terminated, which can happen in the following ways: (i) if  the firm calls the bond at some stopping time
$\sigma$ first (i.e. $\sigma<\tau$), the bondholder will receive a pre-specified
surrender price $K$ at time $\sigma$; (ii) if  the investor chooses to convert their bond at some
stopping time $\tau$ first (i.e. $\tau<\sigma$), the bondholder will obtain $
X_{\tau}^x$ at time $\tau$ from converting their bond to the same\footnote{{Although we assume that the bondholder converts one bond for one share, the results immediately generalise to the case of a conversion rate of one bond to $\gamma$ shares by reinterpreting $X$ as the value of $\gamma$ shares on exercise. Since we model $X$ as an exponential Brownian motion the dynamics of $X$ are unchanged.}} number of shares of firm's stock, where $X^x$ represents the stock price process starting from $X_0^x=x$; (iii) if the bondholder and the firm choose to
stop the contract simultaneously (i.e. $\sigma=\tau$), then the bondholder is allowed to convert a proportion $m\in[0,1]$ of the bond to the firm's stock, and the rest is called by the firm who pays the same surrender price $K$ per share. As a result, the bondholder will obtain $m X^x_\tau+(1-m)K$ in this case.

%\bDGH{
%\begin{remark}
%{In principle, at each opportunity to stop the bondholder and firm face a one-period game with payoff $K$ if (only) the firm stops, payoff {$X^x_\tau$} if (only) the bondholder stops, a payoff $mX^x_\tau+(1-m)K$ if both parties stop, and a (to be determined as part of the problem) continuation value if neither party stops. For a given continuation value such simple one-period games always have a solution, but only if we allow the players to follow randomised strategies. It will follow from our results that the callable convertible bond with liquidity constraint always has a value and that randomised strategies are not necessary -- we will give a precise description of the optimal stopping rule for both players and it does not require randomisation. As a consequence, we may imagine that at the event times of the Poisson process (candidate stopping times) one player (say the bondholder) is informed whether the other player (the firm) intends to stop, and may make decisions conditional on this (extra) information. In the financial context, it is very natural to think of the problem in this way. However, because randomised strategies are not needed, it follows from our results that it does not matter if we think of the problem in this way or not.}
%\end{remark}
%}

Let $r>0$ represent the interest rate, and let $P>0$ be the value of perpetuity if the bond is never called or converted, which is $P=\int_0^{\infty}e^{-rt}cdt=\frac{c}{r}$. To economise on notation, we will use the parameter pair $(P,r)$ instead of $(c,r)$ from now on.
Then the payoff of a perpetual callable convertible bond has the following form
\begin{eqnarray}\label{eq:convertible_bond_payoff_1}
R^m(\tau,\sigma) & = & \int_0^{\sigma\wedge \tau}
e^{-ru}rP\,du+e^{-r\tau}
X^x_{\tau}\mathbbm{1}_{\{\tau<\sigma\}}+e^{-r\sigma}K\mathbbm{1}_{\{\sigma<\tau\}}\\
& & \hspace{10mm} +e^{-r\tau}(m X^x_{\tau}+(1-m)K)\mathbbm{1}_{\{\tau=\sigma\}}\notag.
\end{eqnarray}
Then, using integration by parts we can rewrite $R^m(\tau,\sigma)$ as
\begin{eqnarray*}
    R^m(\tau,\sigma) &=&\
     \left\{ P + e^{-r \tau} \left(  X^x_\tau - P \right) \right\} \1_{ \{ \tau <\sigma \} } + \left\{ P + e^{-r \tau} \left( K - P \right) \right\} \1_{ \{ \tau > \sigma \} }\\
    &&\hspace{10mm} +\left\{P+e^{-r\tau}\left(m X^x_\tau+(1-m)K-P\right)\right\}\1_{\{\tau=\sigma\}}\\
&=&\ L_{\tau} \1_{ \{ \tau <\sigma \} }+ U_{\sigma} \1_{ \{ \tau > \sigma \} }+ M_{\tau}\1_{\{\tau=\sigma<\infty\}}+M_\infty\1_{\{\tau=\sigma=\infty\}},
\end{eqnarray*}
{which is exactly of the form in \eqref{eq:dynkinpayoff} {and Corollary \ref{cor_reductionc1}} with 
$L_t = P+e^{-r t}(X^x_t-P)$, $U_t =P+e^{-r t} (K-P)$, $M_t =P+e^{-r t}( mX^x_t + (1-m) K-P)$,
and $M_\infty=\lim_{t\uparrow \infty}M_t$. Note that $M$ always lies between $L$ and $U$ so the generalised order condition holds, but the order condition $L \leq M\leq  U$ fails when $X^x > K$.}

We now specify the probability space that we will work on, as well as the stock price process $X^x$. Suppose that the complete probability space \((\Omega, \mathcal{F}, \mathbb{P})\) supports both a Brownian motion \(W = \{W_t\}_{t \geq 0}\) with its natural filtration \(\mathbb{F}^W = \{\mathcal{F}_t^W\}_{t \geq 0}\), and an independent Poisson process \(N = \{N_t\}_{t \geq 0}\) with constant intensity \(\lambda > 0\). The jump times of the Poisson process are denoted by \(\{T_k\}_{k \geq 1}\), with \(T_\infty = +\infty\), and the natural filtration of the Poisson process is denoted by \(\mathbb{H} = \{\mathcal{H}_t\}_{t \geq 0}\). We define the filtration \(\mathbb{G} = \{\mathcal{G}_t\}_{t \geq 0}\) as the smallest filtration that contains both \(\mathbb{F}^W\) and \(\mathbb{H}\), i.e., for each \(t \geq 0\),
$
\mathcal{G}_t = \sigma(\mathcal{F}_t^W \cup \mathcal{H}_t).
$
Finally, \(\mathbb{G}\) is augmented to satisfy the usual conditions of right-continuity and completeness with respect to \(\mathbb{P}\).

 We assume that the stock price process $X^x$ follows a geometric Brownian motion.
\begin{assumption}\label{assumption}
The price process $X^x$ of the firm's stock follows
$$X_t^x=x+\int_0^t \mu X_s^xds+\int_0^t \nu X_s^xdW_s,$$
where $\mu\in\mathbb{R}$ and $\nu>0$ are constants which represent the stock price's drift and volatility, respectively.
\end{assumption}

%Note that, we use $\nu$ for the volatility instead of the more commonly used $\sigma$, as $\sigma$ is reserved as the stopping time for the \textsc{inf} player in general and for the firm in the callable convertible bond problem.

Following \cite{Liang_Sun} (and also \cite{dupuis2002optimal} and others) the liquidity constraint is modelled via the jump times of the Poisson process. Instead of allowing $\sigma$ and $\tau$ to be any stopping times, we assume that $\sigma,\tau \in \mathcal{R}(\lambda)$, where
\[{\mathcal{R}}(\lambda)=\{ \mbox{$\xi$: $\xi$ is a $\mathbbm{G}$-stopping time such that $\xi(\omega)=T_k(\omega)$ for some $k\in \{ 1, \dots \infty \}$}\}.\]
%Herein, $\mathbb{G}$ is the filtration generated by the underlying Brownian motion $W$ (with its natural filtration $\mathbb{F}^W$) and an exogenous Poisson process (with its natural filtration $\mathbb{H}$ and jump times $\{T_{N}\}_{N\geq 1}$), i.e. $\mathbb{G}=\mathbb{F}^W \vee \mathbb{H}$. Then, in summary, we work on a filtered probability space $(\Omega, \mathcal{F}, \mathbb{G}, \mathbb{P})$ which supports a Brownian motion $W$ driving the stock price process and a Poisson process, and $\Tau$ is the set of $\{T_{N}\}_{N\geq 1}$-valued stopping times. It should be clear that the set $\Tau$ satisfies the assumptions listed in Theorem \ref{theorem_value_2}, where we use the notation $T_\infty=\infty$.
Observe that, $\mathcal{R}(\lambda)$ {satisfies Assumption \ref{assumption:tau}}, with $\eta = T_\infty=+\infty$ {being} the largest element.

We further introduce the stopping $\sigma$-algebra and its filtration $\Tilde{\mathbb{G}}=(\mathcal{G}_{T_n})_{n\geq 1}$, and let $\mathcal{N}(\lambda)$ be the collection of all the $\Tilde{\mathbb{G}}$-stopping times. This is an equivalent formulation of the Poisson constraint, as any element in $N \in \mathcal{N}(\lambda)$ {corresponds to an element $\gamma$ of $\mathcal{R}(\lambda)$ via $\gamma = T_N$.}
A {useful} element in $\mathcal{N}(\lambda)$ is the following first hitting time
\begin{equation}\label{def_na}
N_y=\inf\{n\geq 1: X^x_{T_{n}}\geq y\}\in\mathcal{N}(\lambda).
\end{equation}
and then $T_{N_y}\in\mathcal{R}(\lambda)$ is the first {event time of the Poisson process} such that the price process $X^x$ {evaluated at this time} exceeds {(or is equal to)} some constant $y$. Such stopping times will play key roles in the convertible bond problem and the construction of optimal stopping strategies. In particular, for the callable convertible bond problem, we have $\gamma(\omega)=\inf\{T_k(\omega):\, 1\leq k\leq \infty, L_{T_k}(\omega)\geq U_{T_k}(\omega)\}\wedge \infty=T_{N_K}(\omega)$, so $\gamma=T_{N_K}$. Note that, on the set $\{\gamma=\infty\}$, we have $X_{T_n}<K$ for all $1\leq n<\infty$, hence $M_{\infty}=P$ holds.

Due to the Markovian setup, it is useful to consider the (upper and lower) value of the convertible bond as functions of the initial value of the stock price. Then, with the superscript $\lambda$ denoting the rate of the Poisson process and the subscript ca showing that we are talking about a callable bond, we define
the upper and lower value of the callable convertible bond via
\begin{eqnarray}
\label{upperValues_1}
\ol{v}_{{ca,m}}^{\lambda}(x) & = & \inf_{\sigma\in\mathcal{R}(\lambda)}\sup_{\tau\in\mathcal{R}(\lambda)}J^{m,x}(\tau,\sigma), \\
\label{lowerValues_1}
\underline{v}_{ca,m}^{\lambda}(x) & = & \sup_{\tau\in\mathcal{R}(\lambda)}\inf_{\sigma\in\mathcal{R}(\lambda)}J^{m,x}(\tau,\sigma),
\end{eqnarray}
where
\begin{equation}
\label{eq:Jdef}
J^{m,x}(\tau,\sigma)=\mathbb{E}[R^m(\tau,\sigma)].
\end{equation}
Note that $J^{m,x} = J^{U,M,L}$.

Our goals are threefold: (i) to show that $\ol{v}^\lambda_{ca,m}= \ul{v}^\lambda_{ca,m}$ as functions of $x$ which implies the existence of the game value $v_{ca,m}^\lambda(x)$; (ii) to find a saddle point $(\tau^*,\sigma^*)$ to the game, and (iii) to find the explicit form of the value function $v_{ca,m}^\lambda$ as a function of $x$.

\begin{remark}\label{remark_large_mu}
When $\mu\geq \lambda+r$ and $m>0$, the convertible bond problem is not well-posed. It is clear that if the bondholder chooses to convert at the first opportunity then their payoff is bounded below by $e^{-rT_1}mX^x_{T_1}$. It follows that
\begin{eqnarray*}
    \mathbb{E}[e^{-rT_1}mX^x_{T_1}]&=&\mathbb{E}\left[\int_0^\infty \lambda e^{-\lambda u}e^{-ru}mX^x_{u}\,du\right]\\
    &=&\int_0^\infty \lambda m x e^{-(\lambda+r)u}e^{\mu u}\,du\\&=&+\infty.
\end{eqnarray*}
Hence, in the rest of the paper we focus attention on the cases (1) $\mu<\lambda+r$, $m\in[0,1]$ and (2) $\mu\geq \lambda+r$, $m=0$. The extant literature has always assumed absolute priority for the bondholder and therefore always assumed $\mu < \lambda + r$.
\end{remark}

\subsection{Some auxiliary results}

Define the operator {$\mathcal{L}_{\lambda}$} and quadratic $Q_\lambda$ by
\begin{eqnarray}
\label{differential_operator}
\mathcal{L}_{\lambda}f: & = &\frac12 \nu^2x^2f''+\mu xf'-(r+\lambda)f,\; f\in C^2(\mathbb{R}_+) \\
Q_{\lambda}(z): &= &\frac12\nu^2z^2+(\mu-\frac12\nu^2)z-(r+\lambda). \label{eq:Q_lambda}
\end{eqnarray}
Recall that $r>0$ and $\lambda>0$ so that $r+\lambda>0$ and $Q_\lambda$ has two real roots $\alpha_\lambda$ and $\beta_\lambda$ where $\beta_\lambda < 0 < \alpha_\lambda$.
Further, the general solution of $\mathcal{L}_\lambda f=0$ is $f(x)=Ax^\al+Bx^\bl$ for some constants $A$ and $B$. For simplicity, we also write
$\alpha:=\alpha_0$, $\beta:=\beta_0$, $\mathcal{L}:=\mathcal{L}_0$ and $Q:=Q_0$. In the rest of this subsection we give some auxiliary results for $\alpha_{\lambda}$ and $\beta_{\lambda}$.

\begin{lemma}\label{lemma_alphabounds}
The following bounds of $\al$ and $\bl$ hold:
\begin{enumerate}[label={(\arabic*)}]
    \item If $\mu<0$, then $\beta_\lambda \in (\frac{\lambda+r}{\mu},0)$;
    \item If $\mu\in(0,\lambda+r)$, then $\alpha_\lambda \in (1, \frac{\lambda+r}{\mu})$;
    \item If $\mu=\lambda+r$, then $\al=1$;
    \item If $\mu> \lambda+r$, then $\al\in (\frac{\lambda+r}{\mu},1)$.
\end{enumerate}
In particular, $\alpha\in(1,\infty)$ if $\mu \leq 0$, $\alpha\in(1,\frac{r}{\mu})$ if $0<\mu<r$, $\alpha=1$ if $\mu=r$, and $\alpha \in (\frac{r}{\mu},1)$ if $\mu> r$.
\end{lemma}
\begin{proof}
The results follow from $Q_\lambda(1) = \mu - (r+\lambda)$ and $Q_\lambda(\frac{\lambda+r}{\mu})=\frac{1}{2}\nu^2\left(\frac{(\lambda+r)^2}{\mu^2}-\frac{\lambda+r}{\mu}\right)$.
\end{proof}
\begin{lemma}\label{lemma_alphaorder_r}
%The following inequality holds:
\( \alpha_\lambda\leq \alpha \frac{\lambda+r}{r}.\)
%\begin{enumerate}
%    \item $\alpha_\lambda\leq \alpha \frac{\lambda+r}{r}$.
%    \item $\beta_\lambda\geq \beta \frac{\lambda+r}{r}$.
%\end{enumerate}
\end{lemma}
\begin{proof}
The result follows by observing $\al\frac{\lambda+r}{r}\geq 0$ and $Q_\lambda(\alpha \frac{\lambda+r}{r})= \frac{1}{2}\nu^2\alpha^2\frac{\lambda(\lambda+r)}{r^2}$.
\end{proof}
\begin{lemma}\label{lemma_alphaorder_mu}
Assume that $\mu\neq r$. Then the following inequalities hold:
\begin{enumerate}[label={(\arabic*)}]
    \item $\alpha_\lambda \leq \alpha \frac{\lambda+r-\mu}{r-\mu}-\frac{\lambda}{r-\mu}$ if $\mu<\lambda+r$;
    \item $\alpha_\lambda \geq \alpha \frac{\lambda+r-\mu}{r-\mu}-\frac{\lambda}{r-\mu}$ if $\mu>\lambda+r$.
\end{enumerate}
\end{lemma}
\begin{proof}
See appendix.
\end{proof}

Finally, for fixed parameters $r$, $\lambda$ and $\mu$, we define a constant $\hat{m}$
{which will later act as a threshold value between different regimes:}
\begin{equation}\label{def_mhat}
    \hat{m}=\frac{(\lambda+r-\mu)(\alpha(\lambda+r)-\bl r)}{\lambda(\lambda+r-\mu \bl)}.
\end{equation}

\begin{lemma}\label{lemma_mhat_alpha}
The following inequalities hold:
{
\begin{enumerate}[label={(\arabic*)}]
    \item  if $\mu<r$ then $\hat{m}>\alpha>1$;
      if $\mu=r$ then $\hat{m}=\alpha=1$;
      and if $\mu>r$ then $\hat{m}<\alpha<1$.
    \item  if $\mu<\lambda+r$ then $\hat{m}>0$, if $\mu=\lambda+r$ then $\hat{m}=0$ and if $\mu>\lambda+r$ then $\hat{m}<0$.
\end{enumerate}
}
\end{lemma}
\begin{proof}
    See appendix.
\end{proof}

\section{Auxiliary optimal stopping problem for the bondholder}\label{sec:large}

\subsection{The auxiliary optimal stopping problem}
Throughout this section the subscript $b$ is intended to convey that the quantity arises in a problem viewed from the perspective of the bondholder.

{Consider the situation faced by the bondholder where the firm has declared a strategy of calling the bond at $T_{N_K}$. The bondholder chooses $\tau$, and receives coupons up to time $\tau \wedge T_{N_K}$ and either $X^x_\tau$ or $(mX^x_\tau +(1-m) K )$ or $K$ (at time $\tau \wedge T_{N_K}$) depending on whether $\tau < T_{N_K}$ or $\tau= T_{N_K}$ or $\tau > T_{N_K}$. Note that on $\tau \geq T_{N_K}$, the payment received at $T_{N_K}$ for a choice with $\tau > T_{N_K}$, namely $K$, is less than or equal to the payment received at $T_{N_K}$ from the choice $\tau = T_{N_K}$, namely $m X^x_{T_{N_K}} + (1-m)K$. Hence, it is never optimal for the bondholder to choose a stopping time $\tau$ for which $\tau > T_{N_K}$ with positive probability, and we may restrict attention to stopping rules with $\tau \leq T_{N_K}$.

The objective of the bondholder is to }
\begin{equation}\label{eq:OSPb}
\mbox{Find \, $\sup_{\tau\in \mathcal{R}(\lambda), \tau\leq T_{N_K}} E^{\mathcal{R}(\lambda),x}_b(\tau)$ and $\tau^*_b$ which achieves the supremum,}
\end{equation}

where
{
\begin{eqnarray*}
E^{\mathcal{R}(\lambda),x}_b(\tau) & = &
%\mathbb{E} \left[ (L\wedge M)_\tau\right]  \\& = &
\mathbb{E} \left[ \int_0^{\tau}
e^{-ru}rP\,du+e^{-r\tau}(m X^x_{\tau}+(1-m)\min\{ X^x_{\tau},K\}) \right],
\\&=&\mathbb{E} \left[ \int_0^{\tau}
e^{-ru}rP + e^{-r\tau}(X^x_{\tau} I_{\{\tau < T_{N_K}\}}  + (mX^x_\tau +(1-m) K)I_{\{ \tau = T_{N_K}\}}) \right].
\end{eqnarray*}}
Focusing on the event times of the Poisson process, the problem can be restated as
\begin{equation}\label{eq:optimal_stopping_bondholdr_1}
\mbox{Find \, $\sup_{N\in \mathcal{N}(\lambda), N\leq {N_K}} E^{\mathcal{N}(\lambda),x}_b(N)$ and $N^*_b$ which achieves the supremum,}
\end{equation}
where
{\footnotesize{
\begin{equation*}
E^{\mathcal{N}(\lambda),x}_b(N)  =\mathbb{E}\left[  \int_0^{T_N} e^{-ru}rP\,du
+e^{-rT_N} \left( (m X^x_{T_{N_{K}}}+(1-m)K)\mathbbm{1}_{\{N={N_{K}}\}}
  +X^{x}_{T_N}\mathbbm{1}_{\{N<{N_{K}}\}} \right) \right].
\end{equation*}
}}

We set $v^\lambda_{b,m}(x) = \sup_{\tau\in\mathcal{R}(\lambda),\tau\leq T_{N_K}} E^{\mathcal{R}(\lambda),x}_b(\tau) = \sup_{N\in\mathcal{N}(\lambda), N\leq {N_K}} E^{\mathcal{N}(\lambda),x}_b(N)$.
Standard arguments imply that we expect that $v^\lambda_{b,m}$ is a solution of the HJB equation
\begin{equation}
\label{eq:HJBgeneral}
\mathcal{L}_{\lambda}V(x)+rP+\lambda(\max\{V(x), x\}\1_{\{x<K\}}+(m x+(1-m)K)\1_{\{x\geq K\}})=0.
\end{equation}

The aim in this section is to solve the above HJB equation \emph{explicitly} subject to {appropriate} boundary conditions, and then to show that the candidate solution we obtain is indeed the value function of the optimal stopping problem.

We consider three cases: the case $m=1$ and $\mu<r$; the case $m \in [0,1)$ and $\mu < \lambda+r$ or $m=0$ and $\mu>\lambda+r$; and the case $m=0$, $\mu = \lambda+r$. In fact we only consider these cases when $K$ is `large' (in a sense to be made clear later) and `$K$ large' does not arise when $m=1$ and $r \leq \mu < \lambda+r$. Recall that we already saw in Remark~\ref{remark_large_mu} that if $\mu \geq \lambda+ r$ and $m>0$ then the problem is ill-posed.

\subsection{Case $m=1$, $\mu<r$}

Recall that in this case, by Lemma~\ref{lemma_mhat_alpha} we have $\hat{m}>1$ where $\hat{m}$ is as given in (\ref{def_mhat}). Suppose that $K>P$, and define $\xscol$ by
\begin{equation}\label{co1}
\xscol= P\frac{\hat{m}}{ \hat{m}-1}.
\end{equation}

\begin{lemma}\label{lemma_xscol1_bounds}
Suppose that $m=1$ and $\mu<r$.
Let $\xscol$ be given as in \eqref{co1}. Then, $\xscol > \max \{ P, \frac{rP}{r-\mu} \frac{\lambda+r-\mu}{\lambda+r} \}$ and $\xscol < P \frac{\alpha}{\alpha-1}$.
\end{lemma}

\begin{proof}
Given the expression \eqref{co1}, $\xscol > P$ follows from the result $\hat{m}>\alpha\geq 1$, which is proved in Lemma \ref{lemma_mhat_alpha}.

Moreover, $\xscol > \frac{rP}{r-\mu} \frac{\lambda+r-\mu}{\lambda+r}$ is equivalent to $r > \alpha \mu$. If $\mu\leq 0$ this is immediate. Otherwise, it follows from Lemma~\ref{lemma_alphabounds}.

Finally, $\xscol < P \frac{\alpha}{\alpha-1}$ is equivalent to
$\frac{\hat{m}}{\hat{m}-1}< \frac{\alpha}{\alpha-1}$,
which holds by Lemma \ref{lemma_mhat_alpha}.
\end{proof}

For $z \in (P,K)$ define $V^\lambda_{z}$ by
\begin{equation}\label{Vz}
V^\lambda_{z}(x) = \left\{ \begin{array}{lll}
P +  \left[ z -P\right]\frac{x^\alpha}{z^\alpha}, & \; & x<z;\\
\frac{rP}{\lambda+r} + \frac{\lambda  x}{\lambda+r-\mu} + \left[ \frac{(r-\mu) z}{\lambda + r-\mu} - \frac{rP}{\lambda+r} \right] \frac{x^\bl}{z^{\bl}}, & \; & x\geq z.
\end{array} \right.
\end{equation}

\begin{lemma} \label{lem:bondholder_original}
Suppose that $m=1$ and $\mu<r$. %Suppose further that $K\geq \xscol$.
%Then, the value function $v^{\lambda}_{b,1}(x)$ of the optimal stopping problem (\ref{eq:optimal_stopping_bondholder}) has the explicit expression

Then $V = V^{\lambda}_z$ satisfies: $V$ is continuous, $V(0)=P$, $V(z)=z$, $V$ is of linear growth at infinity and $\mathcal{L}_{\lambda}V(x)+rP+\lambda V(x)=0$ on $(0,z)$ and
$\mathcal{L}_{\lambda}V(x)+rP+\lambda x=0$ on $(z,\infty)$.
\end{lemma}

\begin{proof}
Immediate.
\end{proof}

Define $V^\lambda_{b,1}$ by $V^\lambda_{b,1}= V^\lambda_{\xscol}$ so that
\begin{equation}\label{V_co_1}
V^\lambda_{b,1}(x) = \left\{ \begin{array}{lll}
P +  \left[ \xscol -P\right]\frac{x^\alpha}{(\xscol)^\alpha}, & \; & x<\xscol;\\
\frac{rP}{\lambda+r} + \frac{\lambda  x}{\lambda+r-\mu} + \left[ \frac{(r-\mu) \xscol}{\lambda + r-\mu} - \frac{rP}{\lambda+r} \right] \frac{x^\bl}{(\xscol)^{\bl}}, & \; & x\geq \xscol.
\end{array} \right.
\end{equation}

\begin{lemma}\label{lem:bondholderoriginalprop}
Suppose that $m=1$ and $\mu<r$. Suppose further that $K\geq \xscol$.

Then $V^\lambda_{b,1}$ is $C^2$ on $(0,\infty)$, and has bounded first derivative.
Moreover {it} solves
\[\mathcal{L}_{\lambda}V(x)+rP+\lambda(\max\{V(x), x\}\1_{\{x<K\}}+x\1_{\{x\geq K\}})=0.\]
which is a special case of \eqref{eq:HJBgeneral} for $m=1$.

Further, $x \leq V^\lambda_{b,1}(x) \leq  \xscol \leq  K$ for $x \leq \xscol$, and $V^\lambda_{b,1}(x)\leq x$ for $x\geq \xscol$.
\end{lemma}

\begin{remark}
In this case, since $V^\lambda_{b,1} \leq x$ for $x \geq K$, $V^\lambda_{b,1}$ satisfies the simpler looking HJB equation,
$\mathcal{L}_{\lambda}V(x)+rP+\lambda \max\{V(x), x\} = 0$. However, we keep the more general expression because it will make it easier to translate proofs from this section to later results.
\end{remark}

\begin{proof}[Proof of Lemma~\ref{lem:bondholderoriginalprop}]
%We need to show that $V=V^\lambda_{\xscol}$ has continuous first derivative at $\xscol$ and to show that $x<V(x)<K$ on $[0,\xscol)$ and $V(x)<x$ on $(\xscol,\infty)$.

The function $V^\lambda_z$ is $C^2$ on the set $(0,\infty)\setminus \{z\}$. It is easy to check that 
$V^\lambda_z$ and its order derivative are continuous at 
$z$ provided $z$ solves:
\[ \frac{\alpha}{z} \left[ z - P \right] = \frac{\lambda }{\lambda + r-\mu} + \frac{\bl}{z}  \left[ \frac{(r-\mu) z}{\lambda + r-\mu} - \frac{rP}{\lambda+r} \right]. \]
This can be rearranged to give $z=\xscol$, and therefore $V^\lambda_{b,1}$ is $C^1$ on $(0,\infty)$. Moreover, its first derivative is easily seen to be bounded.

Given the lower bounds on $\xscol$ of Lemma~\ref{lemma_xscol1_bounds}, it is clear that     
the first order derivative of $V^\lambda_{b,1}$ is continuous  at $\xscol$ which implies that $V=V^\lambda_{b,1}$ is increasing and convex.
Then, since $\lim_{x \downarrow 0} V(x) = P>0$, and  $\lim_{x \uparrow \infty} \frac{V(x)}{x} = \frac{\lambda }{\lambda+r-\mu} < 1$, it follows that $V$ crosses the line $y=x$ exactly once on $(0,\infty)$. In particular, $V(x)> x$ for $x\in(0,\xscol)$ and $V(x) \leq  x$ for $x\in[\xscol, \infty)$. Further, on $(0,\xscol)$, $P < V(x)< V(\xscol) = \xscol < K$.

As a result, $V_{b,1}^\lambda$ satisfies the HJB equation, which then implies that $V_{b,1}^\lambda$ is $C^2$.
\end{proof}

\begin{proposition} \label{prop_bondholder_original}
Suppose that $m=1$ and $\mu<r$. Suppose further that $K\geq \xscol$.

%Then, the value function $v^{\lambda}_{b,1}(x)$ of the optimal stopping problem (\ref{eq:optimal_stopping_bondholder}) has the explicit expression
Consider the optimal stopping problem (\ref{eq:OSPb}): we have $v^\lambda_{b,1} = V^\lambda_{b,1}$ where $V^\lambda_{b,1}$ is given in \eqref{V_co_1}.
%where the optimal threshold $\xscol$ is determined by value matching and first order smooth fit at $\xscol$, given by
Moreover, the optimal stopping time for (\ref{eq:OSPb}) is given by $\tau_b^*=T_{N_{\xscol}}\leq T_{N_K}$ with $N_y$ defined as in (\ref{def_na}).
\end{proposition}

\begin{proof}
First we show that $V_{b,1}^\lambda$ satisfies the recursive equation
    \begin{eqnarray*}
\lefteqn{V^\lambda_{b,1}(x)} \nonumber \\
 & = &  \mathbb{E}\left[\int_0^{T_1}e^{-ru}rP\,du
 + e^{-rT_1}(\max\{V_{b,1}^\lambda(X^x_{T_1}), X^x_{T_1}\}\1_{\{X^x_{T_1}<K\}}+ X^x_{T_1}\1_{\{X^x_{T_1}\geq K\}})\right].
 \label{eq:Vre}
\end{eqnarray*}
Define $Z=(Z_t)_{t \geq 0}$ by
{\footnotesize{
\begin{equation*}
     Z_t=e^{-(r+\lambda)t}V_{b,1}^\lambda(X^x_t)+\int_0^te^{-(r+\lambda)u}\Big(rP+\lambda\max\{V_{b,1}^\lambda(X^x_u), X^x_u\}\1_{\{X^x_u<K\}}+\lambda X^x_u\1_{\{X^x_u\geq K\}}\Big)\,du.
\end{equation*}
}}
By {It\^{o}'s} formula,
{\footnotesize{
\begin{eqnarray*}
    Z_t&=&V_{b,1}^\lambda(x)+\int_0^te^{-(r+\lambda)u}(V_{b,1}^\lambda)'(X^x_u)\nu X^x_u\,dW_u
    \\&& \hspace{10mm}+\int_0^t e^{-(r+\lambda)u}(\mathcal{L}_\lambda V_{b,1}^\lambda(X^x_u)+rP+\lambda \max\{V_{b,1}^\lambda(X^x_u),X^x_u\}\1_{\{X^x_u<K\}}+\lambda X^x_u\1_{\{X^x_u\geq K\}})\,du.
\end{eqnarray*}
}}
By Lemma \ref{lem:bondholderoriginalprop}, $V_{b,1}^\lambda$ satisfies the HJB equation and hence the Lebesgue integral term vanishes. Further, since $V_{b,1}^\lambda$ has bounded first derivative and $\E[(X^x_u)^2]$ is bounded on $[0,t)$ for each $t$, the quadratic variation of the stochastic integral term has finite expectation, and therefore $Z$ is a martingale. Hence, for arbitrary $t>0$,
%\begin{eqnarray*}
%       \lefteqn{V_{b,1}^\lambda(x)}\\&=&\mathbb{E}[Z_0]=\mathbb{E}[Z_t]\\
%    &=&\mathbb{E}\bigg[e^{-(r+\lambda)t}V_{b,1}^\lambda(X^x_t)+\int_0^te^{-(r+\lambda)u}\Big(rP+\lambda\max\{V_{b,1}^\lambda(X^x_u), X^x_u\}\1_{\{X^x_u<K\}}+\lambda X^x_u\1_{\{X^x_u\geq K\}}\Big)\,du\bigg].
%\end{eqnarray*}
$V_{b,1}^\lambda(x) = \mathbb{E}[Z_0] = \mathbb{E}[Z_t] = \lim_{t\uparrow \infty} \mathbb{E}[Z_t]$.

We know that $V^\lambda_{b,1}(x)\leq K+x$ by Lemma \ref{lem:bondholderoriginalprop}, and we have $\mu<r$. Hence the term $\mathbb{E}[e^{-(r+\lambda)t}V_{b,1}^\lambda(X^x_t)]$ vanishes as $t$ tends to infinity. Then, by monotone convergence,
{\footnotesize{
\begin{eqnarray*}
    V_{b,1}^\lambda(x)&=&\mathbb{E}\bigg[\int_0^\infty e^{-(r+\lambda)u}\Big(rP+\lambda\max\{V_{b,1}^\lambda(X^x_u), X^x_u\}\1_{\{X^x_u<K\}}+\lambda X^x_u\1_{\{X^x_u\geq K\}}\Big)\,du\bigg]\\
    &=&\mathbb{E}\left[\int_0^{T_1}e^{-ru}rP\,du    + e^{-rT_1}\left(\max\{V_{b,1}^\lambda(X^x_{T_1}), X^x_{T_1}\}\1_{\{X^x_{T_1}<K\}}+ X^x_{T_1}\1_{\{X^x_{T_1}\geq K\}}\right)\right].
\end{eqnarray*}
}}
Now consider $Y=(Y_n)_{n \geq 1}$ where
$$ Y_n = \int_0^{T_n}e^{-ru}rPdu+e^{-rT_n}\max\{V^\lambda_{b,1}(X^x_{T_n}), X^x_{T_n}\}\1_{\{X^x_{T_n}<K\}}+e^{-rT_n}X^x_{T_n}\1_{\{X^x_{T_n}\geq K\}}.$$
We want to show:
\begin{itemize}
    \item $Y$ is a $\tilde{\mathbb{G}}$-supermartingale;
    \item $Y$ is a uniformly integrable $\tilde{\mathbb{G}}$-martingale when stopped by ${N_{\xscol}}$.
\end{itemize}

First, we prove the supermartingale property for $Y$. Take arbitrary $n\geq 1$. We have:
{\footnotesize{
\begin{eqnarray*}
    \lefteqn{\mathbb{E}[Y_{n+1}|\mathcal{G}_{T_n}]}\\
    &=&\int_0^{T_n}e^{-ru}rP\,du+\mathbb{E}\left[\int_{T_n}^{T_{n+1}}e^{-ru}rP\,du|\mathcal{G}_{T_n}\right]\\
    &&+\mathbb{E}\left[e^{-rT_{n+1}}\left(\max\{V^\lambda_{b,1}(X^x_{T_{n+1}}), X^x_{T_{n+1}}\}\1_{\{X^x_{T_{n+1}}<K\}}+X^x_{T_{n+1}}\1_{\{X^x_{T_{n+1}}\geq K\}}\right)|\mathcal{G}_{T_n}\right]\\
    &=&\int_0^{T_n}e^{-ru}rP\,du+e^{-rT_n}\mathbb{E}\left[\int_{T_n}^{T_{n+1}}e^{-r(u-T_n)}rP\,du|\mathcal{G}_{T_n}\right]\\&&+e^{-rT_n}\mathbb{E}\left[e^{-r(T_{n+1}-T_n)}\left(\max\{V^\lambda_{b,1}(X^x_{T_{n+1}}), X^x_{T_{n+1}}\}\1_{\{X^x_{T_{n+1}}< K\}}+X^x_{T_{n+1}}\1_{\{X^x_{T_{n+1}}\geq K\}}\right)|\mathcal{G}_{T_n}\right]\\
    &=&\int_0^{T_n}e^{-ru}rP\,du+e^{-rT_n}\mathbb{E}^{X^x_{T_n}}\left[\int_{0}^{T_1}e^{-ru}rP\,du\right]
    \\&&+e^{-rT_n}\mathbb{E}^{X^x_{T_n}}\left[e^{-rT_1}\left(\max\{V^\lambda_{b,1}(X_{T_1}), X_{T_1}\}\1_{\{X_{T_1}<K\}}+X_{T_1}\1_{\{X_{T_1}\geq K\}}\right)\right]\\
    &=&\int_0^{T_n}e^{-ru}rP\,du+e^{-rT_n}V_{b,1}^\lambda(X^x_{T_n})\leq Y_n,
\end{eqnarray*}
}}
where the third equality follows by strong Markov property, the final equality follows by the recursive equation, and the inequality holds by the fact that $V^\lambda_{b,1}(x)\leq x$ on the set $(\xscol,\infty)$ and $\xscol< K$.

Second, we prove the martingale property for the stopped process. Observe that:
\begin{eqnarray*}
    \E[Y_{(n+1)\wedge N_{\xscol}}|\mathcal{G}_{T_n}]&=&Y_{ n\wedge N_{\xscol}}\1_{\{N_{\xscol}\leq n\}}+\E[Y_{n+1}\1_{\{N_{\xscol}\geq n+1\}}|\mathcal{G}_{T_n}].
\end{eqnarray*}
So it remains to prove that $\E[Y_{n+1}\1_{\{N_{\xscol}\geq n+1\}}|\mathcal{G}_{T_n}]=Y_{n}\1_{\{N_{\xscol}\geq n+1\}}$. By the same argument as in the proof of supermartingale property:
\[\E[Y_{n+1}\1_{\{N_{\xscol}\geq n+1\}}|\mathcal{G}_{T_n}]=\left(\int_0^{T_n}e^{-ru}rP\,du+e^{-rT_n}V_{b,1}^\lambda(X^x_{T_n})\right)\1_{\{N_{\xscol}\geq n+1\}}.\]
We know that $X^x_{T_n}<\xscol$ holds on the set ${\{N_{\xscol}\geq n+1\}}$, and hence the RHS of the above equation is exactly $Y_{n}\1_{\{N_{\xscol}\geq n+1\}}$ by the result that $V_{b,1}^\lambda(x)\geq x$ on the set $(0,\xscol)$. This implies the martingale property.

Finally, we prove that the stopped process $(Y_{n \wedge N_{x^*_{b,\lambda}}})_{n \geq 1}$ is uniformly integrable. Since the integral term of $Y$ is bounded and $V_{b,1}^\lambda(x)$ is dominated by $x$ on $(\xscol,\infty)$, it suffices to prove that $\left(e^{-r(\eta^* \wedge T_n)}X^x_{\eta^* \wedge T_n}\right)_{n\geq 1}$ is uniformly integrable, where we write $\eta^*=T_{N_{\xscol}}$ to keep the number of subscripts under control.  {In the current setting where $\mu<r$, this follows immediately from the fact that $(e^{-rt}X^x_t)_{t \geq 0}$ is uniformly integrable.}

%{and $e^{-r\eta^*}X^x_{\eta^*}$ is integrable by Lemma \ref{lemma_integrability}. Hence uniform integrability follows.}

%But $e^{-r\eta^*}X^x_{\eta^*}<Y_{N_{\xscol}}$ and $Y_{N_{\xscol}}$ is exactly the a.s. limit of the nonnegative martingale $(Y_{n\wedge N_{\xscol}})_{n\geq 1}$, and hence by the martingale convergence theorem, $\mathbb{E}[Y_{N_{\xscol}}] \leq \E[Y_1]<\infty$.

{
By the optional stopping theorem, for any $N\in\mathcal{N}(\lambda)$, we have that $Y_1\geq \mathbb{E}[Y_{N}|\mathcal{G}_{T_1}]$. This implies that
\begin{eqnarray*}
    e^{-rT_1}\max\{V^\lambda_{b,1}(X^x_{T_1}),
X^x_{T_1}\}\geq \mathbb{E}\left[\int_{T_1}^{T_N}e^{-ru}rP\,du+e^{-rT_N}\max\{V^\lambda_{b,1}(X^x_{T_N}),
X^x_{T_N}\}|\mathcal{G}_{T_1}\right]
\end{eqnarray*}

Hence, by the Markov property:
}
%In turn, for any $N\in\mathcal{N}(\lambda)$,
\begin{eqnarray}
% \nonumber % Remove numbering (before each equation)
\lefteqn{\max\{V^\lambda_{b,1}(X^x_{T_1}),
X^x_{T_1}\} } \nonumber \\
& \geq &  \mathbb{E}^{X^x_{T_1}} \left[ \int_{T_1}^{T_N}e^{-r(u-T_1)}rP\,du+e^{-r(T_N-T_1)}\max\{V^\lambda_{b,1}(X^x_{T_N}),
X^x_{T_N}\} \right] \nonumber \\
& \geq & \mathbb{E}^{X^x_{T_1}} \left[\int_{T_1}^{T_N}
e^{-r(u-T_1)}rP\,du+e^{-r(T_N-T_1)}
X^x_{T_N} \right]. \label{eq:VNinequality}
\end{eqnarray}

Using the fact that $V^{\lambda}_{b,1}$ solves \eqref{eq:Vre} (and $V^{\lambda}_{b,1}(x) \leq x$ for $x \geq K$) we find that
\begin{equation}\label{supermart0}
% \nonumber % Remove numbering (before each equation)
V^{\lambda}_{b,1}(x) = \mathbb{E}^x \left[\int_0^{T_1}
e^{-ru}rP\,du+e^{-rT_1} \max\{ V^\lambda_{b,1}(X^x_{T_1}),X^x_{T_1}\} \right].
\end{equation}
Combining this with \eqref{eq:VNinequality} we find that for any $N \leq N_K$,
\begin{equation}\label{supermart01}
% \nonumber % Remove numbering (before each equation)
V^{\lambda}_{b,1}(x) \geq \mathbb{E}^x \left[\int_0^{T_N}
e^{-ru}rP\,du+e^{-rT_N}
X^x_{T_N} \right].
\end{equation}
Since $N$ is arbitrary, we conclude that $V^{\lambda}_{b,1} \geq v^\lambda_{b,1}$.

It remains to prove equality, and simultaneously to prove the optimality of $\eta^*=T_{N_{\xscol}}$ where
$N_{\xscol}=\inf \{n \geq 1: V^{\lambda}_{b,1}(X^x_{T_n})\leq  X^x_{T_n}\} = \inf \{n \geq 1: X^x_{T_n}\geq \xscol \}$.

Since $\left(Y_{n\wedge {N_{\xscol}}}\right)_{n\geq 1}$ is a uniformly integrable $\tilde{\mathbb{G}}$-martingale, it follows that
\begin{eqnarray*}
% \nonumber % Remove numbering (before each equation)
\lefteqn{\max\{V^{\lambda}_{b,1}(X^x_{T_1}),
X^x_{T_1}\} }\\
& = &  \mathbb{E}^{X^x_{T_1}} \left[ \int_{T_1}^{\eta}
e^{-r(u-T_1)}rP\,du+e^{-r(\eta^*-T_1)}\max\left\{V^\lambda_{b,1}\left(X^x_{\eta^*}\right),
X^x_{\eta^*}\right\} \right]\\
& = & \mathbb{E}^{X^x_{T_1}} \left[\int_{T_1}^{\eta^*}
e^{-r(u-T_1)} rP \,du+e^{-r(\eta^*-T_1)}
X^x_{\eta} \right].
\end{eqnarray*}

Hence, the inequality (\ref{eq:VNinequality}) becomes an equality with $N=N_{\xscol}$.
Combining this with \eqref{supermart0} gives that $V^{\lambda}_{b,1}(x) = \mathbb{E}\left[\int_0^{\eta^*}
e^{-ru}rP\,du+e^{-r \eta^*}
X^x_{\eta^*} \right]$.

This completes the proof and we have $V^\lambda_{b,1}=v_{b,1}^\lambda$.
\end{proof}

Proposition \ref{prop_bondholder_original} implies that the domain $(0,\infty)$ can be divided into the continuation region $(0,\xscol)$ and the stopping region $[\xscol,\infty)$. %Furthermore, we have the following properties of the value function $v^\lambda_{b,1}(x)$ and the optimal stopping time $\tau^*$.

\begin{remark}{
The proof of Proposition \ref{prop_bondholder_original} is included primarily for convenience, as it closely resembles the one found in Dupuis and Wang \cite[Theorem 1]{dupuis2002optimal}. In their work, Dupuis and Wang also addressed optimal stopping problems driven by a geometric Brownian motion with a Poisson constraint, and focused on a perpetual American call option. It is noteworthy that they only considered the case $\mu < r$ because otherwise, the problem is ill-posed.}
\end{remark}

%\begin{corollary}\label{cor_1_original} Suppose that $m=1$ and $\mu<r$. Suppose further that the surrender price satisfies $K\geq  \xscol$. Then, the value %function satisfies $v^\lambda_{b,1}(x)\leq K$ for $x\leq \xscol$.
%\end{corollary}

%\begin{proof} Since $v^\lambda_{b,1}(x)$ is increasing in $x$, we have $v^\lambda_{b,1}(x)\leq v^\lambda_{b,1}(\xscol)$ for $x\leq \xscol$. On the other hand, by the value matching at $\xscol$, we have $v^\lambda_{b,1}(\xscol)= \xscol$. The conclusion then follows from the assumption $ \xscol\leq K$.
%\end{proof}

\subsection{Case $m\in[0, 1)$, $\mu< \lambda+r$ and case $m=0$, $\mu>\lambda+r$}
Now we consider the problem where the bondholder does not have absolute priority. Note that we postpone the study of the case $\mu=\lambda+r$ to a separate section because the formul{\ae} are slightly different in that case.

For the parameter combinations we study in this section the optimal stopping boundary for the optimal stopping problem {(\ref{eq:OSPb})} does not have an explicit form as in (\ref{co1}). Instead, it will be
be determined by the fixed point of the function $f^m$ given by (\ref{f^m}) below.

\subsubsection{The optimal stopping boundary via a fixed point}
We start by defining a {family of functions $f^m:D\rightarrow \mathbb{R}^+$} where the superscript $m$ denotes the parameter in the simultaneous payoff, {and the domain $D \subseteq \mathbb{R}_+$ depends on the value of $\mu$ relative to $r$ and $r+\lambda$.
In particular, $D = [0,P\frac{\hat{m}}{\hat{m}-1}]$ for $\mu<r$, $D = \mathbb{R}_+$ for $r \leq \mu < \lambda + r$ and $D =[P\frac{\hat{m}}{\hat{m}-1},\infty)$ for $\mu \geq \lambda +r$.} Then, on it's domain $f^m$ is given by
\begin{equation}\label{f^m}
f^m(x)=K\frac{1}{(1-m)^{1/\al}}\left(\frac{P}{K}\hat{m}-\frac{\hat{m}-1}{K} x\right)^{1/\alpha_\lambda}
\end{equation}
where $\hat{m}$ is as given in (\ref{def_mhat}).

\begin{lemma}\label{lemma_flm_domain}
The function $f^m$ satisfies the following properties with respect to $x$,
{
\begin{enumerate}[label={(\arabic*)}]
    \item If $\mu<r$, then $f^m$ is decreasing on $D = [0,P\frac{\hat{m}}{\hat{m}-1}]$;
    \item If $\mu=r$, then $f^m$ is constant on $D=\mathbb{R}^+$;
    \item If $r<\mu<\lambda+r$, then $f^m$ is increasing on $D=\mathbb{R}^+$;
    \item If $\mu>\lambda+r$, then $f^0$ is increasing on $D=[P\frac{\hat{m}}{\hat{m}-1},\infty)$.
\end{enumerate}
}
\end{lemma}

\begin{proof}
    If $\mu<r$, then $\hat{m}>1$ by Lemma \ref{lemma_mhat_alpha}. This implies that $f^m(0)>0$, $P\frac{\hat{m}}{\hat{m}-1}>0$, and $f^m(P\frac{\hat{m}}{\hat{m}-1})=0$. Also, $\frac{P}{K}\hat{m}-\frac{\hat{m}-1}{K}x$ is a decreasing function, with zero point $P\frac{\hat{m}}{\hat{m}-1}$. Hence $f^m$ is decreasing on $[0,P\frac{\hat{m}}{\hat{m}-1}]$.

    If $\mu=r$, then $\hat{m}=1$ by Lemma \ref{lemma_mhat_alpha}. Hence $f^m$ is a constant function with $f^m(0)=K\frac{1}{(1-m)^{1/\al}}\frac{P^{1/\alpha_\lambda}}{K^{1/\alpha_\lambda}}>0$.

    If $\mu\in(r,\lambda+r)$, then $\hat{m}\in(0,1)$ by Lemma \ref{lemma_mhat_alpha}. This implies that $f^m(0)>0$ and is increasing.

    Finally, if $\mu>\lambda+r$, then $\hat{m}<0$ by Lemma \ref{lemma_mhat_alpha}. Therefore $\frac{P}{K}\hat{m}-\frac{\hat{m}-1}{K}x$ is increasing with zero point $P\frac{\hat{m}}{\hat{m}-1}$, hence $f^0$ is only well-defined on $[P\frac{\hat{m}}{\hat{m}-1},\infty)$, and $f^0$ is increasing on this interval.
\end{proof}

Next we define an auxiliary function $\hat{K}(m)$. {$\hat{K}(m)$ is a critical threshold for the surrender price $K$ and later we will see different behaviour depending on whether $K \geq \hat{K}(m)$ or $K<\hat{K}(m)$. Set}
\[\hat{K}(m)=P\frac{\hat{m}}{\hat{m}-m}\1_{\{0\leq  m<\hat{m}\}}+P\1_{\{\hat{m}\leq m=0\}}+\infty(\1_{\{0< \hat{m}\leq m\}} + \1_{\{\hat{m}\leq 0< m\}}),\hspace{5mm} m\in[0,1].\]
%\begin{figure}[h!]
%\centering
%\includegraphics[width=0.6\textwidth]{plots/sketch_convertible_small.jpg}
%\caption{Case $\mu<r$}
%\label{fig_sketch1}
%\includegraphics[width=0.6\textwidth]{plots/sketch_convertible_mod.jpg}
%\caption{Case $\mu\in[r,\lambda+r)$}
%\label{fig_sketch2}
%\end{figure}

%Recall that, when $\mu<r$ and $m=1$, Corollary \ref{cor_1_original} shows that the callable feature has no impact when $K$ is above $\xscol$, and one can check that $\hat{K}(1)=\xscol$. In fact, $\hat{K}(m)$ plays the role as the critical threshold for $K$, which suggests that bondholder's problem is related to the convertible bond problem when $K\geq \hat{K}(m)$.

\begin{figure}[H]
  \begin{minipage}[b]{0.5\linewidth}
    \includegraphics[width=\textwidth]{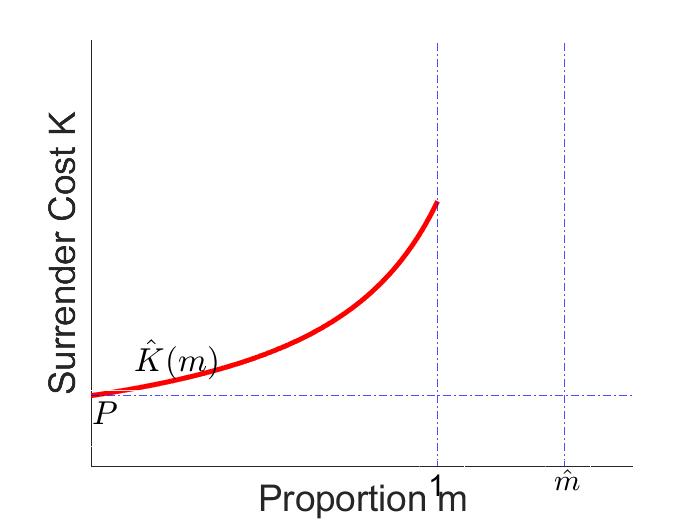}
\caption{The function $\hat{K}(m)$ for $\mu<r$, whence $\hat{m}>1$.}
\label{fig_sketch5}
  \end{minipage}
  \hfill
  \begin{minipage}[b]{0.5\linewidth}
    \includegraphics[width=\textwidth]{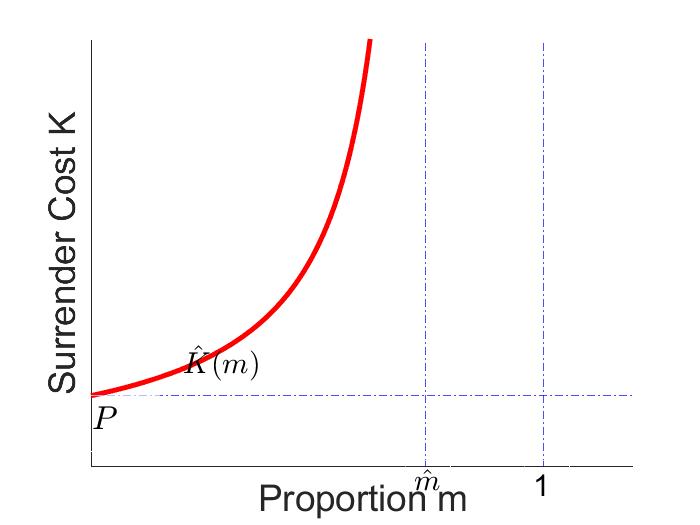}
\caption{The function $\hat{K}(m)$ for $\mu\in[r,\lambda+r)$, whence $\hat{m} \in (0,1]$.}
\label{fig_sketch6}
  \end{minipage}
\end{figure}

Figures \ref{fig_sketch5}-\ref{fig_sketch6} illustrate the critical threshold $\hat{K}(m)$ for the surrender price $K$.

\begin{lemma}\label{lemma_fixedpoint}
Let $\mu$ and $m$ satisfy either $\mu<\lambda+r$ and $m\in[0,1)$, or $\mu>\lambda+r$ and $m=0$. Then, the function $f^m$ has a unique fixed point $\xscol(m)$ in $D$. This fixed point satisfies $\xscol(m)<K$ if and only if $K > \hat{K}(m)$. %$(0,K]$ if and only if  $K\geq \hat{K}(m)$.
Moreover,  $\xscol(m)=K$ if and only if $K=\hat{K}(m)<\infty$.
\end{lemma}

\begin{proof}
See appendix.
\end{proof}

{
The fixed point $\xscol(m)$ {is defined in} the case $m\in[0,1)$. %As this fixed point acts as the optimal stopping boundary for the bondholder's optimal stopping problem (see Proposition \ref{prop_bondholder_1} below),
When $\mu<r$, we may extend its domain to the case $m=1$ by defining  $\xscol(1)=\xscol$ where $\xscol$ is given by (\ref{co1}).
\begin{remark}\label{remark:xscol}
From the expression of $f^m$, $f^m(x)$ is increasing in $m$ for fixed $x\in D$. Therefore the fixed point $\xscol(m)$ is increasing with respect to $m$ on $[0,1)$. In particular, when $\mu<r$, it follows from the domain $D$ of $f^m$ that $\xscol(m)\leq \xscol(1)$, so $\xscol(m)$ is increasing on $[0,1]$.
\end{remark}
}
\begin{corollary}\label{cor:flmproperty}
Let $\mu$ and $m$ satisfy either $\mu<\lambda+r$ and $m\in[0,1)$, or $\mu>\lambda+r$ and $m=0$. Then, $f^m(P)>P$ if $\mu<\lambda+r$ and $f^0(P)<P$ if $\mu>\lambda+r$. As a result, $\xscol(m)>P$.
\end{corollary}

\begin{proof}
This follows by substituting $x=P$ into $f^m(x)$.
\end{proof}

\begin{figure}[H]
  \begin{minipage}[b]{0.5\linewidth}
    \includegraphics[width=\linewidth]{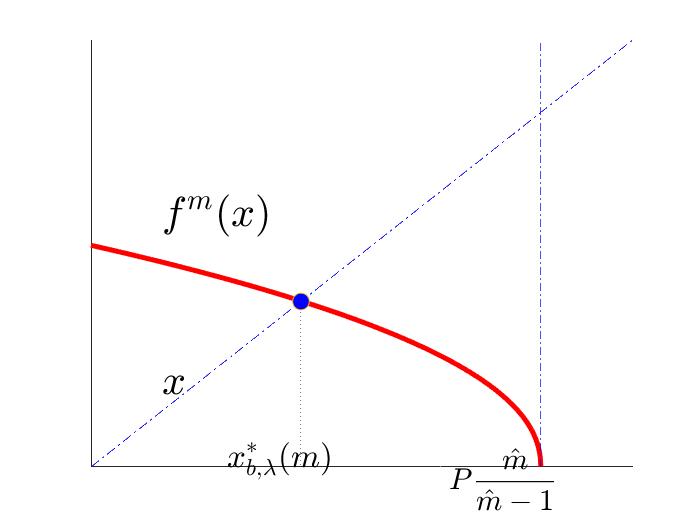}
    \caption{Function $f^m(x)$ for $\mu<r$}
    \label{fig_sketch1}
  \end{minipage}
   \hfill
  \begin{minipage}[b]{0.5\linewidth}
    \includegraphics[width=\linewidth]{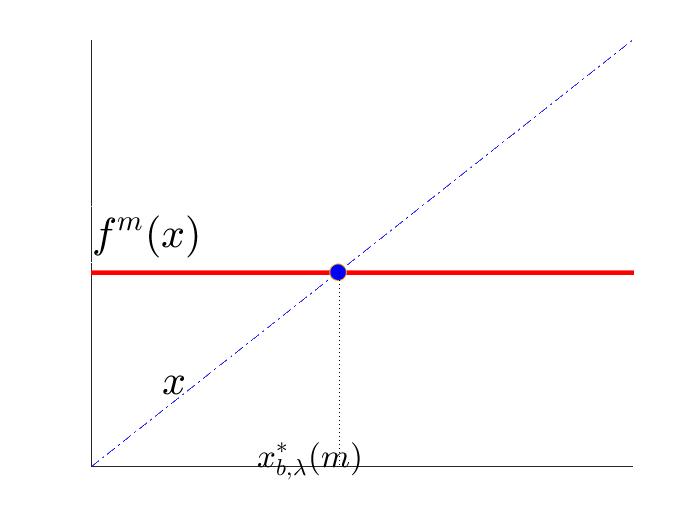}
    \caption{Function $f^m(x)$ for $\mu=r$}
  \end{minipage}
\end{figure}

\begin{figure}[H]
  \begin{minipage}[b]{0.5\linewidth}
    \includegraphics[width=\linewidth]{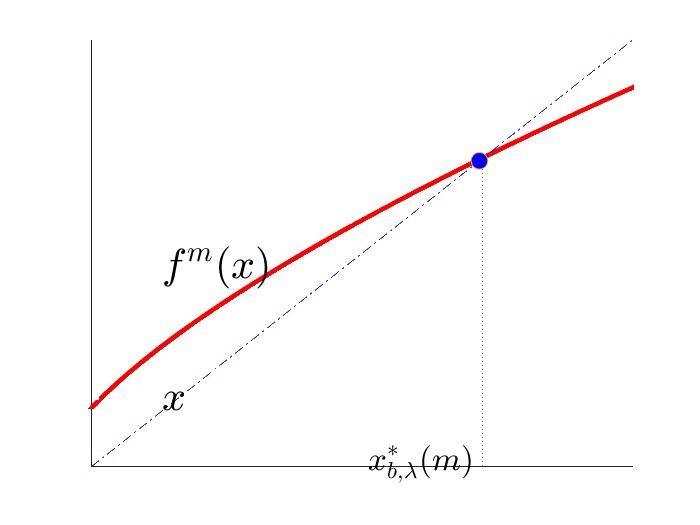}
    \caption{Function $f^m(x)$ for $\mu\in(r,\lambda+r)$}
  \end{minipage}
  \hfill
  \begin{minipage}[b]{0.5\linewidth}
    \includegraphics[width=\linewidth]{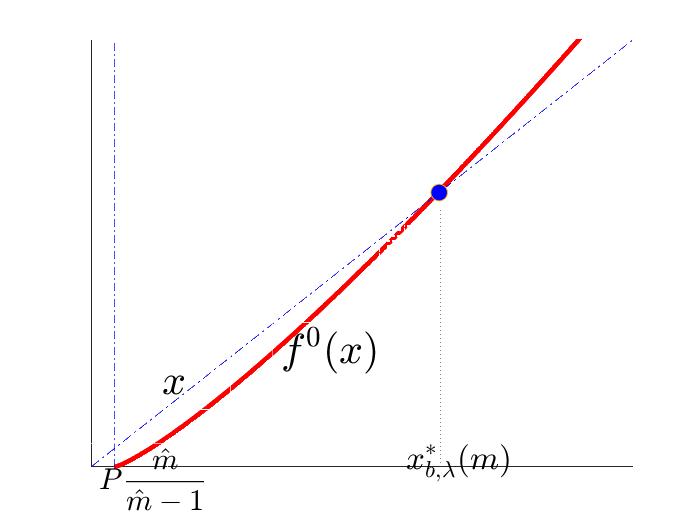}
    \caption{Function $f^m(x)$ for $\mu>\lambda+r$}
    \label{fig_sketch4}
  \end{minipage}
\end{figure}

Figures \ref{fig_sketch1}-\ref{fig_sketch4} illustrate the function $f^m$ with its fixed point $\xscol(m)$.


%Observe the following:
%\begin{itemize}
%    \item If $m=0$, then $\hat{K}(0)=P$ and there are only two cases: $K\leq P$ (`small' $K$) and $K>P$ (`large' $K$).
%    \item If $\mu\in[r,\lambda+r)$ and $m\in[\hat{m},1]$, then $\hat{m}=\infty$ and there are only two cases: $K\leq P$ (`small' $K$) and $K>P$ (`moderate' $K$).
%    \item Otherwise, there are always three cases: $K\leq P$, $K\in(P,\hat{K}(m))$ and $K\geq \hat{K}(m)$.
%\end{itemize}

%The plots Fig \ref{fig_sketch1} and \ref{fig_sketch2} should tell a clearer story. In particular, note that we only consider the bondholder's optimal stopping problem when $K\geq \hat{K}(m)$, therefore cases such as $m=1$, $\mu\in[r,\lambda+r)$ will not be considered in the next section.

%\subsubsection{Solving the optimal stopping problem}
We now define some auxiliary functions of $m$ under the assumption that $K\geq \hat{K}(m)$. These functions will serve as coefficients of the value function in the large $K$ case (which justifies the subscripts):
\begin{eqnarray}
\label{eq:A_{lar}}
A_{lar}(m)&=&\frac{(\alpha-\beta_\lambda \frac{r-\mu}{\lambda+r-\mu}-\frac{\lambda}{\lambda+r-\mu}) \xscol(m)-\alpha P+\beta_\lambda \frac{rP}{\lambda+r}}{\alpha_\lambda-\beta_\lambda};\\
\label{eq:B_{lar}}
 B_{lar}(m)&=&\frac{(\alpha_\lambda \frac{r-\mu}{\lambda+r-\mu}+\frac{\lambda}{\lambda+r-\mu}-\alpha) \xscol(m)+\alpha P-\alpha_\lambda \frac{rP}{\lambda+r}}{\alpha_\lambda-\beta_\lambda}; \\
\label{eq:C_{lar}}
C_{lar}(m)&=&A_{lar}(m)\frac{\alpha_\lambda}{\beta_\lambda}\theta(m)^{\alpha_\lambda}+B_{lar}(m)\theta(m)^{\beta_\lambda}+\frac{\lambda (1-m) K }{\bl(\lambda+r-\mu)},
%\label{eq:theta}
%\theta(m)&=&
%    \left(\frac{1}{A_{lar}(m)}\frac{1}{\beta_\lambda-\alpha_\lambda}\frac{\lambda (1-m) K}{\lambda+r-\mu}\left(1-\frac{\mu \beta_\lambda}{\lambda+r}\right)\right)^{1/\alpha_\lambda},
\end{eqnarray}
where $\theta(m)=\frac{K}{\xscol(m)}$.

The following lemma gives the signs of $A_{lar}$ and $B_{lar}$.
\begin{lemma}\label{lemma_large_coef}
Let $\mu$ and $m$ satisfy either $\mu<\lambda+r$ and $m\in[0,1)$, or $\mu>\lambda+r$ and $m=0$. Suppose further that $K\geq \hat{K}(m)$. Then, the following inequalities hold:
\begin{enumerate}[label={(\arabic*)}]
    \item  if $\mu<\lambda+r$ and $m\in[0,1)$ then $A_{lar}(m)<0$; if $\mu>\lambda+r$ then $A_{lar}(0)>0$;
%    \item if $\mu>\lambda+r$ then $A_{lar}(0)>0$;
    \item $B_{lar}(m) \geq 0$.
\end{enumerate}
\end{lemma}
\begin{proof}
Firstly, observe that $\xscol(m)=f^m(\xscol(m))$ implies that
\[ 0 <  \frac{K^{\alpha_\lambda}}{\xscol(m)^{\alpha_\lambda}}  = \frac{1-m}{\frac{P}{K}\hat{m}-\frac{\hat{m}-1}{K}\xscol(m)}.\]
Substituting the expression for $\hat{m}$ and rearranging, we obtain
\begin{eqnarray*}
    \frac{K^{\alpha_\lambda}}{\xscol(m)^{\alpha_\lambda}}  &=&\frac{\lambda(1-m)K(\lambda+r-\mu\bl)}{(\lambda+r)(\lambda+r-\mu)(\alpha P-\bl\frac{rP}{\lambda+r}+(\bl\frac{r-\mu}{\lambda+r-\mu}+\frac{\lambda}{\lambda+r-\mu}-\alpha)\xscol(m))}\\
    &=&\left(\frac{\lambda(1-m)K(\lambda+r-\mu\bl)}{(\lambda+r)(\lambda+r-\mu)A_{lar}(m)(\bl-\al)}\right).
\end{eqnarray*}
It is immediate that $\lambda(1-m)K$ and $\lambda+r$ are positive, whereas $\bl-\al$ is always negative. If $\mu\geq 0$, then it is clear that $\lambda+r-\mu\bl>0$; if $\mu<0$, then $\lambda+r-\mu\bl>0$ follows by Lemma \ref{lemma_alphabounds}. Hence, the term $(\lambda+r-\mu)A_{lar}(m)$ must be negative. If $\mu<\lambda+r$, then $\lambda+r-\mu>0$ and $A_{lar}(m)$ is required to be negative; if $\mu>\lambda+r$, then $\lambda+r-\mu<0$ and $A_{lar}(0)$ is positive.

To prove $B_{lar}(m) \geq 0$, it suffices to check the sign of its numerator. Note that, $\alpha P-\alpha_\lambda \frac{rP}{\lambda+r}\geq 0$ by Lemma \ref{lemma_alphaorder_r}.

If $\mu<r$, the result $A_{lar}(m)<0$ provides an upper bound for $ \xscol(m)$:
\begin{equation}\label{bound_xscol}
     \xscol(m)<\frac{\alpha P-\beta_\lambda \frac{rP}{\lambda+r}}{\alpha-\beta_\lambda \frac{r-\mu}{\lambda+r-\mu}-\frac{\lambda}{\lambda+r-\mu}}.
\end{equation}
Observe that, under the assumption $\mu<r$, we have $\alpha_\lambda \frac{r-\mu}{\lambda+r-\mu}+\frac{\lambda}{\lambda+r-\mu}-\alpha\leq 0$ by Lemma \ref{lemma_alphaorder_mu}. Hence it suffices to check the following inequality:
\begin{equation}\label{Brearrange}
    \left(\alpha_\lambda \frac{r-\mu}{\lambda+r-\mu}+\frac{\lambda}{\lambda+r-\mu}-\alpha\right)\frac{\alpha P-\beta_\lambda \frac{rP}{\lambda+r}}{\alpha-\beta_\lambda \frac{r-\mu}{\lambda+r-\mu}-\frac{\lambda}{\lambda+r-\mu}}+\alpha P-\alpha_\lambda \frac{rP}{\lambda+r}\geq 0,
\end{equation}
%After some algebra, this is equivalent to:
%\begin{equation}\label{Brearrange2}
%    \alpha_\lambda \frac{r}{\lambda+r}\frac{\lambda}{\lambda+r-\mu}-\alpha \alpha_\lambda \frac{r}{\lambda+r}-\alpha \beta_\lambda\frac{r-\mu}{\lambda+r-\mu}\geq \beta_\lambda \frac{r}{\lambda+r}\frac{\lambda}{\lambda+r-\mu}-\alpha\beta_\lambda \frac{r}{\lambda+r}-\alpha\alpha_\lambda \frac{r-\mu}{\lambda+r-\mu}.
%\end{equation}
which is equivalent to:
\begin{equation}
    \frac{(r -\alpha\mu) \lambda}{(\lambda+r)(\lambda+r-\mu)}\geq 0.
\end{equation}
Hence, under the assumption $\mu<r$, $B_{lar}(m) \geq 0$ is equivalent to $r \geq \alpha\mu$, which holds directly if $\mu\leq 0$, and follows by Lemma \ref{lemma_alphabounds} if $\mu>0$.

If $\mu=r$, $B_{lar}(m)\geq 0$ follows by Lemma \ref{lemma_alphabounds} as $\alpha=1$.

If $\mu\in(r,\lambda+r)$ or $\mu>\lambda+r$, then Lemma \ref{lemma_alphaorder_mu} implies that $\alpha_\lambda \frac{r-\mu}{\lambda+r-\mu}+\frac{\lambda}{\lambda+r-\mu}-\alpha\geq 0 $, and the result for the sign of $B_{lar}$ follows by a similar argument.
\end{proof}

\subsubsection{Solving the optimal stopping problem}
Recall that we are assuming $\mu<\lambda+r$ and $m\in[0,1)$, or assuming $\mu>\lambda+r$ and $m=0$. {In the case $\mu<\lambda + r$ the following result will be useful. The proof is given in the appendix.

\begin{lemma}\label{lemma_integrability}
    Suppose that $\mu<\lambda+r$. Then, for any $y>0$, $e^{-rT_{N_y}}X^x_{T_{N_y}}$ is integrable.

\end{lemma}
}

Define
\begin{equation}\label{V_co_m}
V^\lambda_{b,m}(x) = \left\{ \begin{array}{lll}
P+\frac{x^\alpha}{\xscol(m)^\alpha}( \xscol(m)-P), & \; & x<\xscol(m);\\
A_{lar}(m)\frac{x^{\alpha_\lambda}}{(\xscol(m))^{\alpha_\lambda}}+B_{lar}(m)\frac{x^{\beta_\lambda}}{(\xscol(m))^{\beta_\lambda}}+\frac{\lambda  x}{\lambda+r-\mu}+\frac{rP}{\lambda+r}, & \; &  \xscol(m) \leq x\leq K;\\
C_{lar}(m)\frac{x^{\beta_\lambda}}{K^{\beta_\lambda}}+\frac{\lambda (1-m)K+rP}{\lambda+r}+\frac{\lambda m x}{\lambda+r-\mu},& \; &    x> K,
\end{array} \right.
\end{equation}
where $\xscol(m)\in(0,K]$ is the unique fixed point of $f^m$ given in Lemma \ref{lemma_fixedpoint} and $A_{lar}$, $B_{lar}$ and $C_{lar}$ are defined by (\ref{eq:A_{lar}}-\ref{eq:C_{lar}}).

\begin{lemma} \label{lem:PropV} Suppose $\mu<\lambda+r$ and $m\in[0,\hat{m} \wedge 1)$, or $\mu>\lambda+r$ and $m=0$. Suppose that $K \geq \hat{K}(m)$. 

Then
$V^\lambda_{b,m}$ satisfies: $V^\lambda_{b,m}(0) = P$, $V^\lambda_{b,m}$ is of linear growth at infinity, $V^\lambda_{b,m}$ is $C^2$ with a first derivative which is bounded $(0,\infty)$; %$V^\lambda_{b,m}$ is increasing;
$x \leq V^\lambda_{b,m}(x) \leq \xscol(m) \leq K$ for $x\in(0,\xscol(m))$; $V^\lambda_{b,m}(\xscol)=\xscol$, $V^\lambda_{b,m} \leq x$ for $x\in(\xscol(m),K)$.

Further, $V^\lambda_{b,m}$ solves %the HJB equation
\begin{equation}
\label{eq:bondHJBunmodified}
\mathcal{L}_{\lambda}V(x)+rP+\lambda(\max\{V(x), x\}\1_{\{x<K\}}+(m x+(1-m)K)\1_{\{x\geq K\}})=0.
\end{equation}

\end{lemma}

\begin{proof}
%\bDGH{It turns out that, after substituting $z=\xscol(m)$, $v_{b,m}^\lambda$ defined in (\ref{V_co_m}) is a candidate value function which satisfies the boundary conditions and solves the corresponding HJB equation. In particular, it satisfies value matching and first order smooth fit at $\xscol(m)$ and $K$ via the following equalities, which can be verified by substituting (\ref{eq:A_{lar}}-\ref{eq:C_{lar}}),}
{It is easy to see that $V^\lambda_{b,m}(0) = P$, $V^\lambda_{b,m}(\xscol-) = \xscol$ and $V^\lambda_{b,m}$ is of linear growth at infinity.

Note that $V^\lambda_{b,m}$ and its first order derivative are 
continuous at $\xscol$ are equivalent to
{\footnotesize{
\begin{eqnarray*}
  A_{lar}(m)+B_{lar}(m)+\frac{\lambda  \xscol(m)}{\lambda+r-\mu}+\frac{rP}{\lambda+r}&=&\xscol(m);\label{large_eqn1_bondholder}\\
 \frac{\alpha_\lambda}{\xscol(m)} A_{lar}(m)+\frac{\beta_\lambda}{\xscol(m)} B_{lar}(m)+\frac{\lambda  }{\lambda+r-\mu}
 &=&\frac{\alpha}{\xscol(m)} ( \xscol(m)-P);\label{large_eqn2_bondholder}
\end{eqnarray*}
}}and these equations are easily seen to be satisfied, given the expressions for $A_{lar}$ and $B_{lar}$ in \eqref{eq:A_{lar}} and \eqref{eq:B_{lar}}.
Similarly, note that $V^\lambda_{b,m}$ and its first order derivative are 
continuous at $K$  are equivalent to
{\footnotesize{
\begin{eqnarray*}
 A_{lar}(m)\theta(m)^{\alpha_\lambda}+B_{lar}(m)\theta(m)^{\beta_\lambda}+\frac{\lambda K}{\lambda+r-\mu}&=&C_{lar}(m)+\frac{\lambda (1-m) K}{\lambda+r}+\frac{\lambda m K}{\lambda+r-\mu};\label{large_eqn3_bondholder}
 \\
 \frac{\alpha_\lambda}{K} A_{lar}(m)\theta(m)^\al+\frac{\beta_\lambda}{K} B_{lar}(m) \theta^\bl+\frac{\lambda }{\lambda +r-\mu}&=&\frac{\beta_\lambda}{K} C_{lar}(m)+\frac{\lambda m }{\lambda+r-\mu};\label{large_eqn4_bondholder}
\end{eqnarray*}
}}and again, some algebra and the formul\ae \eqref{eq:A_{lar}}-\eqref{eq:B_{lar}} together with the definition of $\theta(m)$ show that these are satisfied.

Next, we prove that $V^\lambda_{b,m}(x)>x$ on $(0,\xscol(m))$. For this purpose, taking $m$ as fixed, define $g_1 = g_1^m : (0,\infty) \rightarrow \mathbb{R}$ by $g_1(x):=P+\frac{x^\alpha}{\xscol(m)^\alpha}( \xscol(m)-P)- x$.

Suppose $\mu<r$. By Corollary \ref{cor:flmproperty}, $\xscol(m)>P$, and hence $V^\lambda_{b,m}$ is increasing on $(0,\xscol(m))$, which, together with Lemma \ref{lemma_fixedpoint}, implies that $V^\lambda_{b,m}(x)\leq \xscol(m) \leq K$ on $(0,\xscol(m))$. Further, $g_1(x)$ is convex by Lemma \ref{lemma_alphabounds}. By Lemma \ref{lemma_mhat_alpha}, $\hat{m}>\alpha$ and this implies that $\xscol(m)<P\frac{\alpha}{\alpha-1}$, and hence $\xscol(m)g_1'(\xscol(m))=(\alpha-1)\xscol(m)-\alpha P<0$. Therefore $g_1(x)$ must be decreasing on $(0,\xscol(m))$, which implies that $V^\lambda_{b,m}(x)>x$ on this interval since $g_1(\xscol)=0$.

If $\mu=r$, $V^\lambda_{b,m}(x)>x$ follows by direct computation, as we know the value of $\xscol(m)$ in this case. In the case $\mu\in(r,\lambda+r)$ or $\mu>\lambda+r$, $g_1$ is a concave function by Lemma \ref{lemma_alphabounds} and Corollary \ref{cor:flmproperty}, and it is easy to see that $g_1'(y)= - \frac{P}{y}<0$ at any zero point $y$ of $g_1$. Hence $\xscol(m)$ must be the unique zero point, and $V^\lambda_{b,m}(x)>x$ must hold on $(0,\xscol(m))$.

Next we prove $V^\lambda_{b,m}(x)< x$ on $(\xscol(m),K)$. We make use of the signs of $A_{lar}(m)$ and $B_{lar}(m)$ as described in Lemma \ref{lemma_large_coef}. Define $g_2 = g_2^m : (0,\infty) \rightarrow \mathbb{R}$ by $g_2(x)=A_{lar}(m)\frac{x^{\alpha_\lambda}}{(\xscol(m))^{\alpha_\lambda}}+B_{lar}(m)\frac{x^{\beta_\lambda}}{(\xscol(m))^{\beta_\lambda}}+\frac{\lambda  x}{\lambda+r-\mu}+\frac{rP}{\lambda+r}- x$.

If $\mu\leq r$, then it follows from Lemmas \ref{lemma_alphabounds} and \ref{lemma_large_coef} that
$$g_2'(x)=\frac{\al A_{lar}(m)}{(\xscol(m))^\al}x^{\al-1}+\frac{\bl B_{lar}(m)}{(\xscol(m))^\bl}x^{\bl-1}+\frac{\mu-r}{\lambda+r-\mu}\leq 0.$$
Hence, $g_2$ is a decreasing function with a zero point at $\xscol(m)$, which implies that $V^\lambda_{b,m}(x)< x$ on $(\xscol(m),K)$. If $\mu>r$, then at any zero point $y$ of $g_2$ we have
\begin{eqnarray*}
\lefteqn{y g_2'(y)} \\
    &=&\frac{\al A_{lar}(m)}{(\xscol(m))^\al}y^{\al}+\frac{\bl B_{lar}(m)}{(\xscol(m))^\bl}y^{\bl}+\frac{\mu-r}{\lambda+r-\mu} y\\
        &=& \frac{\al A_{lar}(m)}{(\xscol(m))^\al}y^{\al}+\frac{\bl B_{lar}(m)}{(\xscol(m))^\bl}y^{\bl}-\left(\frac{A_{lar}(m) }{(\xscol(m))^{\alpha_\lambda}}y^{\alpha_\lambda}+\frac{B_{lar}(m)}{(\xscol(m))^{\beta\lambda}}y^{\beta_\lambda}+\frac{rP}{\lambda+r}\right)\\
        &=& (\al-1)\frac{A_{lar}(m)}{(\xscol(m))^{\alpha_\lambda}}y^{\alpha_\lambda}+(\bl-1)\frac{B_{lar}(m)}{(\xscol(m))^{\beta\lambda}}y^{\beta_\lambda}-\frac{rP}{\lambda+r}<0,
\end{eqnarray*}
where the final inequality holds by Lemma \ref{lemma_alphabounds} and \ref{lemma_large_coef}. This implies that $g_2$ can only cross 0 in a downward direction (at least on $(\xscol,\infty)$). It follows that $V^\lambda_{b,m}(x)< x$
holds on $(\xscol(m),K)$.

Given that $V^\lambda_{b,m}$ is $C^1$ we immediately see by differentiation on the various regimes that $V^\lambda_{b,m}$ satisfies
\begin{equation}
\label{eq:bondHJBmodified}
\mathcal{L}_{\lambda}V(x)+rP+\lambda(V(x)\1_{\{x<\xscol\}}+ x\1_{\{\xscol\leq x<K\}} + (m x+(1-m)K)\1_{\{x\geq K\}})=0.
\end{equation}
Then, since $V^\lambda_{b,m}(x)>x$ for $x\in(0,\xscol)$ and  $V^\lambda_{b,m}(x)<x$ on $x\in(\xscol,K)$ the fact that $V^\lambda_{b,m}(x)>x$ solves \eqref{eq:bondHJBmodified} means that it also solves \eqref{eq:bondHJBunmodified}.
}
\end{proof}

Under the assumptions that $\mu < \lambda + r$, $0 \leq m \leq \hat{m}$ and $K\geq \hat{K}(m)$ we have that $\frac{\lambda m}{\lambda+r-\mu}<\frac{\lambda \hat{m}}{\lambda+r-\mu}=\frac{\alpha(\lambda+r)-\bl r}{\lambda+r-\mu\bl}$. By Lemma \ref{lemma_mhat_alpha}, $\frac{\alpha(\lambda+r)-\bl r}{\lambda+r-\mu\bl}\geq 1$ if $\mu\leq r$, and $\frac{\alpha(\lambda+r)-\bl r}{\lambda+r-\mu\bl}<1$ otherwise. Hence, even if we have $\mu>r$, $\frac{\lambda m}{\lambda+r-\mu}$ is still strictly dominated by 1. That is, the function $V^\lambda_{b,m}$ is eventually dominated by the identity function. Further, if $\mu \geq \lambda+r$ and $m=0$ then the fact that $V^\lambda_{b,m}$ is bounded follows immediately from the definition in \eqref{V_co_m}.

%(We don't need $V(x)<x$ on $(K,\infty)$.)}

\begin{proposition}\label{prop_bondholder_1}
%Let $\mu$ and $m$ satisfy Suppose further that $K\geq \hat{K}(m)$.
Suppose $\mu<\lambda+r$ and $m\in[0, \hat{m} \wedge 1)$, or $\mu>\lambda+r$ and $m=0$. Suppose $K \geq \hat{K}(m)$.

Then, the value function $v^{\lambda}_{b,m}$ of the optimal stopping problem {(\ref{eq:OSPb})} is given by
\[ v^{\lambda}_{b,m}(x) = V^{\lambda}_{b,m}(x).  \]
Further, the optimal stopping time  is given by $\tau^*_b=T_{N_{\xscol(m)}}\leq T_{N_K}$. % where $N_y$ is defined in (\ref{def_na}).
\end{proposition}

\begin{proof}
The proof is very similar to the proof of Proposition \ref{prop_bondholder_original} and is omitted. The only significant change is in
the proof that the stopped process $(Y_{ n \wedge {N_{x^*_{b,\lambda}}} })_{n \geq 1}$ is uniformly integrable, where in the general case
\begin{eqnarray*}
Y_n &=& \int_0^{T_n}e^{-ru}rPdu+e^{-rT_n}\max\{V^\lambda_{b,1}(X^x_{T_n}), mX^x_{T_n} + (1-m)K \}\1_{\{X^x_{T_n}<K\}} \\
&& \hspace{40mm}
+e^{-rT_n}(mX^x_{T_n} + (1-m)K) \1_{\{X^x_{T_n}\geq K\}}.
\end{eqnarray*}

When $m=0$ the uniform integrability follows from the fact that $V^\lambda_{b,0}$ is dominated by $K$, thanks to (\ref{V_co_m}).

When $\mu < \lambda +r$ the result will follow as in the proof of Proposition \ref{prop_bondholder_original} if we can show the uniform integrability of $\left(e^{-r(\eta^* \wedge T_n)}X^x_{\eta^* \wedge T_n}\right)_{n\geq 1}$ where $\eta^*=T_{N_{\xscol}}$.
Observe that $e^{-r T_n}X^x_{T_n}\leq K$ on the set $\{T_n<\eta^*\}$. Then the uniform integrability of $\left(e^{-r(\eta^* \wedge T_n)}X^x_{\eta^* \wedge T_n}\right)_{n\geq 1}$ follows from Lemma~\ref{lemma_integrability}.
\end{proof}

\begin{remark}\label{Remark_P_1}
If $m=0$ and $K=\hat{K}(0)=P$, then (\ref{V_co_m}) can be simplified to $V_{b,0}^\lambda(x)=P$ for any $\mu\neq \lambda+r$. Indeed, $\xscol(0)=\hat{K}(0)=P$ by Lemma \ref{lemma_fixedpoint}, and one can check directly that $C_{lar}(0)=0$ when $K=P$. The optimal strategy for the bondholder in this case is $\tau^*=T_{N_K}$.
\end{remark}

Proposition \ref{prop_bondholder_1} implies that, if $\mu$ and $m$ either satisfy $\mu<\lambda+r$, $m< 1$ or $\mu>\lambda+r$, $m=0$, and $K\geq \hat{K}(m)$, then the whole region $(0,\infty)$ can be divided into the continuation region $(0,\xscol(m))$ and the stopping region $[\xscol(m),\infty)$, with $V(x)>x$ on $(0,\xscol(m))$ and $V(x)<x$ on $(\xscol(m),K)$.

\begin{comment}Furthermore, we have the following properties of the value function $v^\lambda_{b,m}(x)$ and the optimal stopping time $\tau^*$, which is a counterpart of Corollary \ref{cor_1_original}.

\begin{corollary}\label{cor_1} Let $\mu$ and $m$ satisfy either $\mu<\lambda+r$ and $m\in[0,1)$, or $\mu>\lambda+r$ and $m=0$. Suppose further that $K\geq \hat{K}(m)$. Then the value function satisfies $v^\lambda_{b,m}(x)\leq K$ for $x\leq \xscol(m)$.
\end{corollary}

\begin{proof} Since $v^\lambda_{b,m}(x)$ is increasing in $x$ on $(0,\xscol(m))$, we have $v^\lambda_{b,m}(x)\leq v^\lambda_{b,m}(\xscol(m))$ for $x\leq \xscol(m)$. On the other hand, by the value matching at $\xscol(m)$, $v^\lambda_{b,m}(\xscol(m))= \xscol(m)$, which is dominated by $K$ as we checked in Lemma \ref{lemma_fixedpoint}.
\end{proof}
\end{comment}

%\textcolor{red}{In particular, we do not consider the bondholder's problem when $m\geq \hat{m}$ %and $\mu\in[r,\lambda+r)$, as $\hat{K}(1)=\infty$.}

%\end{document}

\subsection{Case $m=0, \mu=\lambda+r$, $K \geq P$}
The proofs of this section follow along similar arguments as in the previous section. The main difference here is that, the general solution of $\mathcal{L}_\lambda V+rP+\lambda x$ is now (note that $\al=1$ by Lemma \ref{lemma_alphabounds}) $Ax+Bx^\bl+\frac{rP}{\mu}-\frac{\lambda}{\frac{1}{2}\nu^2+\mu}x\log x$, which makes the computation more involved. We only present the final result here, as the proof is similar to that in Proposition \ref{prop_bondholder_1}.

\begin{proposition}\label{prop_bondholder_2}
    Suppose that $\mu=\lambda+r$, $m=0$ and $K\geq \hat{K}(0)=P$. Then, the value function $v_{b,0}^\lambda(x)$ is given by
\begin{equation}\label{V_co_3}
v^\lambda_{b,0}(x) = \left\{ \begin{array}{lll}
P+\frac{x^\alpha}{(\xscol(0))^\alpha}( \xscol(0)-P), & \; & x<\xscol(0);\\
A_{lar}(0)\frac{x}{\xscol(0)}+B_{lar}(0)\frac{x^{\beta_\lambda}}{(\xscol(0))^{\beta_\lambda}}+\frac{rP}{\mu}-\frac{\lambda}{\frac{1}{2}\nu^2+\mu}x\log x, & \; &  \xscol(0) \leq x\leq K;\\
C_{lar}(0)\frac{x^{\beta_\lambda}}{K^{\beta_\lambda}}+\frac{\lambda K+rP}{\lambda+r},& \; &    x> K,
\end{array} \right.
\end{equation}
where $\xscol(0)$ is characterised as the unique zero point of the function $f:\mathbb{R}^+\rightarrow\mathbb{R}$ defined by
$$f(x)=\left(\alpha-\bl\frac{r}{\mu}\right)\left(x-P\right)+(1-\bl)\frac{\lambda}{\frac{1}{2}\nu^2+\mu}x(\log x-\log K),$$
where, in turn, the constants $A_{lar}(0)$, $B_{lar}(0)$ and $C_{lar}(0)$ are given by
\begin{eqnarray*}
A_{lar}(0)&=&\frac{(\alpha-\beta_\lambda +\frac{\lambda}{\frac{1}{2}\nu^2+\mu}) \xscol(0)-\alpha P+\beta_\lambda \frac{rP}{\lambda+r}}{1-\beta_\lambda}+\frac{\lambda \xscol(0)\log(\xscol(0))}{\frac{1}{2}\nu^2+\mu};\\
 B_{lar}(0)&=&\frac{(1 -\frac{\lambda}{\frac{1}{2}\nu^2+\mu}-\alpha) \xscol(m)+\alpha P- \frac{rP}{\lambda+r}}{1-\beta_\lambda}; \\
C_{lar}(0)&=&A_{lar}(0)\frac{1}{\beta_\lambda}\frac{K}{\xscol(0)}+B_{lar}(0)\frac{K^{\beta_\lambda}}{\xscol(0)^\bl}-\frac{1}{\beta_\lambda}\frac{\lambda}{\frac{1}{2}\nu^2+\mu} K(1+\log K).
\end{eqnarray*}
The optimal stopping time for the problem (\ref{eq:OSPb}) is given by $\tau^*=T_{N_{\xscol(0)}}\leq T_{N_K}$. In particular, $\xscol(0)>P$,  $v_{b,0}^\lambda(x)\leq K$ for $x\in(0,\xscol(0))$ and $v_{b,0}^\lambda(x)$ is bounded for $x\geq 0$.
\end{proposition}

\begin{remark}\label{Remark_P_2}
If $m=0$ and $K=\hat{K}(0)=P$, then (\ref{V_co_3}) can be rewritten as: $v_{b,0}^\lambda(x)=P$. Indeed, it is straightforward {to see} that $x=P$ is a zero point of the function $f$, and one can check directly that $C_{lar}(0)=0$ when $K=P$. We also find that an optimal stopping rule is $\tau^*_b=T_{N_K}$.
\end{remark}

%Proposition \ref{prop_bondholder_2} implies that the region $(0,\infty)$ can be divided into the continuation region $(0,\xscol(0))$ and the stopping region $[\xscol(0),\infty)$. %Furthermore, we have the following property of the value function $v^\lambda_{b,0}(x)$ and the optimal stopping time $\tau^*$.

\section{Auxiliary optimal stopping problems for the firm}\label{sec:mod&small}
\subsection{The auxiliary optimal stopping problem}

In this section we consider an optimal problem faced by the firm. We use
the subscript $f$ to convey that a quantity arises from a bond problem faced by the firm.

{Consider the situation faced by the firm considering calling the convertible bond where the bondholder has declared a strategy of converting the bond at $T_{N_K}$. This leads to the following optimal stopping problem for the firm}
{
\begin{equation}\label{eq:OSPf}
\mbox{Find \, $\inf_{\sigma\in\mathcal{R}(\lambda),\sigma\leq T_{N_K}}E^{\mathcal{R}(\lambda),x}_f(\sigma)$ and $\sigma^*_f$ which achieves the infimum,}
\end{equation}
%and the optimiser $\sigma^*_f$ where
}
where
\begin{equation}
E^{\mathcal{R}(\lambda),x}_f(\sigma)  =
\mathbb{E} \left[ \int_0^{\sigma}
e^{-ru}rP\,du+e^{-r\sigma}(m\max\{ X^x_{\sigma},K\}+(1-m)K) \right].
\end{equation}
Focusing on the event times of the Poisson process, the problem can be restated as
{
\begin{equation}\label{eq:optimal_stopping_firm_1}
\mbox{Find \, $\inf_{N\in\mathcal{N}(\lambda),N\leq {N_K}}E^{\mathcal{N}(\lambda),x}_f(N)$ and $N^*_f$ which achieves the infimum,}
\end{equation}
}
where
\begin{equation*}
E^{\mathcal{N}(\lambda),x}_f (N)
= \mathbb{E} \left[ \int_0^{T_N}
e^{-ru}rP\,du+e^{-rT_N}(m\max\{ X^x_{T_N},K\}+(1-m)K) \right].
\end{equation*}

As for the bondholder's problem, we expect that the value function is a solution of the HJB equation
$$\mathcal{L}_{\lambda}V+rP+\lambda(\min\{V(x),K\}\mathbbm{1}_{\{x<K\}}+(m x+(1-m)K)\mathbbm{1}_{\{x\geq K\}})=0.$$
The aim in this section is to solve the above HJB equation \emph{explicitly} subject to {appropriate} boundary conditions, and then to verify that this candidate solution is the actual solution{ of the optimal stopping problem}.

We aim to cover all the cases which are well posed that were not covered in the previous section: in particular, $m \in [(0 \vee \hat{m}) \wedge 1,1]$ or $m \in [0,\hat{m}]$ and $K<\hat{K}(m)$. It turns out that the general situation can be divided into two cases: either $K$ is `small' ($K\leq P$)  or $K$ is `moderate'. When $K$ is `small', we consider any value of $m$ and $\mu$ such that the problem is well-posed. The case `$K$ is moderate' can be understood as `$K>P$ but $K$ is not large'. Since $\hat{K}(0)=P$, the case `$K$ is moderate' does not arise when $m=0$. First we consider the `moderate' case.

\subsection{Case $m\in(0,1]$, $\mu<\lambda+r$ and $K\in(P,\hat{K}(m))$}\label{ssec:firmmod}
Define $\xscal(m)$ by:
\begin{equation}\label{eq:xscal}
\xscal(m) =   K \left( \frac{1}{K-P} \frac{  m K }{\hat{m}}  \right)^{-1/\al}.
\end{equation}
\,
\begin{lemma}\label{lemma_xscal_mod}
Suppose that $\mu<\lambda+r$, $m\in( \hat{m} \vee 0,1]$ and $K\in(P,\hat{K}(m))$.
Then $\xscal(m)\leq K$.
\end{lemma}

\begin{proof}  We have
\begin{equation*}
 \xscal(m)\leq K  \Leftrightarrow
\frac{1}{K-P}\frac{mK}{\hat{m}}\geq 1
\Leftrightarrow  \frac{m}{\hat{m}}\geq \frac{K-P}{K} \Leftrightarrow K \leq \hat{K}(m).
\end{equation*}
\end{proof}

Fix $m$, under the assumption that $K\in(P,\hat{K}(m))$, we now define constants which will serve as coefficients of the value function in the moderate $K$ case:
\begin{eqnarray}
    \label{eq:Amod}
A_{mod} & = & \frac{K-P}{\al - \bl} \left(\alpha - \frac{r \bl}{\lambda + r} \right) > 0;  \\
\label{eq:Bmod}
B_{mod} & = & \frac{K-P}{\al - \bl} \left(\frac{r \al}{\lambda + r}-\alpha \right) < 0; \\
\label{eq:Cmod}
C_{mod}(m) & = & \frac{ A_{mod} \al}{\bl} \tilde{\theta}(m)^{\al} + B_{mod}  \tilde{\theta}(m)^\bl -
\frac{\lambda m K }{\lambda + r-\mu} \frac{1}{\bl},
\end{eqnarray}
where $\tilde{\theta}(m)=\frac{K}{\xscal(m)}$. The sign of $A_{mod}$ is immediate and the sign of $B_{mod}$ follows by Lemma \ref{lemma_alphaorder_r}.

Define:
\begin{equation}\label{V_f_m}
V_{f,m}^\lambda(x) = \left\{ \begin{array}{lll}
P+\frac{x^\alpha}{(\xscal(m))^\alpha}(K-P), & \; & x < \xscal(m); \\
A_{mod}\frac{x^{\alpha_\lambda}}{(\xscal(m))^{\alpha_\lambda}}+B_{mod}\frac{x^{\beta_\lambda}}{(\xscal(m))^{\beta\lambda}}+\frac{\lambda K+rP}{\lambda+r}, & & \xscal(m) \leq x \leq K; \\
C_{mod}(m)\frac{x^{\beta_\lambda}}{K^{\beta_\lambda}}+\frac{\lambda (1-m)K+rP}{\lambda+r}+\frac{\lambda m x}{\lambda+r-\mu}, & &  x>K.
\end{array} \right.
\end{equation}

\begin{lemma}\label{lem:firmprop}
Suppose that $\mu<\lambda+r$, $m\in(\hat{m} \vee 0,1]$ and $K\in(P,\hat{K}(m))$.

Then $V^\lambda_{f,m}$ satisfies: $V^\lambda_{f,m}(0)=P$, $V^\lambda_{f,m}$ is of linear growth at infinity, $V^\lambda_{f,m}$ is $C^2$ with a first derivative which is bounded on $(0,\infty)$; $x \leq V^\lambda_{f,m}(x)<K$ for $x\in(0,\xscal(m))$; $V^\lambda_{f,m}(\xscal(m))=K$; $V^\lambda_{f,m}(x)>K$ for $x\in(\xscal(m),K)$.

Further, $V^\lambda_{f,m}$ solves the HJB equation:
\begin{equation}\label{eq:modHJBunmodified}
    \mathcal{L}_{\lambda}V+rP+\lambda(\min\{V(x),K\}\mathbbm{1}_{\{x<K\}}+(m x+(1-m)K)\mathbbm{1}_{\{x\geq K\}})=0.
\end{equation}
\end{lemma}
\begin{proof}
It is easy to see that $V^\lambda_{f,m}(0)=P$, $V^\lambda_{f,m}(\xscal(m)-)=K$, $V^\lambda_{f,m}$ is of linear growth at infinity, and $V^\lambda_{f,m}$ is $C^2$ with bounded first derivative except possibly at $\xscal(m)$ and $K$.

Note that  $V^\lambda_{f,m}$ and its first order derivative are continuous at $\xscal(m)$ are equivalent to:
\begin{eqnarray}
    K & = & \frac{rP+\lambda K}{\lambda + r} + A_{mod}  + B_{mod}; \label{eq:1}\\
\left( K - P \right) \frac{\alpha}{\xscal(m)} & = & A_{mod} \frac{\al}{\xscal(m)} + B_{mod}  \frac{\bl}{\xscal(m)}. \label{eq:2}
\end{eqnarray}
Given the expressions for $A_{mod}$ and $B_{mod}$ it is easily seen that these equations are satisfied. This shows that $V^\lambda_{f,m}$ is $C^1$ at $\xscal(m)$. Similarly, note that $V^\lambda_{f,m}$ and its first order derivative are continuous at $K$ are equivalent to
\begin{eqnarray}
\label{eq:3}\\
\frac{\lambda K}{\lambda + r} + A_{mod} \tilde{\theta}(m)^\al + B_{mod}  \tilde{\theta}(m)^\bl & = &  \frac{\lambda m K}{\lambda + r-\mu}+\frac{\lambda (1-m)K}{\lambda + r} + C_{mod}(m);\notag \\
\\\label{eq:4}
  \frac{A_{mod}}{K} \al \tilde{\theta}(m)^\al +  \frac{B_{mod}}{K} \bl \tilde{\theta}(m)^\bl & = & \frac{\lambda m }{\lambda + r-\mu} + \frac{C_{mod}(m)}{K} \bl \notag.
\end{eqnarray}
Some algebra shows that the choices $\tilde{\theta}(m) = \frac{K}{\xscal(m)}$ and $C_{mod}$ given by \eqref{eq:Cmod} mean that \eqref{eq:3} and \eqref{eq:4} are satisfied.

Next we prove that $V^\lambda_{f,m}<K$ on $(0,\xscal(m))$ and $V^\lambda_{f,m}>K$ on $(\xscal(m),K)$. The first statement is immediate since $K>P$. For the second, from the signs of $A_{mod}$ and $B_{mod}$ we conclude that $V^\lambda_{f,m}$ is increasing. Since $V^\lambda_{f,m}(\xscal(m))=K$ the result follows.

Now we show that $V^\lambda_{f,m}$ solves the HJB equation \eqref{eq:modHJBunmodified}.
Considering the various regimes separately, $V_{f,m}^\lambda$ satisfies
\begin{equation}\label{eq:modHJBmodified}
    \mathcal{L}_{\lambda}V+rP+\lambda(V(x)\1_{\{x<\xscal(m)\}}+K\mathbbm{1}_{\{\xscal(m)\leq x<K\}}+(m x+(1-m)K)\mathbbm{1}_{\{x\geq K\}})=0.
\end{equation}
Since $V^\lambda_{f,m}(x)<K$ on $(0,\xscal(m))$ and $V^\lambda_{f,m}(x)>K$ on $(\xscal(m),K)$, this implies that $V^\lambda_{f,m}$ solves (\ref{eq:modHJBunmodified}). It further follows that $V^\lambda_{f,m}$ is $C^2$ on $(0,\infty)$.

It remains to prove that $V_{f,m}^\lambda(x)\geq x$ on $(0,\xscal(m))$. Since $V^\lambda_{f,m}(0)=P>0$, $V^\lambda_{f,m}$ is increasing on $(0,K)$ and $V^\lambda_{f,m}(\xscal(m))=K\geq \xscal(m)$, it suffices to show that $V^\lambda_{f,m}$ does not cross the identity function on $(0, \xscal(m)]$. In turn, this will follow if $(V^\lambda_{f,m})'(y) < 1$ at any crossing point $y \in (0, \xscal(m)]$ --- then $V^\lambda_{f,m}$ may cross down below $ x$ but cannot cross back above.

Let $y \in (0, \xscal(m)]$ be such that $V^{\lambda}_{f,m}(y) =  y$. Then
\( P + \left( K - P \right) \left( \frac{y}{\xscal(m)} \right)^\alpha =  y \)
and
\[ (V^\lambda_{f,m})'(y) < 1 \Leftrightarrow \frac{\alpha}{y} \left( y - P \right) <  1 \Leftrightarrow (\alpha-1)<P\frac{\alpha}{y}. \]
If $\mu\geq r$, then $\alpha-1\leq 0$ by Lemma \ref{lemma_alphabounds} and the final inequality is trivial. If $\mu<r$, then the final inequality is equivalent to $y <   P \frac{\alpha}{\alpha-1}$, which holds as $y\leq \xscal(m)< \hat{K}(m)=P\frac{\hat{m}}{\hat{m}-m}\leq P\frac{\alpha}{\alpha-1}$.
\end{proof}
\begin{proposition}\label{prop_firm_mod}
Suppose that $\mu<\lambda+r$ and $m\in(0,1]$. Suppose further that $K\in(P,\hat{K}(m))$.

Then, the value function $v^{\lambda}_{f,m}(x)$ of the optimal stopping problem (\ref{eq:OSPf}) is given by:
\[v_{f,m}^\lambda(x)=V_{f,m}^\lambda(x).\]
Further, the optimal stopping time is given by $\sigma^*_f=T_{N_{\xscal(m)}}\leq T_{N_K}$.
\end{proposition}

\begin{proof}
The proof is similar to the one in Proposition \ref{prop_bondholder_original}, where we use Lemma~\ref{lemma_integrability} to prove uniform integrability, as in the proof of Proposition~\ref{prop_bondholder_1}.
\end{proof}

%\begin{corollary}\label{cor_3} Suppose that $\mu<\lambda+r$ and $m\in(0,1]$. Suppose further that $K\in(P,\hat{K}(m))$. Then, the value function $v_{f,m}^\lambda(x)$  satisfies $v^\lambda_{f,m}(x)\geq   x$ for $x\leq \xscal(m)$.
%\end{corollary}

%\begin{proof} Note that $v^{\lambda}_{f,m}(x)$ is increasing in $x$, $v^\lambda_{f,m}(0) = P>0$ and $v^\lambda_{f,m}(\xscal(m)) = K >  \xscal(m)$, so it is sufficient to show that $v^\lambda_{f,m}$ does not cross $ x$ on $(0, \xscal(m)]$. In turn, this will follow if $(v^\lambda_{f,m})'(y) < 1$ at any crossing point $y \in (0, \xscal(m)]$ --- then $v^\lambda_{f,m}$ may cross down below $ x$ but cannot cross back above.

%Let $y \in (0, \xscal(m)]$ be such that $v^{\lambda}_{f,m}(y) =  y$. Then
%\( P + \left( K - P \right) \left( \frac{y}{\xscal} \right)^\alpha =  y \)
%and
%\[ (v^\lambda_{f,m})'(y) < 1 \Leftrightarrow \frac{\alpha}{y} \left( y - P \right) <  1 \Leftrightarrow (\alpha-1)<P\frac{\alpha}{y}. \]
%If $\mu\geq r$, then $\alpha-1\leq 0$ by Lemma \ref{lemma_alphabounds} and the final inequality is trivial. If $\mu<r$, then the final inequality is equivalent to $y <   P \frac{\alpha}{\alpha-1}$, which holds as $y\leq \xscal(m)\leq \hat{K}(m)=P\frac{\hat{m}}{\hat{m}-m}\leq P\frac{\alpha}{\alpha-1}$. \end{proof}

\subsection{Case $K\leq P$}
We suppose $\mu < \lambda +r$ or $\mu \geq \lambda+r$ and $m=0$.
Define the constants
\begin{eqnarray}
\label{eq:As}
    A_{sma}&=&\frac{\lambda  K}{\alpha_\lambda-\beta_\lambda}\frac{(\lambda+r)-\beta_\lambda\mu}{(\lambda+r-\mu)(\lambda+r)};\\
    \label{eq:Bs}
    B_{sma}&=&\frac{\lambda  K}{\alpha_\lambda-\beta_\lambda}\frac{(\lambda+r)-\alpha_\lambda\mu}{(\lambda+r-\mu)(\lambda+r)},
\end{eqnarray}
and the candidate value function
\begin{equation}\label{V_fs_m}
V_{f,m}^\lambda(x) = \left\{ \begin{array}{lll}\frac{rP+\lambda K}{\lambda+r}+mA_{sma}\frac{x^{\alpha_\lambda}}{K^{\alpha_\lambda}}, & \; & x<K; \\
mB_{sma}\frac{x^{\beta_\lambda}}{K^{\beta_\lambda}}+\frac{\lambda m  x}{\lambda+r-\mu}+\frac{\lambda(1-m) K+rP}{\lambda+r}, & & x \geq  K.
\end{array}\right.
\end{equation}
Note that if $\mu \geq \lambda+r$ then $m=0$ and $A_{sma}$ and $B_{sma}$ do not enter into the definition of $V_{f,m}^\lambda$. Otherwise, if $\mu<\lambda+r$ then $A_{sma}$ and $B_{sma}$ are positive by Lemma~ {(\ref{lemma_alphabounds})}.

\begin{lemma}
 Suppose that $K\leq P$, $m=0$ if $\mu\geq \lambda+r$ and $m\in[0,1]$ if $\mu<\lambda+r$.

 Then $V^\lambda_{f,m}$ satisfies: $V^\lambda_{f,m}(0)=\frac{rP+\lambda K}{\lambda+r}$, $V^\lambda_{f,m}$ is of linear growth at infinity, $V^\lambda_{f,m}$ is $C^2$ with a first derivative which is bounded on $(0,\infty)$; $V^\lambda_{f,m}\geq K$ on $(0,\infty)$.

 Further, $V^\lambda_{f,m}$ solves the HJB equation:
 \begin{equation}\label{eq:smallHJBunmodified}
     \mathcal{L}_{\lambda}V+rP+\lambda(\min\{V(x),K\}\mathbbm{1}_{\{x<K\}}+(m x+(1-m)K)\mathbbm{1}_{\{x\geq K\}})=0.
 \end{equation}

\end{lemma}
\begin{proof}
It is easy to see that $V_{f,m}^\lambda(0)=\frac{rP+\lambda K}{\lambda+r}\geq K$ and $V_{f,m}^\lambda$ is linear growth at infinity. Also, except at $K$, $V_{f,m}^\lambda$ is immediately seen to be increasing and $C^2$ with a first derivative which is bounded.

It remains to check that $V_{f,m}^\lambda$ and its first order derivative are continuous at $K$. This is equivalent to:
\begin{eqnarray*}
  \frac{\lambda K+rP}{\lambda+r}+mA_{sma}  & = & \frac{\lambda(1-m)K+rP}{\lambda+r}+\frac{\lambda m K}{\lambda+r-\mu}+mB_{sma};  \\
 \alpha_\lambda mA_{sma}  & = & \frac{\lambda m K}{\lambda+r-\mu}+\beta_\lambda mB_{sma},
\end{eqnarray*}
and these identities can be shown to be satisfied using the definitions of $A_{sma}$ and $B_{sma}$ and some algebra.
Finally, one can check by taking derivatives on the different regimes that $V^\lambda_{f,m}$ satisfies
$\mathcal{L}_{\lambda}V+rP+\lambda(K \mathbbm{1}_{\{x<K\}}+(m x+(1-m)K)\mathbbm{1}_{\{x\geq K\}})=0.$
Given that $V_{f,m}^\lambda\geq K$, it follows that $V^\lambda_{f,m}$ also satisfies (\ref{eq:smallHJBunmodified}).
\end{proof}
\begin{proposition}\label{lemma_firm_small}
Suppose that $K\leq P$, $m=0$ if $\mu\geq \lambda+r$ and $m\in[0,1]$ if $\mu<\lambda+r$.

Then, the value function $v_{f,m}^\lambda(x)$ of the optimal stopping problem (\ref{eq:OSPf}) is given by:
\[v_{f,m}^\lambda(x)=V^\lambda_{f,m}(x).\]
Further, the optimal stopping time is given by $\sigma^*_f=T_1$.
\end{proposition}
\begin{proof}
The proof is similar to that of Proposition \ref{prop_bondholder_original}, hence is omitted.
\end{proof}
\begin{remark}\label{Remark_P_3}
Note that, if $K=P$ and $m=0$, then the optimal stopping time of the problem (\ref{eq:OSPf}) is not unique, as any $\sigma\leq T_{N_K}$ is optimal. Further, the value (\ref{V_fs_m}) satisfies $v_{f,m}^\lambda(x)=P$. In particular, this agrees with the value of the bondholder's problem in Proposition \ref{prop_bondholder_1} and \ref{prop_bondholder_2}, as shown in Remark \ref{Remark_P_1} and \ref{Remark_P_2}.

\end{remark}

%%%%%%%%%%%%%%%%%%%%%%%%%%%%%%%%%%%%%%%%%
\section{Pricing the callable convertible bond}\label{sec:pricing}

In this section, we will make use of the auxiliary optimal stopping problems we studied in previous sections and conclude that the value of the game agrees with the value of a corresponding optimal stopping problem, via Theorem~\ref{theorem:reduction1} or \ref{theorem:reduction2}.

\begin{theorem}\label{theorem:value}
(Large $K$ case). Suppose that the parameters $\mu$ and $m$ are such that either
\begin{enumerate}[label={(\arabic*)}]
    \item $\mu<r$ and $m=1$; or
    \item $\mu<\lambda+r$ and $m\in[0,\hat{m})$; or
    \item $\mu\geq \lambda+r$ and $m=0$.
\end{enumerate}
Suppose further that $K\geq \hat{K}(m)$. Then,
$(\tau^*_{b},\gamma)=(T_{N_{\xscol(m)}},T_{N_{K}})$
is a saddle point for $J^{m,x}$ in (\ref{upperValues_1})-(\ref{lowerValues_1}), and the game value is given by $v^\lambda_{ca,m}(x)=v^{\lambda}_{b,m}(x)$, with $v^{\lambda}_{b,m}(x)$ given in Proposition \ref{prop_bondholder_original}, \ref{prop_bondholder_1} or \ref{prop_bondholder_2} depending on the assumptions on the parameter $\mu$.

(Moderate $K$ case). Suppose that $m\in(0,1]$ and $\mu<\lambda+r$. Suppose further that $K\in(P,\hat{K}(m))$. Then,
$(\gamma,\sigma^*_{f})=(T_{N_{K}},T_{N_{\xscal(m)}})$
is a saddle point for $J^{m,x}$ in (\ref{upperValues_1})-(\ref{lowerValues_1}), and the game value is given by $v^\lambda_{ca,m}(x)=v^{\lambda}_{f,m}(x)$, with $v^{\lambda}_{f,m}(x)$ given in Proposition \ref{prop_firm_mod}.

(Small $K$ case). Suppose that that $K\leq P$ and the parameters $\mu$ and $m$ satisfy either
\begin{enumerate}[label={(\arabic*)}]
    \item $\mu<\lambda+r$ and $m\in[0,1]$;  or
    \item $\mu\geq \lambda+r$ and $m=0$.
\end{enumerate}
Then,
$(\gamma,\sigma^*_{f})=(T_{N_K},T_1)$
is a saddle point for $J^{m,x}$ in (\ref{upperValues_1})-(\ref{lowerValues_1}), and the game value is given by $v_{ca,m}^\lambda=v_{f,m}^\lambda$, with $v_{f,m}^\lambda(x)$ defined in Proposition \ref{lemma_firm_small}.
\end{theorem}
\begin{proof}
    For the large $K$ case, it suffices to check all the assumptions in Corollary~\ref{cor_reductionc1}. Corollary \ref{cor_reductionc2} can be used to prove the moderate and small $K$ cases.

    The payoff processes $L,M$ and $U$ can be written in the form stated in Corollary~\ref{cor_reductionc1} and \ref{cor_reductionc2}, with $l(x)=x-P$, $m(x)=mx+(1-m)K-P$, $u(x)=K-P$ and $p=P$. The game is driven by a geometric Brownian motion $X$ which is strong Markov, and the generalised order condition is satisfied. We have that $\gamma=T_{N_K}\in\mathcal{R}(\lambda)$, with $M_\infty=P$ on the set $\{\gamma=\infty\}$.

    For the large $K$ case, the corresponding optimal stopping problem mentioned in Corollary \ref{cor_reductionc1} is (\ref{eq:OSPb}), which has been solved in Section \ref{sec:large}, with value $v_{b,m}^\lambda$ and optimal strategy $\tau^*_b$ defined in Proposition \ref{prop_bondholder_original}, \ref{prop_bondholder_1} or \ref{prop_bondholder_2}. Also, we have that $v_{b,m}^\lambda(x)\leq K=u(x)+P$ on the continuation region $(0,\xscol(m))$ by Lemma \ref{lem:bondholderoriginalprop}, \ref{lem:PropV} or Proposition \ref{prop_bondholder_2}, depending on the assumptions on $\mu$.  Hence the game has a value $v_{b,m}^\lambda$ and saddle point $(\tau^*_b,\gamma)$ by Corollary \ref{cor_reductionc1}.

    For the moderate and small $K$ cases, the corresponding optimal stopping problem mentioned in Corollary \ref{cor_reductionc2} is (\ref{eq:OSPf}), which has been solved in Section \ref{sec:mod&small}, with value $v_{f,m}^\lambda$ and optimal strategy $\sigma^*_f$ defined in Proposition \ref{prop_firm_mod} or \ref{lemma_firm_small}. Also, we have that $v_{f,m}^\lambda(x)\geq x=l(x)+P$ on the continuation region by Lemma \ref{lem:firmprop} for the moderate $K$ case, while there is no continuation region in the small $K$ case. Hence game values and saddle points exist by Corollary \ref{cor_reductionc2}.
\end{proof}

\begin{figure}[H]
  \begin{minipage}[b]{0.5\linewidth}
    \includegraphics[width=\textwidth]{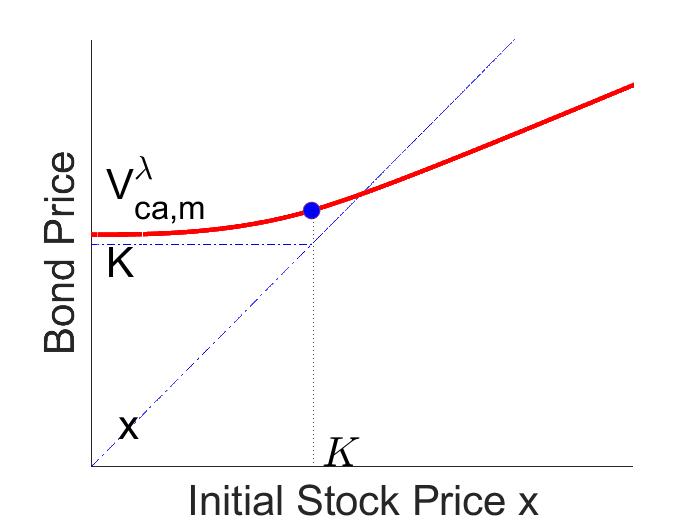}
  \caption{The convertible bond value $v^{\lambda}_{ca,m}$ for the case $K\leq P$, $m\in(0,1]$, $\mu<\lambda+r$.
  %and $\frac{\lambda m}{\lambda+r-\mu}<1$.
  If $m=0$, the value is a constant.
  %If $\frac{\lambda m}{\lambda+r-\mu}\geq 1$, then $x$ stays below $v_{ca,m}^\lambda(x)$.
  }
  \label{fig_sketchv_s}
  \end{minipage}
  \hfill
  \begin{minipage}[b]{0.5\linewidth}
   \includegraphics[width=\textwidth]{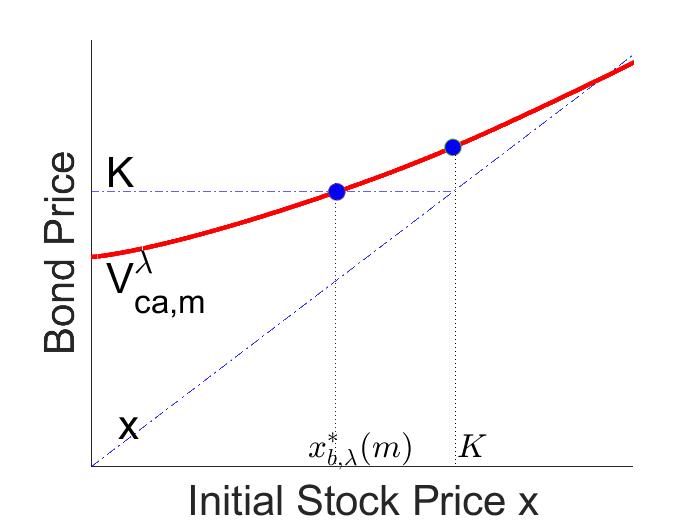}
  \centering
  \caption{The convertible bond value $v^{\lambda}_{ca,m}$ for Case $P<K<\hat{K}(m)$, $\mu<r$.}
  \label{fig_sketchv_mod}
  \end{minipage}
\end{figure}

\begin{figure}[H]
  \begin{minipage}[b]{0.5\linewidth}
 \includegraphics[width=\textwidth]{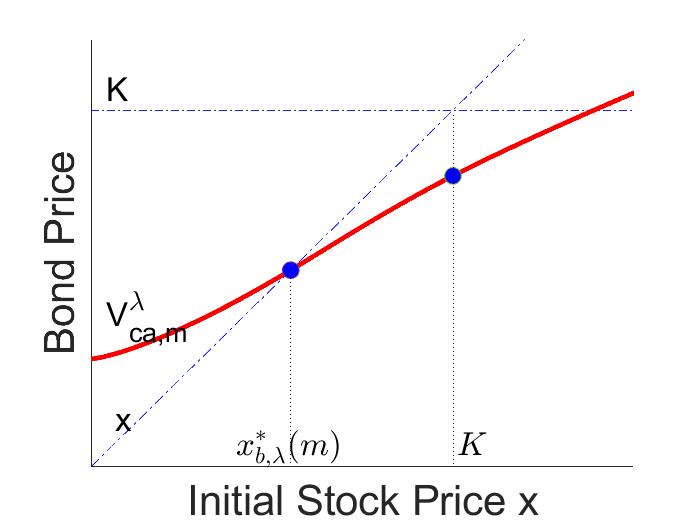}
    \caption{The convertible bond value $v^{\lambda}_{ca,m}$ for the case $K\geq \hat{K}(m)$,
     $m\in(0,1]$ and $\mu<r$. If we take $\mu>r$ and $m\in(0,\hat{m})$ then the value function is concave on $(0,\xscol)$.}
    \label{fig_sketch_lar_01}
  \end{minipage}
   \hfill
  \begin{minipage}[b]{0.5\linewidth}
   \includegraphics[width=\textwidth]{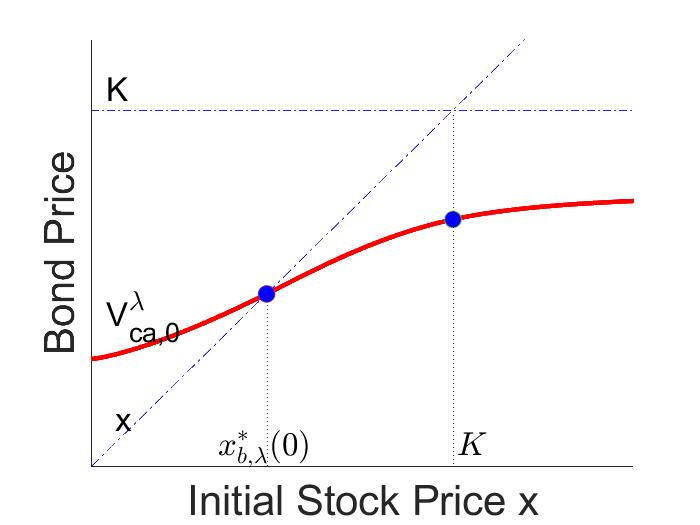}
  \caption{The convertible bond value $v^{\lambda}_{ca,0}$ for Case $K\geq \hat{K}(0)=P$, $m=0$ and $\mu<r$. Note the value is bounded by $K$.}
  \label{fig_sketch_lar_0}
  \end{minipage}
\end{figure}

We summarise all the different cases in our analysis and the corresponding auxiliary optimal stopping problem via the following table: {Note that there is no contradiction for the two cases $m=0, K\leq P$ and $m=0, K\geq P$, as demonstrated in Remark \ref{Remark_P_3}.  Additionally, it's important to recognize that the cases corresponding to the two gray cells do not exist when $m\in[\hat{m},1]$, as $\hat{K}(m)=\infty$.}
\begin{table}[H]
    \caption{The auxiliary optimal stopping problem that is used to find the value}
    \scriptsize\begin{tabular}{c c c c c }
       \hline
           & $\mu<r$    & $r\leq \mu<\lambda+r$  & $\mu=\lambda+r$ &  $\mu>\lambda+r$\cr\cline{1-5}
   \hline
           \multicolumn{5}{l}{\cellcolor{yellow!25}{$m=0$}}\cr
    \hline
            $K\leq P=\hat{K}(0)$ &  \multicolumn{4}{|c}{Firm Proposition \ref{lemma_firm_small}}\cr\cline{1-5}
            $K\geq P=\hat{K}(0)$ &  \multicolumn{2}{|c|}{\shortstack{Bondholder\\ Proposition \ref{prop_bondholder_1}}} & \multicolumn{1}{c|}{\shortstack{Bondholder\\ Proposition \ref{prop_bondholder_2}}} & \multicolumn{1}{c}{\shortstack{Bondholder\\ Proposition \ref{prop_bondholder_1}}}\cr
   \hline
         \multicolumn{5}{l}{\cellcolor{red!25}{$m\in(0,1)$}}\cr
  \hline
           $K\leq P$ & \multicolumn{2}{|c|}{Firm Proposition \ref{lemma_firm_small}} & \multicolumn{2}{c}{ \multirow{3}{*}{\shortstack{Bondholder\\ Remark \ref{remark_large_mu}}}}\\
           $P<K<\hat{K}(m)$ & \multicolumn{2}{|c|}{Firm Proposition \ref{prop_firm_mod}} & \multicolumn{2}{c}{}\\\cline{1-3}
           $K\geq \hat{K}(m)$ & \multicolumn{1}{|c|}{\shortstack{Bondholder\\ Proposition \ref{prop_bondholder_1}}} &\multicolumn{1}{c|}{\cellcolor{gray!25}{\shortstack{\\[0.05ex]Bondholder\\ Proposition \ref{prop_bondholder_1}}}} &\multicolumn{2}{c}{}\\
   \hline
         \multicolumn{5}{l}{\cellcolor{blue!25}{$m=1$}}\cr
   \hline       $K\leq P$ & \multicolumn{2}{|c|}{Firm Proposition \ref{lemma_firm_small}} & \multicolumn{2}{c}{ \multirow{3}{*}{\shortstack{Bondholder\\ Remark \ref{remark_large_mu}}}}\\
          $P<K<\hat{K}(1)=\xscal$ & \multicolumn{2}{|c|}{Firm Proposition \ref{prop_firm_mod}} & \multicolumn{2}{c}{}\\\cline{1-3}
          $K\geq \hat{K}(1)=\xscal$ & \multicolumn{1}{|c|}{\shortstack{Bondholder\\ Proposition \ref{prop_bondholder_original}}} &\multicolumn{1}{c|}{\cellcolor{gray!25}{\shortstack{\\[0.05ex]Bondholder\\ Proposition \ref{prop_bondholder_1}}}}  & \multicolumn{2}{c}{}\\
              \hline
    \end{tabular}  \label{table1}
      \end{table}

\section{Sensitivity of the callable convertible bond to key parameters}
\subsection{Dependence of the game value and the optimal stopping boundary on the proportion $m$}
We can interpret $m$ as a parameter which measures the relative priority of the two agents. Increasing $m$ represents increasing priority for the bondholder. For this reason, intuition would suggest that, {\em ceteris paribus}, the value function should be increasing in $m$ for each $x$.
However, the payoff function is not a monotonically increasing function of $m$, so this is not a trivial result. Nonetheless, the intuition is correct, as argued in the following proposition.

\begin{proposition}\label{prop:increase}
Fix $0\leq m'<m\leq 1$, $\lambda>0, r>0$ and $\mu<\lambda+r$. Then, for any $x>0$:
$$v_{ca,m'}^\lambda(x)\leq v_{ca,m}^\lambda(x).$$
\end{proposition}
\begin{proof}
Recall the definition of the payoff $R^m(\tau,\sigma)$ in (\ref{eq:convertible_bond_payoff_1}). Now we introduce an alternative payoff $R^{c}(\tau,\sigma)$ obtained by modifying the simultaneous payoff $M_\cdot$. In particular, define
{\footnotesize{
\begin{eqnarray*}
R^{c}(\tau,\sigma) & = & \int_0^{\sigma\wedge \tau}
e^{-ru}rP\,du+e^{-r\tau}
X^x_{\tau}\mathbbm{1}_{\{\tau<\sigma\}}+e^{-r\sigma}K\mathbbm{1}_{\{\sigma<\tau\}}\\
& &+e^{-r\tau}(m' X^x_{\tau}+(1-m')K)\mathbbm{1}_{\{\tau=\sigma, X^x_\tau<K\}}+e^{-r\tau}(m X^x_{\tau}+(1-m)K)\mathbbm{1}_{\{\tau=\sigma,X^x_\tau\geq K\}}.\notag
\end{eqnarray*}
}}
Note that $R^{c}(\tau,\sigma)$ and $R^{m}(\tau,\sigma)$ only differ in the simultaneous payoff. We have $R^{c}(\tau,\sigma)\geq R^{m}(\tau,\sigma)$ but the inequality is strict only on the set $\{\tau=\sigma,X^x_\tau<K\}$. We aim to apply Theorem \ref{theorem_value_2} to show that the value $v^\lambda_{ca,c}$ of the Dynkin game with payoff $R^{c}(\tau,\sigma)$ exists and is equal to
$v_{ca,m}^\lambda(x)$. To this end, we take
\begin{equation*}
\left\{ \begin{array}{lll}
U_t=\int_0^te^{-ru}rP\,du+e^{-rt}K ; \\
L_t=\int_0^te^{-ru}rP\,du+e^{-rt}X^x_t; \\
M^1_t=\int_0^te^{-ru}rP\,du+e^{-rt}(mX^x_t+(1-m)K);\\
M^2_t=\int_0^te^{-ru}rP\,du+e^{-rt}((m'X^x_t+(1-m')K)\1_{\{X^x_t<K\}}+(mX^x_t+(1-m)K)\1_{\{X^x_t\geq K\}}),
\end{array} \right.
\end{equation*}
{and define $M^k_\infty=\lim_{t\uparrow \infty} M_t^k$ for $k=1,2$.} Observe that, $\{M^1<M^2\}=\{X^x_.<K\}\subset\{M^2\leq U\}$, i.e. assumption (i) of Theorem \ref{theorem_value_2} is satisfied.

It remains to verify assumptions (ii) or (ii)'. Take $\eta=\infty$. Recall that $$\sigma^*_{sma},\sigma^*_{mod},\sigma^*_{lar}\leq T_{N_{K}}.$$ If $\mu\geq  \frac{\nu^2}{2}$, then $T_{N_{K}}<\infty$ a.s., therefore $\sigma^*_{sma},\sigma^*_{mod}$ and $\sigma^*_{lar}$ are all finite a.s., implying that assumption (ii)' holds. If $\mu< \frac{\nu^2}{2}$, then $e^{-rt}X^x_t$ converges to $0$ almost surely as $t\rightarrow \infty$, therefore $M^1_\infty=M^2_\infty=P$ and assumption (ii) holds.

To finish the proof, observe that $R^{c}(\tau,\sigma)\geq R^{m'}(\tau,\sigma)$, from which trivially we have that
$v_{ca,c}^\lambda(x)\geq v_{ca,m'}^\lambda(x)$. Thus $v_{ca,m}^\lambda(x)\geq v_{ca,m'}^\lambda(x)$.
\end{proof}

{Intuition would suggest that, the bondholder's optimal stopping boundary should be nondecreasing in $m$, and the firm's optimal stopping boundary should be nonincreasing in $m$. This is confirmed in the following result.

\begin{proposition}
    Fix $0\leq m'<m\leq 1$, $\lambda>0$, $r>0$ and $\mu<\lambda+r$. Let $(\tau^*(m'), \sigma^*(m'))$ and $(\tau^*(m),\sigma^*(m))$ be the saddle points of the two games, defined in Theorem \ref{theorem:value}.

    Then we have $\tau^*(m')\leq \tau^*(m)$ and $\sigma^*(m')\geq \sigma^*(m)$.

\end{proposition}
\begin{proof}
We divide the analysis into several cases depending on the value of $K$. Recall that $\hat{K}(m)$ is increasing in $m$.

If $K\leq P$, then $\sigma^*(m')=\sigma^*(m)=T_1$ and $\tau^*(m')=\tau^*(m)=T_{N_K}$ by Theorem \ref{theorem:value}, so the result holds.

If $K\in(P,\hat{K}(m'))$, by Theorem \ref{theorem:value} we have $\sigma^*(m')=T_{N_{\xscal(m')}}$, $\sigma^*(m)=T_{N_{\xscal(m)}}$ and $\tau^*(m')=\tau^*(m)=T_{N_K}$. Observe from (\ref{eq:xscal}) that $\xscal(m)$ is decreasing in $m$, hence the result follows.

If $K\in[\hat{K}(m'),\hat{K}(m))$, then Theorem \ref{theorem:value} implies that $\tau^*(m')=T_{N_{\xscol(m')}}$ and $\sigma^*(m')=T_{N_K}$, while Theorem \ref{theorem:value} implies that $\tau^*(m)=T_{N_K}$ and $\sigma^*(m)=T_{N_{\xscal(m)}}$. Hence the result follows by Lemma \ref{lemma_fixedpoint} and \ref{lemma_xscal_mod}.

Finally, if $K\geq \hat{K}(m)$, then Theorem \ref{theorem:value} implies that $\tau^*(m')=T_{N_{\xscol(m')}}$,$\tau^*(m)=T_{N_{\xscol(m)}}$  and $\sigma^*(m')=\sigma^*(m)=T_{N_K}$. The result therefore follows by Remark \ref{remark:xscol}.
\end{proof}

}

\subsection{Asymptotics in the Poisson rate $\lambda$}
We analyse the asymptotic behaviour of the optimal stopping strategies and the value functions $v_{ca,m}^\lambda$ when we take $\lambda \rightarrow \infty$. Intuitively, they should converge to their counterpart without the liquidity constraint. We start by stating the value of the callable convertible bond without the liquidity constraint. Their proofs are standard and thus omitted.
\begin{proposition}\label{prop:noPoisson1}
Suppose that $m\in[0,1]$ and $\mu<r$. Let the set of admissible stopping times for the firm and the bondholder be the set of all  $\mathbb{F}^W-$stopping times. Define $x^*_{b}:=P\frac{\alpha}{\alpha-1}>0$, and let $H_y=\inf\{t\geq 0:X^x_t\geq y\}$. Then,

(i) If $K\leq P$, then $(\tau^*,\sigma^*)=(H_{K},0)$ is a saddle point and the value is $v_{ca,m}(x)=m\max\{ x,K\}+(1-m)K$.

(ii) If $K\in(P, x^*_{b})$, then $(\tau^*,\sigma^*)=(H_{K},H_{K})$ is a saddle point and the value is
\begin{equation}\label{eq:noPoissonmod}
    v_{ca,m}(x)=\left\{ \begin{array}{lll}P+\frac{x^\alpha}{K^\alpha}(K-P), & \; & x<K; \\
    m x+(1-m)K, & \; & x\geq K\\
\end{array}\right.
\end{equation}

(iii) If $K\geq  x^*_{b}$, then $(\tau^*,\sigma^*)=(H_{x^*_{b}},H_K)$ is a saddle point and the value is
\begin{equation}\label{eq:noPoissonlarge}
    v_{ca,m}(x)=\left\{ \begin{array}{lll}P+\frac{x^\alpha}{(x_{b}^*)^\alpha} (x^*_{b}-P), & \; & x<x^*_{b}; \\
     x, &\;& x^*_{b}\leq x< K;\\
m x+(1-m)K, & & x \geq  K.
\end{array}\right.
\end{equation}
\end{proposition}

%Now we consider the case $\mu\geq r$. By Lemma \ref{lemma_alphabounds}, $\alpha\leq 1$, therefore $\xsco<0$ (or equals $+\infty$ when $\mu=r$) and the bondholder will not stop before $H_{K}$.

\begin{proposition}\label{prop:noPoisson2}
Suppose that $m\in[0,1]$ and $\mu\geq r$. Let the set of admissible stopping times for firm and bondholder be the set of all $\mathbb{F}^W-$stopping times. Then,

(i) If $K\leq P$, then $(\tau^*,\sigma^*)=(H_{K},0)$ is a saddle point and the value is $v_{ca,m}(x)=m\max\{ x,K\}+(1-m)K$.

(ii) If $K\in(P,+\infty)$, then $(\tau^*,\sigma^*)=(H_{K},H_{K})$ is a saddle point and the value is
\begin{equation}
    v_{ca,m}(x)=\left\{ \begin{array}{lll}P+\frac{x^\alpha}{K^\alpha}(K-P), & \; & x<K; \\
    m x+(1-m)K, & \; & x\geq K.\\
\end{array}\right.
\end{equation}
\end{proposition}

{Now we turn to the problem with the liquidity constraint.}
We need the following auxiliary results for our analysis, where we will use notations such as $\hat{m}(\lambda)$, $\hat{K}(m,\lambda)$  and $\xscol(m,\lambda)$ to represent the choice of parameter $\lambda$:

\begin{lemma}\label{lemma:asymptoticlambda}
The following holds as $\lambda\rightarrow \infty$:

(i) $\al,\bl=\mathcal{O}(\sqrt{\lambda})$, where $f(\lambda)=\mathcal{O}(g(\lambda))$ (or $f(\lambda)$ is of order $g(\lambda)$) if there exists some $M> 0$ such that $|f(\lambda)|\leq Mg(\lambda)$ holds for sufficiently large $\lambda$.

    (ii) $\hat{m}(\lambda)\rightarrow \alpha$.

\end{lemma}
\begin{proof}
The results follow from the corresponding expressions.
\end{proof}

\begin{lemma}\label{lemma:asymptoticKhat}
    The following results hold as $\lambda\rightarrow \infty$:

(i) If $\mu<r$ and $m=0$, then $\hat{K}(0,\lambda)=\hat{K}(0)=P$.

(ii) If $\mu<r$ and $m\in(0,1]$, then $\hat{K}(m,\lambda)$ is eventually increasing in $\lambda$ with limit $P\frac{\alpha}{\alpha-m}$.

(iii) If $\mu=r$, then $\hat{K}(m,\lambda)=P\frac{1}{1-m}$, where we abuse the notation when $m=1$.

(iv) If $\mu>r$ and $m\in [ \alpha,1]$, then $\hat{K}(m,\lambda)=+\infty$ for sufficiently large $\lambda$.

(v) If $\mu>r$ and $m\in[0,\alpha)$, then $\hat{K}(m,\lambda)$ is eventually decreasing in $\lambda$ with limit $P\frac{\alpha}{\alpha-m}$.
\end{lemma}

\begin{proof}
See appendix.
\end{proof}

\begin{lemma}\label{Lemma:xstarlimit}
    The following results hold as $\lambda\rightarrow \infty$:

    (i) Suppose that $m\in[0,\alpha)$ and either $\mu>r$, $K>P\frac{\alpha}{\alpha-m}$ or $\mu=r$, $K\geq \frac{\alpha}{\alpha-m}$.  Then $\xscol(m)\rightarrow K$ as $\lambda \rightarrow \infty$.

    (ii) Suppose that $\mu<r$,  $m\in[0,1)$ and $K\in [P\frac{\alpha}{\alpha-m},  x^*_b]$. Then   $\xscol(m,\lambda)\rightarrow K$ as $\lambda \rightarrow \infty$.

    (iii) Suppose that $\mu<r$, $m\in[0,1)$ and $K>x^*_b>P\frac{\alpha}{\alpha-m}$. Then   $\xscol(m,\lambda)\rightarrow x^*_b$ as $\lambda \rightarrow \infty$.

\end{lemma}
\begin{proof}
See appendix.
\end{proof}
%\ECW{Gechun suggested that this should be a lemma instead.}

{

We are ready to state the convergence result for the value function $v_{ca,m}^\lambda$.
\begin{proposition}\label{Prop_convergence}
The value of the convertible bond with the liquidity constraint, $v_{ca,m}^\lambda$, converges to its counterpart without liquidity constraint as $\lambda \rightarrow \infty$.
\end{proposition}
\begin{proof}
It suffices to consider the limit of value functions defined in Proposition \ref{prop_bondholder_original}, \ref{prop_bondholder_1}, \ref{prop_firm_mod} and \ref{lemma_firm_small}, depending on the parameters $\mu,r,m$ and $K$.

If $K\leq P$ and $\lambda$ is sufficiently large, then the assumptions in Proposition \ref{lemma_firm_small} are eventually satisfied, regardless of the value of fixed parameters $\mu,r,m$. By Lemma \ref{lemma:asymptoticlambda}, it is immediate that $A_{sma},B_{sma}\rightarrow 0$ as $\lambda\rightarrow \infty$, which then implies that $v_{ca,m}^\lambda(x)\rightarrow v_{ca,m}(x)$.

If any of the following hold:
\begin{itemize}
    \item $\mu\leq r$, $K\in(P,P\frac{\alpha}{\alpha-m})$;
    \item $\mu>r$, $m\geq \alpha$;
    \item $\mu>r$, $m<\alpha$, $K\in(P,P\frac{\alpha}{\alpha-m}]$,
\end{itemize}
then the assumptions in Proposition \ref{prop_firm_mod} are eventually satisfied by Lemma \ref{lemma:asymptoticKhat}. It follows by Lemma \ref{lemma:asymptoticlambda} that $A_{mod},B_{mod}\rightarrow 0$ and $\xscal(m)\rightarrow K$, which then imply $C_{mod}(m)\rightarrow 0$. Hence $v_{ca,m}^\lambda(x)\rightarrow v_{ca,m}(x)$.

If $\mu<r$, $m\in[0,1)$ and $K\in[P\frac{\alpha}{\alpha-m},x^*_b]$, then the assumptions in Proposition \ref{prop_bondholder_1} are eventually satisfied by Lemma \ref{lemma:asymptoticKhat}. By Lemma \ref{lemma:asymptoticlambda}, $A_{lar}(m),B_{lar}(m), C_{lar}(m)\rightarrow 0$. By Lemma \ref{Lemma:xstarlimit}, $\xscol(m)\rightarrow K$. Therefore, in this case, $v_{ca,m}^\lambda(x)\rightarrow v_{ca,m}(x)$, where $v_{ca,m}(x)$ is defined in (\ref{eq:noPoissonmod}). Note that, if $K=x^*_b$, then the value functions (\ref{eq:noPoissonmod}) and (\ref{eq:noPoissonlarge}) agree with each other.

If $\mu<r$, $m=1$, $K\geq x^*_b$, then the assumptions in Proposition \ref{prop_bondholder_original} are satisfied by Lemma \ref{lemma:asymptoticKhat}, hence $v_{ca,1}^\lambda(x)\rightarrow v_{ca,m}(x)$ by Lemma \ref{lemma:asymptoticlambda}.

Finally, if any of the following holds:
\begin{itemize}
    \item $\mu<r$, $m\in[0,1)$, $K>x^*_b$;
    \item $\mu=r$, $m\in[0,1)$, $K\geq P\frac{\alpha}{\alpha-m}$;
    \item $\mu>r$, $m\in[0,\alpha)$, $K>P\frac{\alpha}{\alpha-m}$,
\end{itemize}
then the assumptions in Proposition \ref{prop_bondholder_1} are satisfied by Lemma \ref{lemma:asymptoticKhat}. We still have $A_{lar}(m),B_{lar}(m), C_{lar}(m)\rightarrow 0$ by Lemma~\ref{lemma:asymptoticlambda}. If $\mu<r$, then $\xscol(m)\rightarrow x^*_b$ by Lemma \ref{Lemma:xstarlimit}, hence $v_{ca,m}^\lambda(x)\rightarrow v_{ca,m}(x)$ which is defined in (\ref{eq:noPoissonlarge}). If $\mu\geq r$, then $\xscol(m)\rightarrow K$ by Lemma \ref{Lemma:xstarlimit} and we conclude.
\end{proof}

}

\bibliographystyle{siamplain}
%\bibliography{callablereference}

%%%%%%%%%%%%%%%%%%

%%%%%%%%%%%%%%

\appendix
\section{Proofs}

%\end{appendix}

[Proof of Lemma \ref{lemma_alphaorder_mu}]
    Firstly we consider the sign of $\alpha \frac{\lambda+r-\mu}{r-\mu}-\frac{\lambda}{r-\mu}$.  If $\mu<r$, then $\alpha>1$ by definition, hence $\alpha(\lambda+r-\mu)-\lambda> (\lambda+r-\mu)-\lambda>0$. If $\mu\in(r,\infty)$, then $\alpha<1$, hence  $\alpha(\lambda+r-\mu)-\lambda< (\lambda+r-\mu)-\lambda<0$, but the denominator $r-\mu$ is also negative. Hence $\alpha \frac{\lambda+r-\mu}{r-\mu}-\frac{\lambda}{r-\mu}>0$, and it therefore suffices to check the sign of $Q_\lambda(\alpha \frac{\lambda+r-\mu}{r-\mu}-\frac{\lambda}{r-\mu})$.
{After some algebra and the use of $Q(\alpha)=0$, we find
\[ Q_\lambda\left(\alpha \frac{\lambda+r-\mu}{r-\mu}-\frac{\lambda}{r-\mu}\right) = \frac{1}{2}\nu^2\lambda \frac{\lambda+r-\mu}{(r-\mu)^2}\left(\alpha^2-2\alpha+1\right). \]}
\begin{comment}
\begin{eqnarray*}
&&Q_\lambda\left(\alpha \frac{\lambda+r-\mu}{r-\mu}-\frac{\lambda}{r-\mu}\right)\\&=&\frac{1}{2}\nu^2\left(\alpha \frac{\lambda+r-\mu}{r-\mu}-\frac{\lambda}{r-\mu}\right)^2+\left(\mu-\frac{1}{2}\nu^2\right)\left(\alpha \frac{\lambda+r-\mu}{r-\mu}-\frac{\lambda}{r-\mu}\right)-(\lambda+r)\\
&=&\frac{1}{2}\nu^2\alpha^2+\left(\mu-\frac{1}{2}\nu^2\right)\alpha-r\\&&+\frac{1}{2}\nu^2\alpha^2\frac{\lambda^2+2\lambda(r-\mu)}{(r-\mu)^2}-\frac{1}{2}\nu^2\alpha\frac{2(\lambda+r-\mu)\lambda}{(r-\mu)^2}+\frac{1}{2}\nu^2\frac{\lambda^2}{(r-\mu)^2}\\
&&+\left(\mu-\frac{1}{2}\nu^2\right)\alpha \frac{\lambda}{r-\mu}-\mu\frac{\lambda}{r-\mu}+\frac{1}{2}\nu^2\frac{\lambda}{r-\mu}-\lambda
\\&=&\frac{\lambda}{r-\mu}\left(\frac{1}{2}\nu^2\alpha^2\frac{2(r-\mu)+\lambda}{r-\mu}+\left(\mu-\frac{1}{2}\nu^2\right)\alpha-(r-\mu)-\mu\right)\\
&&+\frac{\lambda}{r-\mu}\left(\frac{1}{2}\nu^2\left(\frac{\lambda}{r-\mu}-2\alpha\frac{\lambda+r-\mu}{r-\mu}+1\right)\right)\\
&=&\frac{\lambda}{r-\mu}\left(\frac{1}{2}\nu^2\left(\alpha^2\frac{\lambda+r-\mu}{r-\mu}+\frac{\lambda}{r-\mu}-2\alpha\frac{\lambda+r-\mu}{r-\mu}+1\right)\right)\\
&=&\frac{1}{2}\nu^2\lambda \frac{\lambda+r-\mu}{(r-\mu)^2}\left(\alpha^2-2\alpha+1\right),
\end{eqnarray*}
where the third and fourth equality follows by $Q(\alpha)=0$.
\end{comment}
It is clear that $Q_\lambda\left(\alpha \frac{\lambda+r-\mu}{r-\mu}-\frac{\lambda}{r-\mu}\right)$ is positive if $\mu<\lambda+r$, and is negative if $\mu>\lambda+r$. Hence the conclusion follows.
$\square$

[Proof of Lemma \ref{lemma_mhat_alpha}]
    First observe that $\lambda+r-\mu\bl>0$. Indeed, this is immediate if $\mu\geq 0$ and follows by Lemma \ref{lemma_alphabounds} if $\mu<0$. Also recall that the order between $\alpha$ and $1$ in Lemma \ref{lemma_alphabounds}. Hence, to prove (1), it suffices to check the sign of $(\hat{m}-\alpha)\lambda(\lambda+r-\mu\bl)=\alpha(\lambda+r)(r-\mu)+\bl(\alpha \mu\lambda+(\mu-\lambda-r)r)$.

If $\mu\leq 0$, then it is immediate that both $\alpha(\lambda+r)(r-\mu)$ and $\bl(\alpha \mu\lambda+(\mu-\lambda-r)r)$ are strictly positive. If $\mu\in(0,r)$, then $\alpha<\frac{r}{\mu}$ by Lemma \ref{lemma_alphabounds}, and hence $\bl(\alpha \mu\lambda+(\mu-\lambda-r)r)>\bl(\mu-r)r>0$, which implies $\hat{m}>\alpha$.

If $\mu=r$, then $\alpha=1$ by Lemma \ref{lemma_alphabounds}, hence $\alpha \mu \lambda+(\mu-\lambda-r)r=0$ and it follows that $\hat{m}=\alpha=1$. If $\mu>r$, then $\alpha>\frac{r}{\mu}$ by Lemma \ref{lemma_alphabounds}, hence $\bl(\alpha \mu\lambda+(\mu-\lambda-r)r)<\bl(\mu-r)r<0$, which then implies $\hat{m}<\alpha$. The proof of (1) is now complete.

Finally, observe that $\alpha(\lambda+r)-\bl r>0$. Therefore the sign of $\hat{m}$ is equivalent to that of $\lambda+r-\mu$, which is exactly (2).
$\square$

[Proof of Lemma \ref{lemma_fixedpoint}]
We divide the analysis into four cases depending on value of $\mu$.

 If $\mu<r$, then by Lemma \ref{lemma_flm_domain} the function $f^m(x)$ is decreasing on $[0,P\frac{\hat{m}}{\hat{m}-1}]$ with $f^m(0)>0$ and $f^m(P\frac{\hat{m}}{\hat{m}-1})=0$, so it is immediate that $f^m$ has a unique fixed point on $[0,P\frac{\hat{m}}{\hat{m}-1}]$.
 %If $K\geq P\frac{\hat{m}}{\hat{m}-1}$, then $[0,P\frac{\hat{m}}{\hat{m}-1}]\subset [0,K]$ and this unique fixed point must be in $[0,K]$. Therefore we assume $K<P\frac{\hat{m}}{\hat{m}-1}$.
As $f^m$ is decreasing, a sufficient and necessary condition for the fixed point to be in $[0,K]$ is $f^m(K)\leq K$. Observe that,
\begin{equation}\label{flm_rearrange_1}
    f^m(K)=K\left(\frac{1}{1-m}\left(\frac{P}{K}\hat{m}-\hat{m}+1\right)\right)^{1/\al}.
\end{equation}
Therefore, $f^m(K)\leq K$ if and only if  $\frac{1}{1-m}(\frac{P}{K}\hat{m}-\hat{m}+1)\leq 1$. By Lemma \ref{lemma_mhat_alpha}, $\hat{m}>1$, hence $\frac{1}{1-m}(\frac{P}{K}\hat{m}-\hat{m}+1)\leq 1$ is equivalent to $K\geq P\frac{\hat{m}}{\hat{m}-m}=\hat{K}(m)$. In particular, the fixed point is exactly at $K$ if and only if $\frac{1}{1-m}(\frac{P}{K}\hat{m}-\hat{m}+1)= 1$, and this is equivalent to $K=\hat{K}(m)$.

If $\mu=r$, then $f^m$ is a constant function by Lemma \ref{lemma_flm_domain} with $$f^m(x)=K\frac{1}{(1-m)^{1/\al}}\frac{P^{1/\alpha_\lambda}}{K^{1/\alpha_\lambda}},$$ therefore a necessary and sufficient condition for the fixed point to be in the interval $[0,K]$ is $K\frac{1}{(1-m)^{1/\al}}P^{1/\alpha_\lambda}{K^{1/\alpha_\lambda}}\leq K$, which is equivalent to $K\geq P\frac{1}{1-m}=\hat{K}(m)$.

If $\mu\in(r,\lambda+r)$, then $\al>1$ by Lemma \ref{lemma_alphabounds}, which implies $f^m(x)-x\rightarrow -\infty$ as $x\rightarrow \infty$ and that $f^m$ is a concave function. Also, $f^m(0)>0$ by Lemma \ref{lemma_flm_domain}.  Hence, there exists a unique zero point on the positive real line for the function $f^m(x)-x$ and, therefore $f^m$ has a unique fixed point. A  sufficient and necessary condition for the fixed point to be in $[0,K]$ is that $f^m(K)\leq K$, which is equivalent to $\hat{m}(1-\frac{P}{K})\geq m$. If $m\geq \hat{m}$, this inequality always fails, and this agrees with the statement in the lemma as $\hat{K}(m)=\infty$ in such a case. If $m<\hat{m}$, then we can conclude by a similar argument as the $\mu<r$ case.

If $\mu>\lambda+r$, then $\al<1$ by Lemma \ref{lemma_alphabounds}, which implies $f^0(x)-x\rightarrow +\infty$ as $x\rightarrow \infty$ and that $f^m$ is convex on $[P\frac{\hat{m}}{\hat{m}-1},\infty]$. Also, observe that $f^0(P\frac{\hat{m}}{\hat{m}-1})-P\frac{\hat{m}}{\hat{m}-1}<0$ by Lemma \ref{lemma_flm_domain}. Hence, by a similar argument as in the $\mu\in(r,\lambda+r)$ case, $f^0$ has a unique fixed point on $[P\frac{\hat{m}}{\hat{m}-1},\infty)$, and the fixed point is in $[0,K]$ if and only if $f^0(K)\geq K$. The remaining argument is similar to the cases $\mu<r$ and $\mu\in(r,\lambda+r)$ and thus omitted. $\square$

[Proof of Lemma \ref{lemma_integrability}]
    For fixed x, let $S_t:=e^{-rt}X^x_t$. The process $S$ is a geometric Brownian motion with drift $\mu-r$ and volatility $\nu$, and it suffices to prove that $S_{T_{N_y}}$ is integrable.

We aim to find a suitable upper bound for $\mathbb{E}[S_{T_{N_y}}\1_{\{N_y=k\}}]$. By the assumption of the lemma, $S_{T_1}$ is integrable, so it suffices to consider $k\geq 2$.

For each $k\geq 2$, we define $W^k_t(\omega):=W_{T_{k-1}+t}(\omega)-W_{T_{k-1}}(\omega)$ for any $\omega$. By the strong Markov property, $W^k$ is a Brownian motion conditioning on the set $\{T_{k-1}<\infty\}$, with $W^k_0=0$, and is independent of the Poisson process and $W_{T_{k-1}}$.

On the set $\{S_{T_{k-1}}<y\}$, we have
\begin{eqnarray*}
    S_{T_{k-1}+t}&=&S_{T_{k-1}}\exp\left((\mu-r-\frac{\nu^2}{2})t+\nu(W_{T_{k-1}+t}-W_{T_{k-1}})\right)\\ &=& S_{T_{k-1}}\exp\left((\mu-r-\frac{\nu^2}{2})t+\nu W^k_t\right)\\ &\leq&  y\exp\left((\mu-r-\frac{\nu^2}{2})t+\nu
 W^k_t\right):=S^k_t.
\end{eqnarray*}

By taking $t=T_k-T_{k-1}$, we can conclude that $S_{T_k}\1_{\{S_{T_1}<y,...,S_{T_{k-1}}<y,S_{T_k}\geq y\}}\leq S^k_{T_k-T_{k-1}}\1_{\{S_{T_1}<y,...,S_{T_{k-1}}<y,S_{T_k}\geq y\}}$ holds $\omega$-wise. Therefore:
\begin{eqnarray*}
    {\mathbb{E}[S_{T_{N_y}}\1_{\{N_y=k\}}]}&=&\mathbb{E}[S_{T_k}\1_{\{S_{T_1}<y,...,S_{T_{k-1}}<y,S_{T_k}\geq y\}}]
    \\    &\leq & \mathbb{E}[S^k_{T_k-T_{k-1}}\1_{\{S_{T_1}<y,...,S_{T_{k-1}}<y,S_{T_k}\geq y\}}]\\
    &\leq & \mathbb{E}[S^k_{T_k-T_{k-1}}\1_{\{S_{T_1}<y,...,S_{T_{k-1}}<y,S^k_{T_k-T_{k-1}}\geq y\}}]\\
    &=&\mathbb{E}[S^k_{T_k-T_{k-1}}\1_{\{S^k_{T_k-T_{k-1}}\geq y\}}]\mathbb{P}(S_{T_1}<y,...,S_{T_{k-1}}<y)\\
    &\leq&\mathbb{E}[S^k_{T_k-T_{k-1}}\1_{\{S^k_{T_k-T_{k-1}}\geq y\}}]\mathbb{P}(S_{T_1}<y)(\mathbb{P}^y(S_{T_1}<y))^{k-2},
\end{eqnarray*}
where the second equality follows by independence.

Note that, for any $k\geq 2$, $S^k$ is a geometric Brownian motion with the same drift and volatility as $S$, but with $S^k_0=y$. Therefore, $(S^k_{T_k-T_{k-1}})_k$ are identically distributed, and are integrable by the assumption that $\mu<\lambda+r$. The expectation $\mathbb{E}[S^k_{T_k-T_{k-1}}\1_{\{S^k_{T_k-T_{k-1}}\geq y\}}]$ is therefore the same for every $k$ and the sum $\mathbb{E}[S_{T_{N_y}}]$ is therefore finite. As $S_{T_{N_y}}$ is nonnegative, the proof of integrability is therefore complete.
$\square$

[Proof of Lemma \ref{lemma:asymptoticKhat}]
(i) and (iii) follow from the definition of $\hat{K}$. We need to consider the asymptotic behaviour of $\hat{m}(\lambda)$ for (ii),(iv) and (v).

Firstly, observe that $\bl-(\lambda+r)\frac{d\bl}{d\lambda}=\mathcal{O}(\sqrt{\lambda})$ and is negative.

This implies the following:
\begin{enumerate}
    \item If $\mu<r$, then $\hat{m}(\lambda)$ is eventually decreasing with respect to $\lambda$.
    \item If $\mu>r$, then $\hat{m}(\lambda)$ is eventually increasing with respect to $\lambda$.
\end{enumerate}
Indeed, we have:
{\footnotesize{
\begin{equation}\label{hatm2}\lambda^2(\lambda+r-\mu\bl)^2\frac{d\hat{m}}{d\lambda}=\lambda(\lambda+r-\mu)(r-\alpha \mu)(\bl-(\lambda+r)\frac{d\bl}{d\lambda})-(r-\mu)(\lambda+r-\mu\bl)(\alpha(\lambda+r)-\bl r).\end{equation}
}
}
The first term in the RHS of (\ref{hatm2}) is of order $\lambda^{2.5}$ and the second term is only of order $\lambda^2$, so the sign of $\frac{d\hat{m}}{d\lambda}$ agrees with the sign of $\alpha\mu-r$ when $\lambda$ is sufficiently large. Hence the property of $\hat{m}(\lambda)$ follows by Lemma \ref{lemma_alphabounds}.

Under the assumptions in (ii), $\hat{m}(\lambda)$ is increasing, therefore so is $\hat{K}(m,\lambda)=P\frac{\hat{m}(\lambda)}{\hat{m}(\lambda)-m}$, and the limit follows by Lemma~\ref{lemma:asymptoticlambda}; under the assumptions in (iv), eventually we have $m\geq \alpha>\hat{m}(\lambda)$, $\hat{K}(m,\lambda)=+\infty$; $m<\hat{m}(\lambda)$ holds eventually under the assumptions in $(v)$, hence $\hat{K}(m,\lambda)=P\frac{\hat{m}(\lambda)}{\hat{m}(\lambda)-m}$, which is therefore increasing and tends to $P\frac{\alpha}{\alpha-m}$.
$\square$

[Proof of Lemma \ref{Lemma:xstarlimit}] Lemma \ref{lemma:asymptoticKhat} guarantees that, under the assumptions in (i)-(iii), $\xscol(m,\lambda)$ is eventually well-defined as $K$ is always `large'.

Proof of (i): Recall that, $f^m$ is an increasing function in $x$ by Lemma \ref{lemma_flm_domain} for $r\leq \mu<\lambda+r$.  As a result, $\xscol(m,\lambda)=f^m(\xscol(m,\lambda),\lambda)\geq f^m(0,\lambda)$. By Lemma \ref{lemma:asymptoticlambda}, $f^m(0,\lambda)\rightarrow {K}$ as $\lambda\rightarrow \infty$. On the other hand, $\xscol(m,\lambda)\leq K$ for sufficiently large $\lambda$ by Lemma \ref{lemma_fixedpoint}. Therefore the conclusion follows by the sandwich rule.

To prove (ii) and (iii), observe that $\xscol(m,\lambda)$ is eventually increasing under the assumption that $\mu<r$, $m\in[0,1)$ and $K\geq P\frac{\alpha}{\alpha-m}$. Indeed, by taking sufficiently large $\lambda_1>\lambda_2$ such that $\hat{m}(\lambda)$ is decreasing on $[\lambda_2,\infty)$, we have:
\begin{eqnarray*}
    f^m(\xscol(m,\lambda_1),\lambda_1)&=&K\left(\frac{1}{1-m}\left(\frac{P-\xscol(m,\lambda_1)}{K}\hat{m}(\lambda_1)+\frac{\xscol(m,\lambda_1)}{K}\right)\right)^{1/\alpha_{\lambda_1}}\\
    &\geq&K\left(\frac{1}{1-m}\left(\frac{P-\xscol(m,\lambda_1)}{K}\hat{m}(\lambda_1)+\frac{\xscol(m,\lambda_1)}{K}\right)\right)^{1/\alpha_{\lambda_2}}\\
    &\geq&K\left(\frac{1}{1-m}\left(\frac{P-\xscol(m,\lambda_1)}{K}\hat{m}(\lambda_2)+\frac{\xscol(m,\lambda_1)}{K}\right)\right)^{1/\alpha_{\lambda_2}}\\
    &=&f^m(\xscol(m,\lambda_1),\lambda_2),
\end{eqnarray*}
where the first inequality is by the result that $\xscol(m,\lambda)\leq K$ and $\alpha_{\lambda_1}>\alpha_{\lambda_2}$, and the second inequality is by the result that $\xscol(m,\lambda)>P$. Therefore, we have that $\xscol(m,\lambda_1)\geq f^m(\xscol(m,\lambda_1),\lambda_2)$. Given that $f^m$ is decreasing in $x$ and has a unique fixed point, $\xscol(m,\lambda_1)\geq\xscol(m,\lambda_2)$ holds.

Proof of (ii): By Lemma \ref{lemma_fixedpoint}, $\xscol(m,\lambda)$ is bounded above by $K$ and eventually increasing, hence it must converge to some $x_\infty\leq K$, with  $\xscol(m,\lambda)\leq x_\infty$ holds eventually. Hence, by the fact that $f^m$ is decreasing with respect to $x$, the following holds eventually:
$$f^m(K,\lambda)\leq f^m(\xscol(m,\lambda),\lambda)=\xscol(m,\lambda)\leq K.$$
 Observe that, under the assumption $K\leq x^*_b$, $f^m(K,\lambda)\in[0,\infty)$ and converges to $K$ as $\lambda\rightarrow \infty$, hence the claim holds by the sandwich rule.

Proof of (iii): Recall that, under the assumption $\mu<r$, $\xscol(m,\lambda)\leq P\frac{\hat{m}(\lambda)}{\hat{m}(\lambda)-1}$ by definition of $f^m$. By the result that $\hat{m}(\lambda)$ is eventually decreasing and $\hat{m}(\lambda)\geq \alpha$,  $P\frac{\hat{m}(\lambda)}{\hat{m}(\lambda)-1}\leq P\frac{\alpha}{\alpha-1}$. As a result, eventually, $\xscol(m,\lambda)\leq P\frac{\alpha}{\alpha-1} $ and $\xscol(m,\lambda)$ is increasing with respect to $\lambda$, so $\xscol(m,\lambda)$ must converge to some $x_\infty\leq P\frac{\alpha}{\alpha-1}$.

To seek a contradiction, assume that $x_\infty<P\frac{\alpha}{\alpha-1}$. Then, by the result that $f^m(x,\lambda)$ is decreasing with respect to $x$ on $[0,P\frac{\alpha}{\alpha-1}]$, the following inequality holds for sufficiently large $\lambda$:
$$x_\infty\geq \xscol(m,\lambda)=f^m(\xscol(m,\lambda),\lambda)\geq f^m(x_\infty,\lambda).$$
However, if we then take $\lambda \rightarrow \infty$, given that $x_\infty<P\frac{\alpha}{\alpha-1} $, $f^m(x_\infty,\lambda)\rightarrow K>P\frac{\alpha}{\alpha-1}$, which leads to a contradiction. Hence $x_\infty=P\frac{\alpha}{\alpha-1}$ must hold.
$\square$
%\bibliographystyle{unsrtnat}

%%%%%%%%%%%%%%%%%%%%%%%%%%%%%%%%%%%%%

\end{document}